\documentclass[twocolumn,amsmath,amssymb,longbibliography]{revtex4-1}
\usepackage{dcolumn}% Align table columns on decimal point
\usepackage{bm}% bold math
\usepackage{physics}
\usepackage{graphicx,color}% Include figure files
\usepackage{bbm}% bold math
\usepackage{amsmath,amssymb,mathrsfs}
\usepackage{url}
\usepackage{hyperref}
\newcommand{\R}{\mathbb{R}}

\newcommand{\set}[1]{\mathsf{#1}}

\newcommand{\spc}[1]{\mathcal{#1}}

\def\>{\rangle}
\def\<{\langle}
\def\kk{\>\!\>}
\def\bb{\<\!\<}

\newcommand{\map}[1]{\mathcal{#1}}
\newcommand{\St}{{\mathsf{St}}}

\newcommand{\Pur}{{\mathsf{Pur}}}

\newcommand{\Chan}{{\mathsf{Chan}}}

\newcommand{\Map}{{\mathsf{Map}}}
\newcommand{\TS}{{\mathsf{TS}}}
\newcommand{\TP}{{\mathsf{TP}}}

\newcommand{\Choi}{{\mathsf{Choi}}}
\newcommand{\Op}{{\mathsf{Op}}}

\newcommand{\hS}{{\widehat {\map S} }}

\newtheorem{theo}{Theorem}

\newtheorem{lemma}{Lemma}
\newtheorem{prop}{Proposition}
\newtheorem{cor}{Corollary}
\newtheorem{defi}{Definition}

\def\Proof{{\bf Proof.~}}
\def\qed{$\blacksquare$ \newline}

\usepackage{qcircuit}

\begin{document}

%\preprint{}

%   \title{Symmetries of quantum channels and the impossibility of a  universal time reversal}

%   \title{Channel automorphisms and the impossibility of a universal time reversal of open quantum systems as a linear operation}

%\title{Symmetries of quantum processes and the impossibility of a linear time reversal}

\title{{Symmetries of quantum evolutions}} %A Wigner theorem for quantum evolutions}

    \author{Giulio Chiribella}
\email{giulio@cs.hku.hk}
    \affiliation{QICI Quantum Information and Computation Initiative, Department of Computer Science, The University of Hong Kong, Pok Fu Lam Road, Hong Kong}
    \affiliation{Department of Computer Science, University of Oxford, Wolfson Building, Parks Road, Oxford, United  Kingdom}
    \affiliation{Perimeter Institute for Theoretical Physics, 31 Caroline Street North, Waterloo,  Ontario, Canada}

%\title{Time reversal of Open Quantum Systems: Algebraic Approach}% Force line %breaks with \\

%\author{The Collaboration}
%\affiliation{KTH, Stockholm University, Jagiellonian University \& Stockholm University}

\author{Erik Aurell}
\email{eaurell@kth.se}
\affiliation{KTH -- Royal Institute of Technology, AlbaNova University Center, SE-106 91 Stockholm, Sweden}%
\author{Karol \.{Z}yczkowski}
\email{karol.zyczkowski@uj.edu.pl}
\affiliation{Institute of Theoretical Physics, Jagiellonian University, 30-348 Krak\'ow, Poland
}
\affiliation{Center for Theoretical Physics, Polish Academy of Science,
Al Lotnik{\'o}w 32/44,  02-668 Warszawa, Poland}

\date{\today}% It is always \today, today,

\begin{abstract}
A cornerstone of quantum mechanics is the characterisation of  symmetries  provided by Wigner's theorem. Wigner's theorem establishes that every symmetry of the quantum state space must be either a unitary transformation, or an antiunitary transformation. Here we extend Wigner's theorem from quantum states to quantum evolutions, including both the deterministic evolution associated with the dynamics of closed systems, and the stochastic evolutions associated with the outcomes of quantum measurements. We prove that every symmetry  of the space of quantum evolutions can be decomposed into two  state space symmetries that are either both unitary or both antiunitary.  Building on this result, we show that it is impossible to extend the time reversal symmetry of unitary quantum dynamics to a symmetry of the full set of quantum evolutions. Our no-go theorem implies that any time symmetric formulation of quantum theory must either restrict the set of the allowed evolutions, or modify the operational interpretation of quantum states and processes.  Here we propose a time symmetric formulation of quantum theory where  the allowed  quantum evolutions are restricted to a suitable set, which includes  both unitary evolution and projective  measurements, but excludes the deterministic preparation of pure states. The standard  operational formulation of quantum theory can  be retrieved from this time symmetric version by introducing an operation of conditioning on the outcomes of past experiments.
% thus resulting into a  pre-selection of quantum systems that breaks the time symmetry.  
\end{abstract}

%\pacs{03.67.Lx, 42.50.Dv}% PACS, the Physics and Astronomy
\maketitle

\section{Introduction}

 Symmetries  play a central role in the modern approach to quantum mechanics~\cite{weyl1950theory,wigner1959group}. 
  They  provide   powerful methods for solving  problems in atomic physics, condensed matter, high energy physics, and quantum information science~\cite{cornwell1997group,tasaki2020physics,hayashi2017group}. In addition, they  offer guidance to the construction of new physical theories, such as Yang-Mills theories and other theories of fundamental interactions~\cite{weinberg1995quantum}.      The starting point of all these investigations is Wigner's theorem~\cite{wigner1959group,Wi60},  which characterises the symmetries of the state space of any given quantum system.  

 In general, a symmetry is a one-to-one transformation that preserves a certain structure.  For example, 
%rotations about the centre of a sphere are symmetries,  because they preserve its shape. 
canonical transformations are symmetries in Hamiltonian mechanics, because they preserve the form of the Hamilton's equations.   
Wigner's theorem refers to the symmetries of the set of quantum states of a given quantum system. The structure of interest here is the probabilistic structure of quantum theory: when a symmetry is applied, the statistics of measurement outcomes should not change.  

The original formulation of Wigner's theorem refers to  pure states.   Wigner established  that every symmetry of the set of pure states 
%Wigner's result is that every  transformation that preserves the outcome probabilities of quantum measurements
 can be represented by a transformation  that is either unitary or anti-unitary.  Wigner's theorem can be equivalently  formulated in terms of mixed states, represented by density matrices.  In this formulation, the theorem states that every symmetry of the set of density matrices  is either a unitary transformation, of the form  $\rho\mapsto U\rho U^\dag$ for some unitary operator $U$, or an antiunitary transformation, of the form $\rho  \mapsto  U\rho^T  U^\dag$, where $U$ is still a unitary operator, and $\rho^T$ is  the transpose of the density matrix $\rho$ with respect to a fixed but otherwise arbitrary basis.  
 %Operationally, the fundamental structure of the set of mixed states is the possibility of randomisation: for every pair of states $\rho$ and $\sigma$, and for every probability $p$,  the matrix $p\, \rho  +  (1-p)  \,\sigma$ represents a valid mixed state, resulting from the random choice of state $\rho$ with probability $p$, or state $\sigma$ with probability $1-p$.  Since  symmetries can be interpreted  as mere changes of description, it is  natural to require that any symmetry of the set of mixed states  be consistent with  randomisation:   if the symmetry transformation maps transforms $\rho$ into $\rho'$ and $\sigma$ into $\sigma'$, then it should transform the random mixture $p\, \rho  +  (1-p) \, \sigma$ into the random mixture $p\,\rho'  +  (1-p) \, \sigma'$.  Under this requirement,  every symmetry of the set of density matrices  must be either a unitary transformation, of the form  $\rho\mapsto U\rho U^\dag$ for some unitary operator $U$, or an antiunitary transformation, of the form $\rho  \mapsto  U\rho^T  U^\dag$, where $U$ is still a unitary operator, and $\rho^T$ is  the transpose of the density matrix $\rho$ with respect to a fixed (but otherwise arbitrary) basis. 
%    Mathematically, Wigner's theorem provides a characterisation of the automorphisms of the convex set of all density operators.   

In recent years, there has been a growing  interest in the extension of static notions, associated to  the quantum state space, to dynamical notions, associated to the space of  quantum evolutions.   A series of works characterised the possible transformations that map quantum evolutions into quantum evolutions, known as quantum supermaps~\cite{chiribella2008transforming,chiribella2009theoretical,chiribella2013quantum,chiribella2013normal,bisio2019theoretical}. A related notion   was explored in Ref.~\cite{Zy08}, which characterised the transformations of a certain set of  completely positive  maps other than the set of quantum evolutions.  Higher-order transformations were applied to the study of causality in quantum theory~\cite{chiribella2009beyond,oreshkov2012quantum,perinotti2017causal,castro2018dynamics},  to analyze extensions of quantum theory \cite{Zy08},
 quantum dynamical resource theories  \cite{rosset2018resource,ebler2018enhanced,gour2019quantify,wang2019resource,xu2019coherence,theurer2019quantifying,liu2020operational,takagi2020application,kristjansson2020resource,saxena2020dynamical}, and a variety of quantum information processing tasks, such as process tomography~\cite{bisio2009optimal}, cloning~\cite{chiribella2008optimal,bisio2011cloning}, learning~\cite{bisio2010optimal,sedlak2020probabilistic}, and other conversions of quantum evolutions~\cite{chiribella2016optimal,miyazaki2019complex,quintino2019probabilistic}. The lifting from states to evolutions has also led to new dynamical analogues of  the notions of quantum entanglement~\cite{gour2020dynamical,chen2020entanglement} and quantum coherence~\cite{dana2017resource,saxena2020dynamical}.

In this paper we characterise the dynamical symmetries of quantum theory by extending  Wigner's theorem from quantum states to quantum evolutions. 
%The most general evolutions allowed by quantum mechanics are the so-called {\em quantum operations}~\cite{kraus1983states,heinosaari2011mathematical}, which include both the deterministic evolution associated to the dynamics of closed systems, and the stochastic  evolutions associated to the outcomes of quantum measurements.  %, mathematically corresponding to completely positive, trace-non-increasing maps.    
The paper contains three main contributions, highlighted in the next section.  The first contribution is a complete characterisation of the symmetries of the space of quantum evolutions. We provide a Wigner-like theorem showing that every symmetry of quantum evolutions can be decomposed into two symmetries of quantum states, these two symmetries being either both unitary or both antiunitary. The second contribution is a rigorous proof that the set of all quantum evolutions admits no time reversal symmetry. The third contribution is a time symmetric variant of the quantum framework, obtained by restricting the set of allowed quantum evolutions to a suitable subset that preserve the maximally mixed state.  
% state preparations, evolutions, and measurements.  

The paper is organised as follows.  In  Sec.~\ref{sec:overview} we provide an overview of the main results of the paper. In   Sec. \ref{sec:symmetry} we review Wigner's theorem and provide its mixed state version. In  Sec. \ref{sec:symmetryQO} we introduce the notion of symmetry of quantum evolutions, and we derive a Wigner theorem for quantum evolutions.  In  Sec. \ref{sec:imp}, we show that the set of all quantum operations is incompatible with time symmetry.  A way around this no-go result is provided in  Sec.~\ref{sec:tsop}, where we propose time symmetric variant of quantum theory, obtained by restricting the set of allowed quantum evolutions.  Other ways around the no-go result  are discussed in  Sec.~\ref{sec:other}.   
Finally, the conclusions are provided in Sec.~\ref{sec:conclusion}. 

\section{Overview of the main results}\label{sec:overview}

\subsection{Wigner's theorem for quantum evolutions}
Our first contribution is a  complete characterisation of the symmetries of  the space  of quantum evolutions. The most general evolutions allowed by quantum mechanics are the so-called {\em quantum operations}~\cite{kraus1983states,heinosaari2011mathematical}, which include both the deterministic evolution associated to the dynamics of closed systems, and the stochastic  evolutions associated to the outcomes of quantum measurements.  %, mathematically corresponding to completely positive, trace-non-increasing maps.   
 Mathematically, a quantum operation  is a linear, completely positive map  $\map Q$ transforming density matrices on an input Hilbert space $\spc H_{\rm in}$ into (generally subnormalised) density matrices  on an output Hilbert space $\spc H_{\rm out}$. The map $\map Q$  satisfies the trace-non-increasing condition  
\begin{align}\label{qod}
\Tr [\map Q  (\rho)]  \le \Tr[\rho] \, ,
\end{align}  
for every operator $\rho$  acting on  $\spc H_{\rm in}$.  When $\rho$ is a density matrix, the trace $\Tr[\map Q  (\rho)]$ is interpreted as the probability that the quantum operation $\map Q$ takes place on the state $\rho$ in a suitable experiment.  If this probability is zero for every quantum state, then we say that the operation $\map Q$ is impossible.    
%If $\Tr[\map Q  (\rho)]=0$ for every density matrix $\rho$, we say that $\map Q$ is an impossible operation. 

%A limiting case of Eq.~(\ref{qod}) is the null operation $\map Q  = 0$, which never takes place on any input.   

A symmetry of the set of quantum operations is any one-to-one transformation   that preserves the probabilistic structure, by {\em(1)} mapping random mixtures of quantum operations into random mixtures with the same probabilities  %(mathematically, a symmetry $\map S$ should satisfy the condition  $\map S   (\sum_i \, p_i \,  \map Q_i)  =  \sum_i  p_i\, \map S (\map Q_i)$ for every set of quantum operations $(\map Q_i)$ and for every probability distribution $(p_i)$), 
and  {\em (2)}  mapping impossible operations into impossible operations.  %(mathematically, $\map S  (0)  =  0$).    

\begin{figure}
	\includegraphics[width=0.5\textwidth]{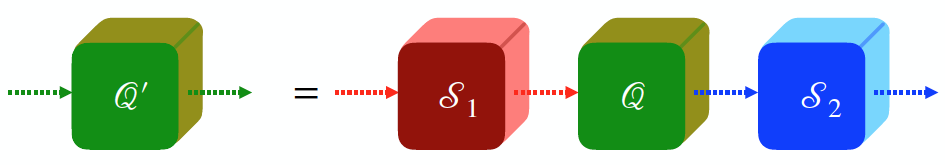}
	\caption{\label{fig:sandwich} {\bf Symmetry of  quantum evolutions.}  The symmetry transformation turns a given quantum evolution $\map Q$ into a new quantum evolution $\map Q'$, obtained by concatenating $\map Q$ with two transformations of quantum states (denoted by $\map S_1$ and $\map S_2$), which are either both unitary or both antiunitary.}
\end{figure}

 We show that the symmetries of the full set of quantum operations have a very rigid structure: every symmetry of quantum operations  can be broken down into two symmetries of quantum states,  one symmetry for the input system, and one symmetry for the output system, as illustrated in Figure~\ref{fig:sandwich}. 
The two symmetries must be either both unitary, or both antiunitary, while all the other  combinations are forbidden.

The Wigner theorem for quantum evolutions provides a rigorous framework for studying the  dynamical symmetries in quantum theory. The theorem   can be used to detect  the presence  or absence of certain symmetries, and  to guide the formulation of new variants of quantum theory exhibiting some desired symmetry features, such as {\em e.g.} time symmetry.   The other contributions of our paper explore these two directions.  
%~\cite{capellini2008}.    
 %A general process in quantum mechanics is described by a {\em quantum operation}~\cite{kraus1983states,heinosaari2011mathematical},  that is, a completely positive, trace non-increasing linear map transforming input density matrices into output density matrices.    Quantum operations  

%Our result shows that all the symmetries of quantum can be reduce  to static symmetries, that is, symmetries of the quantum state space.  

\subsection{No-go theorem for time reversal symmetry}

Our second result is a proof  that the set of all quantum evolutions is incompatible with time reversal symmetry.  

   Microscopic time reversal is a cornerstone
of quantum mechanics, and is
inherited from the charge-time-parity symmetry
of quantum field theory \cite{schwinger1951theory,luders1954equivalence,pauli1955niels}.  For closed quantum systems, the evolution is time symmetric,  in the sense that every unitary evolution $U$ admits  a time reversal  (for example, its inverse $U^\dag$ or its transpose $U^T$), which is also a valid quantum evolution.
  A fundamental question is whether the time reversal symmetry of unitary quantum dynamics can be extended to a symmetry of the  set of all possible quantum evolutions. 
  
 In physics,   symmetries are often viewed as changes of reference frame, and can be used to translate  a description of phenomena made by  one observer to a description of the same phenomena made of another observer \cite{bartlett2007reference}.     
   Operationally, the question of time reversal  is  whether it is possible to find a ``change of reference frame" that provides the description of a generic quantum evolution from the point of view of an observer whose  time coordinate $t$ is replaced by $-t$.   Of course, since most quantum evolutions are irreversible, the time reversal of a given evolution should not be expected to be a physical inversion.  It could be an  approximate inversion \cite{AZZ},
a quasi-inversion \cite{KBF20,SSMZK21},
 or, more generally, any evolution that is in one-to-one correspondence with the original evolution via a suitable change of description.  
 
 %In other words, the problem of time reversal is to find out whether there exists a one-to-one transformation of the set of quantum evolutions, with the property that such transformation coincides with the inverse in the special case of unitary evolutions. 
   
%The problem of time reversing stochastic
%differential equations, the equations
%of physical kinetics, was first posed
%by Schr\"odinger~\cite{Schr31,Aebi96}.
%It is well known that in this restricted
%class of classical diffusion equations
%there are many possible time reversals
%corresponding to different levels 
%of (classical) control~\cite{ChetriteGawedzki}. 
% In quantum information, the issue of finding  optimal approximate time reversals
%of a given quantum operation is deeply connected with tasks in   quantum error correction 
%\cite{Sh96,St96,BDSW97,KL97,KLV00,LB13,GDB20} and quantum Shannon theory~\cite{barnum2002reversing,hayden2004structure,ng2010simple,fawzi2015quantum,beigi2016decoding,junge2018universal}.

We  consider the set of quantum operations from a given quantum system to itself. In this case, the set of quantum operations includes  the set of all unitary transformations.  While the set of all unitary transformations is invariant under the standard time reversal transformation $U \mapsto  U^\dag$, we show that  such time reversal cannot be extended  to the whole set of quantum operations.    {   The same conclusion applies  for the transformation $U \mapsto  U^T$, which also cannot be extended to a transformation of the whole set of quantum operations.  
   
Our result goes beyond the mere observation that the definition of quantum operation (\ref{qod}) is time-asymmetric because it makes a distinction between the input and the output.   What we show is  that it is {\em in principle} impossible to  map the set of quantum operations into itself in a way that agrees with the transformation $U\mapsto  U^\dag$  (or with the transformation $U\mapsto  U^T$) for all unitary evolutions. }  

Our general no-go theorem highlights the time-asymmetric nature of the set of general quantum evolutions, and provides insights into the problem of time symmetry in quantum theory \cite{aharonov1964time,reznik1995time,aharonov2010time}, which has recently been the object of renewed interest in quantum foundations \cite{coecke2012time,oreshkov2015operational,leifer2017time,coecke2017time,di2020quantum,chiribella2020quantum,hardy2021time}.   Our contribution to the debate is a rigorous definition of time symmetry of quantum evolutions, and a proof that the set of all  quantum operations is incompatible with any notion of time symmetry that agrees with the  probabilistic structure of quantum theory.

\subsection{A time symmetric variant of quantum theory}
The last  contribution of this paper is  a candidate for a time symmetric variant of quantum theory. We propose a restricted set of quantum evolutions, named {\em time symmetric quantum operations}, which include  unitary dynamics, and also all projective  measurements described by von Neumann's and L\"uders' update rules.   Interestingly, our time symmetric quantum operations also include all the state transformations allowed by the consistent history framework for  closed quantum systems~\cite{griffiths2003consistent}.

Crucially, our time symmetric variant restricts the set of states that can be prepared deterministically:  while it allows to prepare every quantum state with non-zero probability, the only state that can be prepared deterministically is the maximally mixed state.  

The restriction to a single deterministic state is consistent with general considerations about causality in quantum theory. Quantum theory, in its standard operational formulation, satisfies the causality axiom  \cite{chiribella2010probabilistic,chiribella2011informational,chiribella2016quantum,d2017quantum},  stating that it is impossible to send signals from the future to the past. In general, a physical theory satisfies the causality axiom if and only if, for every physical system described by the theory,  there exists one and only one  deterministic process that discards that   system.   The time reversal of the causality axiom, explored  in  \cite{coecke2017time}, is that the theory contains only one deterministic state preparation  for every  system.    It is natural to expect that a time symmetric version of quantum  theory should possess only   a single deterministic preparation process and a single deterministic discarding process for every system.  In this respect, our choice of time symmetric quantum operations is a natural way to satisfy both conditions.

{ The standard quantum framework and its time symmetric variant are different operational theories.  In the standard framework,  an agent can deterministically prepare any pure quantum state, while in the time symmetric variant the agent can generate pure states with probability at most $1/d$, where $d$ is the system's dimension.  The difference  affects which tasks can be implemented efficiently according to  one theory or  the other.

Despite the aforementioned difference,  the standard operational framework of quantum theory  can be retrieved from the time symmetric version by introducing  an operation of conditioning on state preparations. 
%Conditioning on the preparations amounts to ignoring the fact that pure states can only prepared with probability $1/d$. 
Operationally,  the idea is that  an agent who operates in the forward time direction can observe the preparation of a pure state,  put the system aside, and store it in a memory for later use.   Using a previously stored pure state, the agent can then  implement  any quantum operation on a given system, by letting the  system interact with the memory via  a suitable unitary dynamics. In this way, the full set of evolutions allowed in standard quantum theory is retrieved (see also the recent work by  Hardy~\cite{hardy2021time} for a similar argument, albeit with a different interpretation).

The idea that the standard quantum framework emerges from conditioning on state preparations is compatible with a few empirical observations. First, state preparations are often obtained by performing measurements,  as in a Stern-Gerlach experiment. In those setups, the  role of conditioning in the preparation of pure states (or more generally, non-maximally mixed states) is immediately evident. Second,  even those procedures that  appear to deterministically generate pure states   ultimately rely on some implicit form of conditioning on the outcomes of experiments done in the past. For example, the preparation of a ground state through successive cooling operations requires the experimenter to have access to low temperature reservoirs, corresponding to sets of particles in nearly pure  states. Producing such reservoirs from scratch would require the ability to filter particles based on their quantum state, which again comes down  to  conditioning on the outcomes of certain quantum experiments.   In this respect, our time symmetric version of quantum theory is closely related to the resource theory of purity~\cite{horodecki2013fundamental,gour2015resource,chiribella2017microcanonical},   where every deviation from the maximally mixed state is regarded as a resource. 
 
 The above considerations suggest that one could regard the time-symmetric variant as fundamental, and the standard quantum framework as an effective theory, describing the time-asymmetric abilities of a certain type of agents. This idea agrees with similar considerations made in a series of recent works \cite{di2020quantum,chiribella2020quantum,hardy2021time}, which reexamined the question of time symmetry in quantum theory from an operational perspective.

%The role of conditioning can be regarded as a physical explanation of our no-go theorem for time symmetry:  since the ability to condition on the outcomes of previous experiments is specific to  agents with a given direction of the time arrow, there is no reason why agents with an inverted time arrow should be able to describe effective  quantum operations resulting from such conditioning.    Of course, a time-reversed agent may be in principle able to condition on  future outcomes, and therefore to reproduce all experiments described by the standard framework of quantum theory. However, the crucial point is that the ability to translate back and forth between the description of one agent to the description of the other agent is limited to the subset of experiments that do not include time-asymmetric conditioning operations. 

}

\section{Symmetries of  quantum states}
\label{sec:symmetry}

Here we  first review the classic formulation of Wigner's theorem,   and then provide an equivalent formulation in terms of mixed states.

\subsection{Wigner's theorem}
  
The original formulation of Wigner's theorem \cite{wigner1959group,Wi60} is in terms of pure states.   
 The pure states of a system with Hilbert space $\spc H$ are described by unit rays, that is, equivalence classes of unit vectors that are equal  up to a global phase.  Specifically, for a unit vector $\psi  \in  \spc H$, the corresponding unit ray is the set $\underline{\psi}   :  = \{    e^{i \gamma }  \,  \psi  ~|~   \gamma \in  \R \}$.     
%$ [  |\psi\> ]   :  = \{  |\psi'\>  \in  \spc K ~|~  |\psi'\>  =  e^{i \gamma }  \,   |\psi\>  \, ,   \gamma \in  \R \}$.     

The outcome probabilities  of basic quantum measurements are computed through Born's statistical formula. The fundamental quantity appearing in Born's formula is  the    
  ray product 
\begin{align}
%\nonumber 
\underline{\psi} \cdot \underline{\phi}   
%& : =  \sup_{  \psi'  \in  \underline{\psi} } ~ \sup_{  \phi' \in  \underline{\phi}}  ~   \< \psi'  |\phi'\> \\
%\underline{\psi} \cdot \underline{\phi}   & : =  \sup_{  |\psi'\>  \in  [|\psi\>] } ~ \sup_{  |\phi'\> \in  [|\phi\>]}  ~   \< \psi'  |\phi'\> \\
&  =  |\<\psi  |\phi\>| \, ,  
\end{align}
whose value is independent of the choice of unit vectors  $|\psi\>$ and $|\phi\>$ used to represent  the sets $\underline \psi$ and $\underline \phi$, respectively.

 A symmetry of the set of pure states is a one-to-one transformation that preserves the outcome probabilities in all possible experiments. Mathematically, the set of pure states is the so-called
 {\em ray space}  $\underline {\spc H}$, that is,  the set  of unit rays, equipped with the ray product.    A symmetry is  is a transformation of the ray space into itself, or, more generally, a transformation of a ray space  $\underline {\spc H}$  into another  ray space $\underline {\spc K}$:   
 
 \begin{defi} 
 A symmetry transformation from the ray space $\underline {\spc H}$  to the ray space $\underline {\spc K}$ (a {\em ray space symmetry}, for short)     is a bijective function  $\underline S : \underline {\spc H}      \to \underline {\spc K}  $ that preserves the ray product,  namely
\begin{align}
\underline S  (\underline \psi)   \cdot  \underline S(\underline \phi)   =   \underline \psi \cdot  \underline \phi  \qquad \forall \underline  \psi,  \underline \phi \in \underline{\spc H} \, .
\end{align}    
\end{defi}

We recall that a bijective function $f$  from set $\set A$ to set $\set B$ is  a transformation that maps distinct  elements of $\set A$ into distinct elements of $\set B$,  in such a way that every element of $\set B$ is obtained by applying $f$ to some element of $\set A$.    Intuitively, the property of being bijective is necessary for the two ray spaces  $\underline {\spc H}$  and $\underline {\spc K}$  to be regarded as different descriptions of the same physical system.    
 
% from the point of view of two different observers.
%Note that the preservation of the ray product automatically guarantees that $\underline S$ is injective.  

Wigner's theorem provides a complete characterisation  of the ray space symmetries. Specifically, it shows that every ray space symmetry   is induced by a transformation on the underlying Hilbert space:
% and that such transformation is either unitary or anti-unitary.     
% for every pair of vectors $\psi$ and $\phi$  in $\spc H$.
\begin{theo}[Wigner's theorem]\label{theo:wigner}    For every ray space symmetry $\underline S :  \underline{\spc H} \to  \underline{\spc K}$ there exists a  Hilbert space transformation $S: \spc H\to \spc K$ such that 
\begin{enumerate}
\item  $\underline S ( \underline \psi)  =   \underline {  S  ( \psi)}$ for every $\psi  \in  \spc H$, and  
\item the transformation $S$ is either unitary or anti-unitary.
\end{enumerate}
\end{theo}

%In the following we will provide an equivalent formulation of Wigner's theorem, expressed in terms of density matrices. This equivalent formulation will be used later in the paper to prove a Wigner theorem for quantum operations.   
We recall that a bijective transformation $S  : \spc H \to \spc K$ is  {\em unitary}  if $\< S (\psi )  |   S(\phi) \>  =  \<  \psi|\phi\> $    for every pair of vectors $\psi$ and $\phi$  in $\spc H$, and {\em antiunitary} if $\< S (\psi )  |   S(\phi) \>  =  \<  \phi|\psi\>$.

The prototype of an antiunitary transformation is the transformation  $S: \psi \to  \psi^*$,    where  $\psi^*$  is the complex conjugate of the vector $\psi$  with respect to a fixed  (but otherwise arbitrary) orthonormal basis.
In general, the action of a generic antiunitary transformation $S$  can be represented as $S  (\psi)    =    U  \psi^*$,  where $U :  \spc H \to \spc K$ is a unitary operator.  A general treatment of antiunitary transformations can be found  in the recent work by  Uhlmann \cite{Uhl16}.
%Using this fact, Wigner's theorem can be restated as 
%\begin{cor}    Every ray space symmetry $\underline S :  \underline{\spc H} \to  \underline{\spc K}$ is either of the form   $\underline S ( \underline \psi)  =   \underline {   U   \psi}$ or of the form  $\underline S ( \underline \psi)~=~\underline {   U   \psi^*}$,   where $U  : \spc H \to \spc K$ is a  unitary operator.  
%\end{cor}

\subsection{Density matrix formulation of Wigner's theorem}  

Wigner's theorem can  be equivalently formulated as a characterisation of the symmetries of the full state space, including mixed states described by density matrices.    In this case, the probabilistic interpretation of quantum states implies the possibility of considering random mixtures of different density matrices. Operationally, the randomisation can be due to ignorance about the state of the quantum system, or also to an actual physical mechanism that lets  the choice of state be controlled by some random event, such as the result of coin toss.    

  In the following we will denote  the set of all density matrices on the Hilbert space $\spc H$ by  
   \begin{align}\St (\spc H)  :=  \{  \rho  \in  L(\spc H)~|~ \rho\ge 0  \, , \Tr[\rho] =1\} \, ,
   \end{align} $L(\spc H)$ denoting the set of linear operators on $\spc H$.  A symmetry of state spaces is then defined to be as a one-to-one transformation that preserves random mixtures:  

\begin{defi}  A symmetry  transformation from the state space $\St (\spc H)$ to the state space $\St (\spc K)$   (a {\em state space symmetry}, for short)    is a bijective transformation $\map S :  \St (\spc H)   \to \St (\spc K)$ that is consistent with randomisations, namely
\begin{align}
\map S \big(   p \rho  + (1-p)   \sigma\big)    =  p  \, \map S  (\rho)  +  (1-p)\,  \map S(\sigma)  \, ,
\end{align}
for every pair of density matrices $\rho$ and $\sigma$, and for every probability $p \in  [0,1]$.  
\end{defi}

Wigner's original theorem is equivalent to the following statement:
% that every state space symmetry is either of the unitary form $\map S (\rho)  =  U \rho U^\dag$ for some unitary operator $U$,  or of the antiunitary form   $\map S (\rho)  =  U \rho^T U^\dag$, where  $U$ is still a unitary operator, and $\rho^T$ is the transpose of $\rho$ with respect to a fixed orthonormal basis.  Explicitly, 

\begin{theo}\label{theo:wignerdensity}   {\bf (Density matrix formulation of Wigner's theorem)} Every state space symmetry $\map S:  \St (\spc H)   \to \St (\spc K) $ is either a unitary transformation, of the form $\map S (\rho)  =  U \rho U^\dag$ for some unitary operator $U: \spc H \to \spc K$, or an antiunitary transformation,  of the form $\map S(\rho)  =  U  \rho^T  U^\dag$, where $\rho^T$ denotes the transpose of $\rho$ with respect to a fixed orthonormal basis.   
%In particular, $\map J$ is unital, namely $\map J (I)  =  I$. 
\end{theo}

 Theorem \ref{theo:wignerdensity}   can be  derived from the original Wigner's theorem, as shown in Appendix \ref{app:wignerdensity}. 
%We have seen that the characterisation of state space symmetries in  Theorem \ref{theo:wignerdensity} follows from the characterisation of ray space symmetries in the original Wigner's theorem. 
Conversely, it is also possible to show that Theorem \ref{theo:wignerdensity} implies the original Wigner's theorem, as shown in Appendix \ref{app:equivalence}.    In summary,   Theorem \ref{theo:wignerdensity} is an equivalent formulation of Wigner's original result in the language of density matrices.

\section{Symmetries of quantum evolutions}\label{sec:symmetryQO}
In this section we derive a Wigner theorem for quantum evolutions,  showing that every symmetry of the set of quantum operations can be broken down into two state space symmetries, which are either both unitary or both antiunitary. 
%all arise from the  composition of unitary channels and the transpose map.    

\subsection{Quantum operations}  

Quantum operations \cite{kraus1983states,heinosaari2011mathematical} provide a unified framework for treating all state evolutions  in quantum mechanics. 
% including the unitary evolution associated to Schr\"odinger's equation, as well as the  stochastic evolutions associated to the outcomes of quantum measurements.  
In general, a quantum state change can be stochastic, meaning that an initial state $\rho$ is transformed into a final state $\rho'$ with some probability $p$, generally smaller than 1.   The  non-unit probability of obtaining the output state can be accounted for by considering subnormalised states, corresponding to non-negative operators  $\rho$ with trace $\Tr[\rho]\le 1$.  The trace $\Tr[\rho]$ is  then interpreted as the probability that  the state is prepared  by some suitable stochastic process. When such probability is nonzero, the state of the system conditional to the occurrence of the process is the normalised state $\rho/\Tr[\rho]$. 
    %We denote the set of subnormalised states on a generic Hilbert space $\spc H$ as 
%\begin{align}
%  \Sub\St (\spc H)   =  \{  \rho  \in L (\spc H) ~|~  \rho  \ge 0  \, ,  \, \Tr [\rho]  \le 1 \} \, .
%  \end{align}
%Note that this set is convex, and contains the null state $\rho = 0$ corresponding to the limit case  where no state is being prepared. 

Mathematically, a quantum operation is a completely positive, trace-non-increasing linear map $\map Q$, transforming an initial state $\rho$ into a subnormalised state $\map Q  (\rho)$.  The probability that the state change takes place on the state $\rho$  (in a stochastic process including the quantum operation $\map Q$ among its possible events)  is the trace $\Tr[\map Q(\rho)]$. When such a state change occurs, the state of the system becomes $\rho'  =  \map Q  (\rho)/\Tr[\map Q(\rho)]$.

 Kraus' theorem \cite{kraus1983states} shows that the action of a quantum operation $\map Q$ on a generic density matrix $\rho$ can be expressed as $\map Q (\rho) =  \sum_i  \,  Q_i  \rho  Q_i^\dag$, where the operators $(Q_i)$, called {\em Kraus operators},  satisfy the condition $\sum_i Q_i^\dag Q_i  \le I_{\rm in}$.  When the equality condition $\sum_i Q_i^\dag Q_i  = I_{\rm in}$ is satisfied, the  quantum operation is called a {\em quantum channel}~\cite{holevo2012quantum}.

  In the following, we will denote the set of all quantum operations  by $\Op (\spc H_{\rm in},  \spc H_{\rm out})$ and the subset of all quantum channels by  $\Chan (\spc H_{\rm in},  \spc H_{\rm out})$, where $\spc H_{\rm in}$ and $\spc H_{\rm out}$ are the Hilbert spaces of the input and output system, assumed to be finite dimensional for simplicity of presentation.   We will call $\Op (\spc H_{\rm in},  \spc H_{\rm out})$ and   $\Chan (\spc H_{\rm in},  \spc H_{\rm out})$ the {\em operation space} and the {\em channel space}, respectively.

%Quantum operations can be equivalently defined in an intrinsic way, that refers only to the set of subnormalised quantum states.   We denote the set of subnormalised states on a generic Hilbert space $\spc H$ as 
%\begin{align}
 % \Sub\St (\spc H)   =  \{  \rho  \in L (\spc H) ~|~  \rho  \ge 0  \, ,  \, \Tr [\rho]  \le 1 \} \, .
%  \end{align}
%Note that this set is convex, and contains the bottom element $\rho = 0$ corresponding to the situation where no state is being prepared. 
%A quantum operation can be defined as a map $\map Q:   \Sub\St(\spc H_{\rm in}) \to  \Sub\St(\spc H_{\rm out})$, transforming subnormalised states into subnormalised states in a way that 
%\begin{enumerate}
%\item preserves convex combinations,  and 
%\item preserves the bottom element, namely $\map Q(0)  =  0$.  
%\end{enumerate} 
%Indeed, it is straightforward to see that such map can be uniquely extended to a linear map from $L(\spc H_{\rm in})$ to $L(\spc H_{\rm out})$.   

Like the quantum state space, the  quantum operation space is convex, meaning that the random mixture of two quantum operations is also a quantum operation.   Several aspects of the convex structure of the quantum operation space were investigated in Ref.~\cite{capellini2008}.   

The quantum operation space also contains a special element, namely the null quantum operation $\map Q  =  0$, describing the limit  case of a stochastic evolution that has zero probability to take place.  Convexity and the presence of a null element are the two key elements of the probabilistic structure of the set of quantum operations.

\subsection{Symmetries of the set of quantum operations}  

A symmetry of quantum operations is a reversible transformation that it is consistent with randomisations and maps impossible events into impossible events.  
% transforms the null operation into itself. 
%   Unlike the set of quantum states, however, the set of quantum operations has a richer structure: not only it is convex, but also it contains a  bottom element, corresponding to the null quantum operation  $\map Q  =  0$.  
\begin{defi}
A {\em symmetry transformation from the operation space $\Op (\spc H_{\rm in},  \spc H_{\rm out})$ to the operation space $ \Op (\spc K_{\rm in},  \spc K_{\rm out})$} (an {\em operation space symmetry}, for short)  is a bijective transformation $\map S  :  \Op (\spc H_{\rm in},  \spc H_{\rm out})  \to  \Op (\spc K_{\rm in},  \spc K_{\rm out})$ that  
\begin{enumerate}
\item is consistent with randomisations, namely  $  \map S   \big(   p   \, \map Q   +  (1-p)\,  \map R   \big)  =  p\,  \map S (\map Q)  +  (1-p)  \, \map S  (\map R)$, for every pair of quantum operations  $\map Q$ and $\map R$, and for every probability $p \in  [0,1]$, and 
\item transforms the null operation of $\Op (\spc H_{\rm in},  \spc H_{\rm out})$    into the null operation of $ \Op (\spc K_{\rm in},  \spc K_{\rm out})$. 
% namely $\map S  ( 0)  = 0$.  
\end{enumerate}
%A {\em symmetry of quantum operations} is  a bijective  state space  homomorphism. 
\end{defi} 
%The first condition states that the symmetry transformation is consistent with  the notion of statistical mixture. The second condition means that the symmetry transformation is consistent with the notion of zero probability, as it maps impossible events into impossible events. 

It is not difficult  to see that every symmetry of quantum operations should map quantum channels into quantum channels:   
\begin{prop}\label{prop:channel2channel} 
For every operation space symmetry  $\map S  :   \Op  (\spc H_{\rm in}, \spc H_{\rm out})    \to  \Op  (\spc K_{\rm in}, \spc K_{\rm out})$, one has $\map S \left(  \Chan (\spc H_{\rm in}, \spc H_{\rm out})\right)  =   \Chan(\spc K_{\rm in}, \spc K_{\rm out})$. 
\end{prop}
This result, proved in Appendix \ref{app:channel2channel}, is conceptually important because it shows that every operation space symmetry  induces a channel space symmetry.

\subsection{Wigner theorem for quantum operations}\label{sec:Wigner}

 We now provide a complete characterisation of the symmetries of quantum operations, in analogy to Wigner's theorem for quantum states. Our result shows that every operation space symmetry can be decomposed into two state space symmetries, of which  one transforms the input, and the other transforms the output: 

 \iffalse  Like Wigner's theorem, our result establishes that the possible symmetries are of two types: unitary symmetries, and antiunitary symmetries.  A unitary symmetry  transforms an input quantum operation $\map Q$ by concatenating it with a unitary transformation  $\map V$ on the input space, and with a unitary transformation $\map W$ on the output space, thus obtaining a new quantum operation $\map Q'  =  \map W  \circ \map Q \circ \map V$.  An antiunitary symmetry transforms quantum operations in the same way, except that now both transformations $\map V$ and $\map W$ are antiunitary.  
\fi  

\begin{theo}\label{theo:wignerqo}
{\bf (Wigner's theorem for quantum operations)} Every operation space symmetry  $\map S :  \Op  (\spc H_{\rm in}, \spc H_{\rm out}) \to \Op (\spc K_{\rm in},  \spc K_{\rm out})$  is of the  form $\map S (\map Q)    =    \map S_2  \circ \map Q \circ \map S_1$,   where  $\map S_1 :  \St (\spc K_{\rm in}) \to \St(\spc H_{\rm in})$ and $\map S_2:  \St(\spc H_{\rm out})\to \St(\spc K_{\rm out})$ are  state space symmetries that  are either both unitary or both antiunitary. 
% unitary channels in $\Chan (\spc H)$, and, for every map $\map M \in  \Map (\spc H)$,   $\widetilde {\map M}$ is the map defined by the relation  $\widetilde {\map M}  (X)   :=    \left(\map  M  (X^T) \right)^T$.    
\end{theo}
The proof of Theorem~\ref{theo:wignerqo} is provided in Appendix \ref{app:wignerqo}, using technical lemmas established in Appendices~\ref{app:symm2symm},~\ref{app:break}, and~\ref{app:Wpsi}.   Like Wigner's original theorem, our result establishes that the possible symmetries are of two types. We call these two types  {\em double unitary symmetries} and {\em double antiunitary symmetries}.

Double unitary symmetries can be physically implemented by inserting the input quantum operation between two reversible quantum processes.  Such symmetries are compatible with an active interpretation, in which an agent implements the symmetry by engineering the system's dynamics~\cite{chiribella2008transforming,chiribella2009theoretical}.

 In contrast, double antiunitary transformations are not compatible with an active implementation, at least not in standard quantum mechanics, where antiunitary operations are excluded from the set of  allowed evolutions.  
 
 The prototype of a double antiunitary symmetry is the ``double transpose" transformation $\tau_{\rm in/out}: \map Q  \mapsto    \map Q'  = \tau_{\rm out} \circ \map Q\circ\tau_{\rm in}$,   where   $\tau_{\rm in}$   ($\tau_{\rm out}$) is the transpose on the input (output) system with respect to a fixed basis.    In the Kraus representation, the double transpose corresponds to the complex conjugate of the Kraus operators, mapping a quantum operation with Kraus operators $\{ Q_i\}$ into a new quantum operation with Kraus operators $\{Q_i^*\}$.   The double transpose transformation appeared  in the seminal work by Holevo and Werner on the capacity of bosonic Gaussian channels \cite{holevo2001evaluating}, where it was used to show that entanglement-breaking channels have zero quantum capacity.    Approximate physical realisations of the double transpose were studied in Refs.~\cite{chiribella2016optimal,yang2017units,miyazaki2019complex,quintino2019probabilistic}, which focussed on approximating the complex conjugation of unknown unitary dynamics.

In general, every double antiunitary transformation  has the form   $\map S    (\map Q )  =     \map U  \circ \tau_{\rm in/out} (\map Q)  \circ \map V$, where $\map U: \St (\spc K_{\rm in})\to \St (\spc H_{\rm in})$ and $\map V: \St (\spc H_{\rm out}) \to \St (\spc K_{\rm out})$ are unitary symmetries.    In other words, the double transpose is the seed  of all possible double antiunitary symmetries.  

%We conclude by noting a consequence of Wigner's theorem for quantum operations, namely that every operation symmetry is also a symmetry of the set of unitary dynamics. More generally, the operation space symmetries transform single-Kraus quantum operations $\map Q:  \rho  \mapsto  Q\rho Q^\dag$ into single-Kraus quantum operations $\map Q'  :  \rho  \mapsto Q'\rho Q^{\prime \dag}$.    
%\begin{cor}
%Every operation space symmetry  $\map S :  \Op  (\spc H_{\rm in}, \spc H_{\rm out}) \to \Op (\spc K_{\rm in},  \spc K_{\rm out})$ maps single-Kraus operations into single-Kraus operations.  When the spaces $\spc H_{\rm in}$,  $\spc H_{\rm out}$, $\spc K_{\rm in}$, $\spc K_{\rm out}$ are isomorphic, $\map S$ transforms unitary channels into unitary channels. 
% \end{cor} 
% of the set of quantum operations.  

Further discussion on the differences between double unitary and double antiunitary symmetries, in relation to the notion of complete positivity,  is provided in Appendix~\ref{app:dt}.

\section{No   time reversal of general quantum  evolutions}  
\label{sec:imp}

%We now provide an application of Wigner's theorem for quantum operations, showing that  the set of all stochastic quantum evolutions admits no time reversal symmetry.  

\subsection{Time reversal symmetry  of unitary evolutions and bistochastic channels}\label{subsec:bisto}

The set of unitary evolutions exhibits an obvious time reversal symmetry: for every unitary evolution $U$, the inverse matrix $U^\dag$ is also unitary and describes a valid quantum evolution.  At the level of quantum channels, the standard time reversal maps the unitary channel $\map U:  \rho \mapsto  U \rho U^\dag$ into the inverse channel $\map U^\dag:  \rho  \mapsto  U^\dag \rho  U$.

 For certain irreversible quantum evolutions, it is still possible to define a meaningful notion of time reversal.
Indeed, the time reversal symmetry of the set of unitary channels  can be uniquely extended to a larger set of quantum evolutions, known as the set of bistochastic (or doubly stochastic) quantum channels  \cite{landau1993birkhoff,mendl2009unital}.   A bistochastic channel is a completely positive trace-preserving map $\map C  : \St (\spc H) \to \St (\spc H)$ with the additional property that  $\map C (I)  =  I$.    
%  In the Kraus representation, a bistochastic channel $\map C:  \rho  \mapsto  \map C (\rho)   =  \sum_i C_i \rho  C_i^\dag$ satisfies the conditions $\sum_i C_i C_i^\dag  =  I$ and $\sum_i C_i^\dag C_i  =  I$.    Note that the set of bistochastic channels contains the set of unitary channels as a subset, as every unitary channel $\map U:  \rho \mapsto U \rho U^\dag$ satisfies both conditions $U^\dag U  = I$ and $UU^\dag = I$.  
  
The set of bistochastic channels  contains highly irreversible evolutions, such as  the evolution  that takes every quantum state  to the maximally mixed state. For such evolutions, the time reversal is not  the process that brings back the system to the initial state, for such a process %does not exist. 
 cannot be defined without information about  the environment with which the system interacted.  
 Instead,   the time reversal of a bistochastic channel describes how the same physical device would respond to a preparation of states in the future, rather than a preparation of states in the past~\cite{chiribella2020quantum}.   To clarify this point, consider    the qubit channel $\map C_0:  \rho \mapsto \map C_0 (\rho) =  I/2$,  which maps every state to the maximally mixed state. This channel can be realised as a uniform random mixture of four unitary channels, corresponding to the four Pauli matrices $I,X,Y,$ and $Z$.  To be consistent with random mixtures, the  time reversal of channel $\map C_0$ must be  the uniform mixture of the inverses of the four unitary channels, which again correspond to the four Pauli matrices $I,X,Y,$ and $Z$.  In summary, the time reversal of channel $\map C_0$ is channel $\map C_0$ itself:   in this example, both an ordinary agent and an agent with an inverted time arrow  would describe the overall input-output transformation as a channel that maps every state into the maximally mixed state.   
  
The  standard time reversal of the bistochastic channel $\map C$ is  the channel $\map C^\dag$ defined by $\map C^\dag  (\rho)  =   \sum_i  C_i^\dag  \rho  C_i$.  Note that, in general, the channel $\map C^\dag$ is not the inverse of the channel $\map C$. 
 %$\map C^\dag \circ \map C  \not  =  \map I$. 
   %in general, $\map C$ represents an irreversible process, which cannot be inverted by any physical process.   
In a sense, however, $\map C^\dag$ can be thought as an {\em approximate} inversion, known as Petz' recovery map~\cite{petz1988sufficiency}, and has many applications in quantum information~\cite{barnum2002reversing,hayden2004structure,ng2010simple,fawzi2015quantum,beigi2016decoding,junge2018universal}.    At the structural level, the transformation $\Theta: \map C  \mapsto  \map C^\dag$ is a symmetry of the set of bistochastic channels: it is a bijective transformation, and it is compatible with randomisations.   
%In addition, it maps unitary channels into unitary channels. 

The set of unitary evolutions also admits an alternative time reversal, given by the map $\map U \mapsto \map U^T$, where $\map U^T$ is the {\em transpose unitary channel}, defined by $\map U^T(\rho)  =  U^T  \rho U^*$~\cite{chiribella2020quantum}.   
  This alternative  time reversal can also be extended uniquely to the set of bistochastic channels, by mapping the channel $\map C$ into the channel $\map C^T$ defined by $\map C^T(\rho)   =  \sum_i C_i^T  \rho  C_i^* $.  The  alternative time reversal   $\Theta':  \map C\mapsto \map C^T$ is also a symmetry of the set of bistochastic channels: it is bijective and compatible with randomisations.  

The maps $\Theta$ and $\Theta'$   can be characterised as the two canonical time reversals on the set of bistochastic channel~\cite{chiribella2020quantum}.  More information about their structure can be found in Appendix \ref{app:bisto}, where we characterise the action of  $\Theta$ and $\Theta'$ in terms of the Choi representation~\cite{Cho75a} (see also \cite{BZ} and~\cite{heinosaari2011mathematical}).    

\subsection{No-go theorem for time reversal symmetry of the set of quantum operations}

A natural question is whether the time reversal symmetries of the set of bistochastic channels can be extended to symmetries of the whole set of quantum operations.  
    
The obvious candidates of  time reverals are the maps $\Theta: \map Q  \mapsto  \map Q^\dag$ and $\Theta'  :  =  \map Q \mapsto  \map Q^T$, now defined  on arbitrary quantum operations $\map Q$.      It is relatively easy to see, however, that these two maps fail to be  symmetries of the full set of quantum operations. 
%  For example, consider the quantum operation $\map Q $ defined by $\map Q(\rho)  : =  |0\>\<0|  \,  \Tr[\rho]$, which erases the initial state  of the system and output the pure state $|0\>\<0|$ independently of $\rho$. It is easy to see that the maps $\map Q^\dag$ and $\map Q^T$ are not quantum operations.  Indeed, they act as $\map Q^\dag (\rho) =  \map Q^T   (\rho)  =  I\,  \<0|\rho  |0\> $. The maps $\map Q^\dag$ and $\map Q^T$ transform the density matrix $\rho  =  |0\>\<0|$ into the overnormalised matrix  $\map Q^\dag (|0\>\<0|)   =  \map Q^T(|0\>\<0|)  =  I$.  Hence, $\map Q^\dag$ and $\map Q^T$ fail to be trace non-increasing, and therefore fall  outside the set of  quantum operations.   
%We have seen that the obvious candidate of time reversal symmetries of quantum operations do not work.  
In principle, however, there could exist  other ways to extend the time reversal from the set  of bistochastic channels to the set of all quantum operations.  This possibility is ruled out by the following general theorem:

\begin{theo}{\bf (No time reversal of arbitrary quantum operations)}\label{theo:nogo}
No symmetry  $\map S$ of the set of quantum operations satisfies the condition  $\map S  (\map U)   = \map U^\dag$ for every unitary channel $\map U$, or the condition   $\map S  (\map U)   = \map U^T$ for every unitary channel $\map U$.     
\end{theo}
The proof can be found in Appendix~\ref{app:nogo}.  
  
Theorem \ref{theo:nogo}  shows that the structure of set of all quantum  evolutions is fundamentally incompatible with time reversal symmetry:  informally, this means that it is impossible to define a ``change of reference frame'' corresponding to the transformation $t\mapsto -t$, and compatible with the probabilistic interpretation of quantum evolutions.  

It is worth noting that our no-go result applies to the whole set of quantum operations.
% (trace non-increasing completely positive maps), which can be interpreted as open system
%evolutions where the environment
%undergoes a measurement, and some specific measurement outcome is selected.  
%This measurement does not need to  be
%accessible to the agent implementing the 
%evolution, but it will be recorded somewhere, 
%and possibly used by another agent.  
An  interesting open question is whether our no-go theorem  remains valid for the subset of quantum {channels}.
% ({trace-preserving} completely positive maps), which describe the effective evolution on average over all possible measurement outcomes.  
A no-go theorem for time reversal symmetries of quantum channels has been recently proven in Ref.~\cite{chiribella2020quantum}, under the additional assumption that time reversal should reverse the order in which processes take place in any time sequence. We conjecture that this assumption can be lifted, and that the no-go theorem to time reversal applies also to symmetries of the set of quantum channels.   Such extension of the no-go theorem, however, is beyond the scope of the present paper, which focusses on the symmetries of the full set of quantum evolutions. 
%The reason why we focus on the full set of quantum evolutions is that is  includes  the stochastic evolutions arising from quantum measurements, which are  important in order to analyse  the operational framework of quantum theory, in which measurements play a fundamental role.  

Our no-go result on time reversal poses a strong constraint on any attempt to formulate quantum theory as a time symmetric operational theory.     To better understand the nature of this constraint, it is useful to consider possible relaxations of our no-go theorem. In the following we discuss three such relaxations: the first relaxation consists in restricting the set of allowed quantum evolutions to a time-symmetric subset, the other two relaxations consist in weakening the notion of symmetry in a way that evades our no-go theorem.

\section{A time symmetric variant of quantum theory}\label{sec:tsop} 

{ In this section we provide a time symmetric variant of quantum theory, in which the allowed channels transform maximally mixed states into maximally mixed states.  We show that the standard formulation of quantum theory can be obtained from the time symmetric variant through a suitable operation of conditioning. Then, we show that the time symmetric variant is maximal among the time symmetric quantum theories in which all unitary channels are regarded as allowed dynamics. }

\subsection{Time-symmetric quantum operations}\label{subsec:tsop}
One way to circumvent our no-go theorem on time symmetry is to restrict the attention to a subset of quantum evolutions  on which a time reversal symmetry can be defined.  
%Here we propose the set of quantum operations $\map Q$ with Kraus operators $(  \map Q_i)$ satisfying the conditions 
%\begin{align}\label{tsop}
%\sum_{i}  Q_i^\dag Q_i  \le I_{\spc H_{\rm in}} \quad {\rm and} \quad  \sum_{i}  Q_i Q_i^\dag  \le I_{\spc H_{\rm out}} \, ,
%\end{align}
%where $\spc H_{\rm in}$ and $\spc H_{\rm out}$ are the Hilbert spaces of the input and output system, respectively. 
%We name the  maps satisfying Equation~(\ref{tsop})  {\em time symmetric quantum  operations}, and denote the corresponding set  by  $\TS\Op(\spc H_{\rm in},  \spc H_{\rm out})$.   
Here we propose the set of quantum operations  satisfying the conditions 
 \begin{align}\label{tsopfinal}
  \map Q^\dag (I_{\rm out})   \le I_{\rm in}   \quad {\rm and} \quad 
%\frac{\sum_{i}  Q_i Q_i^\dag}{d_{\rm in}}   \le   \frac  {I_{\spc H_{\rm out}}}{d_{\rm out}}
\map Q\left( \frac{  I_{\rm in}}{d_{\rm in}}\right)  \le  \frac  {I_{\spc H_{\rm out}}}{d_{\rm out}}
 \, ,
\end{align}  
where $d_{\rm in}$ and $d_{\rm out}$ are the dimensions of the Hilbert spaces $\spc H_{\rm in}$ and $\spc H_{\rm out}$.   We name the  maps satisfying Equation~(\ref{tsopfinal})  {\em time symmetric quantum  operations}, and denote
the corresponding set  by  $\TS\Op(\spc H_{\rm in},  \spc H_{\rm out})$.     When the inequalities in Equation~(\ref{tsopfinal})  are satisfied with the equality sign, we call the maps {\em time symmetric quantum channels}, and denote the corresponding set by  $\TS\Chan(\spc H_{\rm in},  \spc H_{\rm out})$.    Note that for $d_{\rm in} = d_{\rm out}$, the set of time symmetric quantum channels coincides with the set of bistochastic channels, and enjoys time symmetry  as discussed earlier in Subsec.~ \ref{subsec:bisto}.

Like the full set of quantum operations, the set of time symmetric quantum operations is convex and contains the null operation $\map Q  =  0$.   It is rather straightforward to see that the linear map 
\begin{align}\label{tradj}
\Theta:  \map Q \mapsto \frac {d_{\rm out}}{d_{\rm in}} \,\map Q^\dag  
\end{align} 
 is a symmetry transformation from the set $\TS\Op(\spc H_{\rm in},  \spc H_{\rm out})$ to the set $\TS\Op(\spc H_{\rm out},  \spc H_{\rm in})$.  Indeed,  it is bijective, it maps  convex combinations into convex combinations with the same probabilities, and it transforms  null operations into null operations. Likewise,  the map 
 \begin{align}\label{trtr}
 \Theta' :    \map Q \mapsto  \frac {d_{\rm out}}{d_{\rm in}} \ \map Q^T
 \end{align} 
 is also a symmetry transformation from the set $\TS\Op(\spc H_{\rm in},  \spc H_{\rm out})$ to the set $\TS\Op(\spc H_{\rm out},  \spc H_{\rm in})$.  
 
 Both maps $\Theta$ and $\Theta'$ transform time symmetric quantum channels in  $\TS\Chan(\spc H_{\rm in},  \spc H_{\rm out})$ into time symmetric quantum channels in   $\TS\Chan(\spc H_{\rm out},  \spc H_{\rm in})$.    For $d_{\rm in}  = d_{\rm out}$,  the maps  $\Theta$ and $\Theta'$ coincide with the standard time reversal $\map U \mapsto U^\dag$ and with the alternative time reversal $\map U \mapsto \map U^T$, respectively.

\subsection{Time-symmetric quantum instruments}\label{subsec:tsinst} 
The set of time symmetric quantum operations can be used to define a time symmetric variant of quantum theory.    In this variant,  the allowed experiments are described by {\em time symmetric quantum instruments}, that is,  collections of time symmetric quantum operations $(\map Q_n)_{n=1}^N$ with the property that the sum $\sum_{n=1}^N  \map Q_n $ is a time symmetric quantum channel, that is, a quantum channel satisfying Eq.~(\ref{tsopfinal}) with the equality sign.    When an experiment is performed on a (possibly subnormalised) state $\rho$, the probability of the  outcome  $n$ is given by the trace $\Tr [ \map Q_n  (\rho)]$, as in standard quantum theory.  
% The time symmetric quantum channels correspond to experiments with only one possible outcome, in which case the evolution is deterministic. %In this respect, the relation between time symmetric quantum channels and time symmetric quantum operations is the same as the relation between ordinary quantum channels and ordinary quantum operations.   

Interestingly, the time symmetric variant  still permits all von Neumann measurements, whose stochastic evolutions are described by  quantum operations   of the form $\map Q_n:  \rho  \mapsto  |n\>\<n|  \,  \<n|\rho |n\>$, where the vectors $ (|n\>)_{n=1}^d$ form an orthonormal basis.     More generally, the set of time symmetric quantum instruments contains also the set of L\"uders measurements, corresponding to quantum operations of the form $\map Q_n  :  \rho \mapsto  P_n  \rho P_n$, where $(P_n)_{n=1}^N$ is a complete set of orthogonal projectors.   In addition, it contains all quantum instruments resulting from sequences of unitary dynamics  interspersed with  L\"uders measurements. These instruments  describe the possible closed system evolutions in the framework of  consistent histories~\cite{griffiths2003consistent}.

The time symmetric variant also permits all demolition measurements, corresponding to positive operator-valued measures (POVMs) and represented by collections of positive operators $(P_n)_{n=1}^K$ satisfying the condition $\sum_{n=1}^K  P_n  =  I$.    Indeed, every quantum operation $\map Q_n$  defined by the relation  $\map Q_n (\rho):  =\Tr [  P_n  \rho]$ is a valid time symmetric operation with one-dimensional output system, according to definition~(\ref{tsopfinal}).

\iffalse
On the other hand, the set of time symmetric quantum operations is  strict subset of the set of quantum operations allowed in quantum mechanics. For example, the quantum operation that discards the input system and produces a fixed state of the output system is not a valid time symmetric operation.  Such operation is represented by the map  $\map Q (\rho) =  \Tr [\rho]  \,  |0\>\<0|$ and has  Kraus operators $ Q_i =  |0\>\< i|$ which clearly fail to satisfy the second condition in Eq. (\ref{tsop}).

When the input and output systems have the same dimension, the set of time symmetric quantum operations has a clear operational interpretation:  the elements of $\TS\Op  (\spc H_{\rm in},  \spc H_{\rm out})$ represent  evolutions that can take place stochastically     in some experiment, whose  outcome probabilities sum up to 1.  Indeed, every time symmetric operation $\map Q$ can be completed by another time symmetric quantum operation  $\map R $, defined by  $\map R  (\rho)  =  R \rho R^\dag$ with $R  =   U \sqrt{  I_{\rm in}  -   \sum_i   Q_i^\dag Q_i}$, where $U$ is any unitary operator from $\spc H_{\rm in}$ to $\spc H_{\rm out}$.      It is immediate to see that 
\fi

 \subsection{Relation between standard  quantum theory and its time symmetric variant}
 
A crucial difference between the time symmetric  variant and standard  quantum theory concerns  the set of allowed states.  In the time symmetric variant,  the only state that can be prepared with unit probability is the maximally mixed state $I/d$.  This is because the  preparation of a $d$-dimensional quantum system is a quantum  operation with one-dimensional input space and $d$-dimensional output space, and setting $d_{\rm in}= 1$ and $d_{\rm out}  =  d$ in Eq.(\ref{tsopfinal}) yields the conditions $\Tr[\rho]  \le  1$ and $\rho  \le I/d$.    In other words, the possible stochastic preparations of the system correspond to subnormalised density matrices satisfying the inequality $\rho  \le I/d$.  The only normalised state is $\rho  =  I/d$, while all the other states can be prepared only stochastically, with some probability less than one.    For example, a pure quantum state can be prepared with probability at most $1/d$.  
%Equivalently, the only time symmetric quantum instrument that prepares a single state $\rho$ is the preparation of the maximally mixed state 
  
%The  restriction on the allowed state preparations is consistent with certain experimental situations, such as the preparation of a single photon's polarisation using a polarising filter. When no prior information about the photon is available, this state preparation procedure yields one of two orthogonal polarisation states with equal probabilities.   
%This observation suggests that the time symmetric variant of quantum theory, besides being of fundamental interest, could also be useful as the   description of a class of real experiments.  

%The  restriction on the allowed state preparations is consistent with certain experimental situations, such as the preparation of a single photon's polarisation using a polarising filter. When no prior information about the photon is available, this state preparation procedure yields one of two orthogonal polarisation states with equal probabilities.   
%This observation suggests that the time symmetric variant of quantum theory, besides being of fundamental interest, could also be useful as the   description of a class of real experiments.  

The standard formulation of quantum theory can be retrieved from the time symmetric variant by introducing an operation of conditioning, whereby  probabilistic state preparations are turned into deterministic ones. Mathematically, conditioning amounts to the non-linear mapping $\rho  \mapsto  \rho/\Tr[\rho]$. Operationally, it arises from the ability of an agent to observe the outcome of a probabilistic state preparation, and to feed-forward  the knowledge of the outcome into future experiments.  

{ Since conditioning enables the deterministic preparation of arbitrary pure states, it also enables the deterministic realisation of arbitrary quantum operations. Indeed, every quantum operation $\map Q$ can be realised as  
\begin{align}\map Q (\rho)  =   \Tr_{\rm aux'} [  (\map I_{\rm out}  \otimes  P)\map U  ( \rho\otimes \psi_0 )] \,,
\end{align} 
where $\psi_0$ is a (normalised) pure state of an auxiliary system  $\rm aux$,  $\map U$ is a joint unitary evolution, transforming states on the Hilbert space $\spc H_{\rm in}  \otimes \spc H_{\rm aux}$ into states on the Hilbert space $ \spc H_{\rm out}\otimes \spc H_{\rm aux'}$, for some output auxiliary system $\rm aux'$, and $0\le P\le I_{\rm aux'}$ is a measurement operator, corresponding to the outcome of a measurement on $\rm aux'$.  
Since both the unitary evolution $\map U$ and the measurement operator $P$ are allowed in the time-symmetric variant, the addition of all possible states preparations via conditioning yields back the full set of quantum operations in the standard formulation of quantum theory.  }

 {\subsection{Maximality  of the time symmetric variant}\label{subsec:maximal}  

We provided a candidate of time symmetric variant of quantum theory. A natural question is whether there exist other candidates, and, in the affirmative case, how they would look like.  
We start by observing that our time symmetric variant  is maximal among all time symmetric variants of quantum theory in which 
\begin{enumerate}
\item for every pair of quantum systems $A$ and $B$, the set of allowed operations from system $A$ to system $B$ is a convex subset of the set of quantum operations allowed in quantum theory, 
\item for every pair of systems $A$ and $B$, the time reversal of the set of operations from  $A$ to  $B$ is the set of operations from $\overline B$ to $\overline A$, where $\overline A$  ($\overline B$) is a suitable quantum system of dimension equal to the dimension of $A$  ($B$), 
\item every unitary channel  is an allowed evolution. 
\end{enumerate} 

The first two requirements imply that every  system  has a unique deterministic state. 
\begin{lemma}\label{prop:unique}
In a time symmetric  quantum theory satisfying Requirements (1) and (2),  the set of subnormalised states of a generic system $B$ contains exactly one density matrix $\omega_B$ with $\Tr[\omega_B]=1$. 
\end{lemma} 

\Proof  Applying  Requirement 2 in the special case where $A$ is one-dimensional, we obtain that the time reversal of  the set of quantum states of system $B$ is  the set of measurement operators of some other system $\overline B$.    By Proposition~\ref{prop:channel2channel},  this implies that the set of deterministic states of system $B$ coincides with the time reversal of the set of deterministic measurement operators on $\overline B$.   By Requirement 1, the set of deterministic measurement operators on $\overline B$ is a subset of the set of deterministic measurement operators in quantum theory. Since  the only deterministic measurement operator in quantum theory  is the identity, we conclude that system $B$ has  only one deterministic state.  \qed

\smallskip 

An immediate consequence of the existence of a unique deterministic state is the following characterisation of the allowed channels: 

\begin{lemma}\label{prop:channelset}
In a time symmetric  quantum theory satisfying Requirements (1) and (2),  the set of allowed channels with generic  input system $A$ and a generic  output system $B$ contains only quantum channels $\map C$ satisfying the condition $\map C (\omega_A)  = \omega_B$, where $\omega_A$  ($\omega_B$) is the unique deterministic state of system $A$  ($B$).  
\end{lemma}
The proof is immediate from Lemma~\ref{prop:unique}, which 
 implies that the subset of allowed channels should map the unique deterministic state of system $A$ into the unique deterministic state of system $B$.

The last step of our argument is to use Requirement 3, namely that all  unitary channels are allowed operations.  Combined with Lemma~\ref{prop:channelset}, Requirement 3 implies that the unique deterministic state of system $A$ should satisfy the condition $\map U (\omega_A)  =  \omega_A$ for every unitary channel $\map U$ acting on the system.   But the only unitarily invariant state is the maximally mixed state $\omega_A  =  I_{A}/_{d_A}$.      Hence, Lemma~\ref{prop:channelset} implies that the time symmetric variant can only contain quantum channels that map maximally mixed states into maximally mixed states. The largest set of such channels  is the set of time symmetric channels defined by Eq.~(\ref{tsopfinal}).  

Summarising, we obtained the following result
\begin{theo}
Every  time symmetric variant of  quantum theory satisfying Requirements (1)-(3)  is contained in the time symmetric theory defined in Subsecs.~\ref{subsec:tsop} and~\ref{subsec:tsinst}.
\end{theo}

In passing, we note that a generalisation of our time symmetric theory   can be obtained by lifting the requirement that all unitary channels are allowed operations.  In this case,  a time reversal of the   allowed quantum operations  can be defined in terms of  Petz' recovery map~\cite{petz1988sufficiency,hayden2004structure}:   for an allowed quantum operation $\map Q$ from system $A$ to  system $B$, the time reversal is  
\begin{align}\label{tradj1}
\Theta  (\map Q)  :  =   \omega_A^{1/2}  \,  \map Q^\dag  (\omega_B^{-1/2} \, \rho \,  \omega_B^{-1/2})  \,   \omega_A^{1/2}\, ,
\end{align}
where $\omega_A$  ($\omega_B$) is the unique deterministic state of system $A$  ($B$), and $\omega_B^{-1/2}$ denotes the inverse of $\omega_B$ on its support.   

Alternatively, one could define the time reversal in terms of the transpose map $\map Q^T$, as  
\begin{align}\label{trtr1}
\Theta'  (\map Q)  :  =   {\omega_A^*}^{1/2}  \,  \map Q^T  ({\omega_B^*}^{-1/2} \, \rho \,  {\omega_B^*}^{-1/2})  \,   {\omega_A^*}^{1/2}\, ,
\end{align}
where $\omega_A^*$ and $\omega_B^*$ are the complex conjugates of $\omega_A$ and $\omega_B$, respectively. 

The time reversal symmetries considered earlier in Eqs.~(\ref{tradj}) and~(\ref{trtr})  are a special case of the symmetries~(\ref{tradj1}) and~(\ref{trtr1}), corresponding to the choice maximally mixed states $\omega_A  =  I_{\rm in}/_{\rm in }$ and $\omega_B  =  I_{\rm out}/_{d_{\rm out }}$ as the unique deterministic states of systems $A$ and $B$. }

\section{ Relaxations of the notion of symmetry}\label{sec:other}
Here we briefly discuss two relaxations of the notion of operation symmetry, and we show that the set of all quantum operations admits a time symmetry in these two relaxed senses.  

\subsection{Non-bijective symmetries}    One way to evade Theorem~\ref{theo:nogo} is to relax the idea that a symmetry should be  bijective.     In particular, for finite dimensional systems one could consider the transformation $\map S:    \Op(\spc H_{\rm in}, \spc H_{\rm out}) \to  \Op(\spc H_{\rm out}, \spc H_{\rm in})  \,  ,    \map Q  \mapsto  \map Q^\dag  /d_{\rm in}$.  This transformation $\map S$ is  maps quantum operations  into quantum operations, is compatible with randomisation, and maps the null operation into the null operation.  It is injective, meaning that it maps distinct quantum operations into distinct quantum operations.  However, it falls short of being surjective: there exist quantum operations that are not of the form $\map S (\map Q)$  for any quantum operation $\map Q$.   If the map $\map S$ were used to describe time reversal, then the set of evolutions observed by an agent with inverted time arrow would be strictly smaller than the set of all quantum evolutions.   Basically, the time-reversed agent would only observe a scaled down version of the set of all possible quantum evolutions. Adopting the standard interpretation that the trace of the output state corresponds to the probability that a transformation occurs, this would mean that a time reversed transformation  $\map S(\map Q)$ would take place with probability at most $1/d_{\rm in}$. 
%A related issue is that the time reversal of a quantum channel would not be a quantum channel, since the map $\map S(\map C)  =  \map C/d$ is not trace-preserving. 

Note that the symmetry transformation $\map S$ is probabilistically reversible: if applied twice, it returns the original quantum operation $\map Q$, scaled down by a factor $1/d_{\rm in}^2$.  If this factor is ignored, one may say that  the set of quantum operations enjoys a symmetry under time reversal. We call such symmetries {\em weak symmetries}, to stress that they only hold up to  an overall scaling factor.  Most of the existing proposals of  time symmetric formulations of quantum theory \cite{aharonov1964time,reznik1995time,aharonov2010time,oreshkov2015operational} exhibit this type of weak symmetry.    Physically, the relevance of weak symmetry is argued based on post-selection:  if an agent postselects on the occurrence of a quantum operation, then any scaling factor becomes irrelevant.  

\medskip 

\subsection{Non-linear symmetries}   

In this paper we defined  operation symmetries as  transformations that preserve the convex structure and  the notion of zero probability, by mapping the null operation into itself.  These two requirements imply that every  symmetry transformation acts  linearly on the vector space spanned by quantum operations (see Proposition \ref{prop:linearsup}  in Appenrdix\ref{app:wignerqo}).

A way to evade Theorem~\ref{theo:nogo} is  to give up the above requirements and to consider non-linear symmetries, such as those considered in non-linear modifications of quantum mechanics~\cite{doebner1992general,doebner1999gauge,goldin2008nonlinear}.    These modifications, however, generally come with a significant change of the operational content of quantum theory.  For example, the state space in non-linear quantum mechanics is generally different from the space of density matrices, and ensembles of pure states that are indistinguishable in ordinary quantum mechanics may become distinguishable in  non-linear modifications~\cite{cavalcanti2012preparation}.   

{ We now show that, if one gives up the requirement of linearity,  a time-reversal can be defined for all possible quantum operations.  A non-linear time reversal was  introduced by Crooks in the case of quantum channels   with $\spc H_{\rm in}  =  \spc H_{\rm out}$~\cite{crooks2008quantum}.  Crooks' definition of time reversal coincides with Petz' recovery map~\cite{petz1988sufficiency,hayden2004structure}:  given a quantum channel $\map C$ and a quantum state $\rho_0$ such that $\map C (\rho_0)  =  \rho_0$,  Crooks' time reversal is the recovery  map  $\map C_{\rho_0}$  defined as  $\map C_{\rho_0} (\rho)  : =   \rho_0^{1/2}\,   \map C^\dag\,  (\rho_0^{-1/2} \, \rho \, \rho_0^{-1/2}) \, \rho_0^{1/2}$, where $\rho_0^{-1}$ is  inverse of $\rho_0$ on its support.  For bistochastic channels, one can choose $\rho_0$ to be the maximally mixed state, and this definition of time-reversal coincides with the definition in Eq.~(\ref{tradj}).  In general, the map $  \Theta  :  \map C  \mapsto  \map C_{\rho_0}$ is  non-linear, because it depends on the choice of a fixed point $\rho_0$ for the channel $\map C$.   

More generally, Crooks' definition can be extended to  channels with different input and output spaces, by fixing an arbitrary state $\rho_0 \in \St (\spc H_{\rm in})$ and defining the state-dependent time reversal $\map C_{\rho_0}$    as 
\begin{align}\label{petzinout}\map C_{\rho_0}  (\rho)  :  =  \rho_0^{1/2}   \map C^\dag  \left(  [\map C  (\rho_0)]^{-1/2} \, \rho\,  [\map C  (\rho_0)]^{-1/2}   \right)  \rho_0^{1/2} \,.
\end{align} 
  Note, however, that  the above definition is non-canonical, due to  the arbitrariness in the choice of the  state $\rho_0$.

Similarly, a time reversal could be defined for arbitrary quantum operations.   Given a quantum operation $\map Q$, one could {\em (1)} fix a complementary quantum operation $\map Q'$ such that the map $\map C_0:  = \map Q+  \map Q'$ is a quantum channel, and  {\em (2)} fix  an arbitrary  quantum state $\rho_0\in \St (\spc H_{\rm in})$.  Then, a time reversal  $\map Q_{\map C_0,\rho_0}$ could be defined via the relation 
 \begin{align}\label{petzqo}
 \map Q_{\map C_0, \rho_0}  (\rho)  :=   \rho_0^{1/2} \,   \map Q^\dag   \left(  \,   [\map C  (\rho_0)]^{-1/2}  \, \rho \,    [\map C  (\rho_0)]^{-1/2} \right) \,   \rho_0^{1/2} \, .
 \end{align}    
 This definition would guarantee that $\map Q_{\map C_0, \rho_0}$  is a valid quantum operation.  Still, the map $\Theta:  \map Q\mapsto  \map Q_{\map C_0, \rho_0}$ is non-linear and non-canonical, due to the arbitrariness in the choice  of channel $\map C_0$ and   state $\rho_0$.}

\section{Conclusions}  
\label{sec:conclusion}

In this work we introduced a notion of symmetry for quantum evolutions, defined as a one-to-one transformation that is consistent with randomisations and with the notion of zero probability. We  established a Wigner theorem for quantum operations,  showing that every symmetry of quantum operations can be broken down into two symmetries of quantum states, which are either both unitary or both antiunitary.

We then showed that the time reversal symmetry of unitary quantum dynamics cannot be extended to a symmetry of the full set of stochastic quantum evolutions.  In other words, the set of all quantum evolutions is incompatible with time symmetry.   Our no-go theorem implies  that it is impossible to translate the  quantum description of experiments made by an ordinary agent into a quantum description of experiments  made by a hypothetical agent with time-reversed arrow of time: if the ordinary agent can perform all possible quantum experiments (and therefore all possible quantum operations), then  its experiments cannot be described by the time-reversed agent.   

It is worth noting that the impossibility of defining a time reversal symmetry is not necessarily unique to quantum theory.  As a matter of fact, most of the the steps in the proof of our no-go result  can be reproduced also in classical probability theory, leading to a no-go theorem for time reversal symmetries of classical stochastic evolutions.  What makes the quantum case more interesting, however, is that in the classical world there is a natural notion of time symmetric  theory, namely {\em deterministic} classical theory, with only pure states and  reversible classical evolutions. 
 In quantum theory, instead, defining a time symmetric theory is more problematic, at least in the standard operational interpretation where  measurements play a central role.  One could,  of course, consider the fragment of quantum theory containing only pure states and unitary evolutions, as it is done in Everett's interpretation of quantum mechanics. If this route is taken, the remaining difficulty would be to give a consistent account of how the theory is to be used to make predictions about the outcomes of experiments \cite{saunders2010many}.    On the other hand, if even just a single projective measurement is  added on top of pure states and unitary evolutions, then all quantum operations can be generated,  and time symmetry is broken.

After establishing our no-go theorem for time symmetry in the standard framework, we formulated a time symmetric variant of quantum theory in which the deterministic operations are required to preserve the maximally mixed state.  We showed that our time symmetric variant is maximal among all time symmetric variants where all unitary channels are allowed evolutions, and that the standard operational framework of quantum theory can be retrieved from the time symmetric variant through an operation of conditioning on the outcomes of state preparations.    

The results of this work provide a rigorous framework for the analysis of  dynamical symmetries in quantum theory.  We hope that the framework and the analysis initiated in this work will inform future developments  in the foundations of quantum mechanics and in  quantum information theory.

    %no symmetry of the set of quantum operations can reverse the action of arbitrary unitary dynamics.

%Furthermore, we conjecture that the largest subset of quantum channels admitting a time reversal symmetry is the set of bistochastic channels, that is, channels that map the maximally mixed state into itself.
%Support to this conjecture comes again from Ref.~\cite{chiribella2020quantum}, which characterised the set of bistochastic channels as the maximal set of channels admitting a time reversal that reverses the order of processes.   
%Extending this characterisation to more general notions of time reversal remains as an open problem for future research.  

%Finally, as   time reversal  symmetry cannot be defined for open quantum systems
%in a way consistent with the quantum theory (at least not on the level of quantum operations), one may look for possible approximate solutions
%to this problem \cite{AZZ}.   
%The characterisation of the optimal approximate time reversals remains an open problem, whose answer however is likely to depend on the choice of  optimality criteria.

\section*{Acknowledgments}
  It is a pleasure to thank Ingemar Bengtsson, Vern Paulsen, Mizanur Rahaman, and Erling St\o rmer for 
  a fruitful correspondence.   GC acknowledges  A Winter for an insightful  discussion on Petz' recovery map that led to Equations~(\ref{tradj}),~(\ref{trtr}), and~(\ref{petzinout}),  and to J Barrett for helpful comments on an earlier version of the results. 
  This work was supported by the National Natural Science Foundation of China through grant 11675136, by the Hong Kong Research Grant Council through grant 17307719 and though the Senior Research Fellowship Scheme SRFS2021-7S02, by the Croucher Foundation,  by the John Templeton Foundation through grant  61466, The Quantum Information Structure of Spacetime  (qiss.fr), by the Swedish Research Council through grant 2020-04980,  by the Foundation for Polish Science through TEAM-NET project (contract no. POIR.04.04.00-00-17C1/18-00), and  
  by the Polish National Science Centre 
  under the project number DEC-2015/18/A/ST2/00274. Research at the Perimeter Institute is supported by the Government of Canada through the Department of Innovation, Science and Economic Development Canada and by the Province of Ontario through the Ministry of Research, Innovation and Science. The opinions expressed in this publication are those of the authors and do not necessarily reflect the views of the John Templeton Foundation.

\appendix

\section{Proof of Theorem \ref{theo:wignerdensity}}\label{app:wignerdensity}

The proof of Theorem \ref{theo:wignerdensity} is based on the following argument:  
\begin{enumerate}
\item every state space symmetry must map pure states into pure states, and therefore it induces a transformation of ray space
\item the induced transformation of ray space preserves the ray product,
\item hence, every state space symmetry induces a ray space symmetry, which can be characterised using (the original formulation of) Wigner's theorem. 
\end{enumerate}

In the following we provide the proof of all the steps used in this argument.  Let us start by showing that every state space symmetry maps pure states into pure states.   For a given vector $\psi  \in \spc H$, we will use the notation $P_\psi  :=   |\psi\>\<\psi|$.
\begin{lemma}
\label{lem:pure2pure}
Every state space symmetry  $\map  S $ maps pure states into pure states.  
 The correspondence $P_\psi  \mapsto \map S(P_\psi)$ is bijective. 
\end{lemma}  
\Proof The proof is standard and is reproduced here just for completeness. For a generic unit vector $\psi  \in \spc K$,   the   operator  $\map S (P_\psi)$ is a density matrix.   Hence, $\map S (P_\psi)$ can be decomposed as as $\map S (P_\psi)  =  \sum_i\,  p_i\, P_{\psi_i}$ where $\{ p_i\}$ are probabilities and $\{|\psi_i\> \}  \in  \spc H$ are unit vectors.   Since $S$ is a state space symmetry, one has $ P_\psi    =   \map S^{-1}\circ \map S (P_\psi)   = \sum_i \, p_i \, \map S^{-1}  (P_{\psi_i})$, and, therefore $  \map S^{-1}  (  P_{\psi_i} )   =  P_\psi$ for all $i$.   Applying $\map S$ on both sides of the equality, one obtains $P_{\psi_i}    =  \map S  (P_\psi)$, $\forall i$.  Hence, $\map S (P_\psi )$ is a pure state.  

The correspondence $P_\psi\mapsto  \map S (P_\psi)$ is  injective, because $\map S$ is injective on the whole space of density matrices.    Moreover, $\map S$ is surjective: since the inverse transformation $\map  S^{-1}$  maps pure states into pure states, every  pure state  $P_\phi $  is the image of the pure  state $\map S^{-1}(P_{\phi})$ under the map $\map S$.   \qed

Now, recall that the set of rank-one projectors is in one-to-one correspondence with the set of unit rays. For every rank-one projector $P$, let ${\rm ray}  (P)$ be the corresponding unit ray, defined as 
\begin{align}\label{raydef}
{\rm ray}  (P) :  = \left.\left\{\frac{P \psi  }{ \|    P\psi\|}  ~\right|~  \psi  \in  \spc H  \, , P\psi  \not =  0  \right\} \, .
\end{align} 
Owing to the one-to-one correspondence between rank-one projectors and unit rays, we have the following  lemma. 
\begin{cor}\label{cor:underlineS}
Every state space symmetry $\map S: \St (\spc H) \to \St(\spc K)$ induces a bijective  transformation $\underline S  :  \underline {\spc H}\to \underline {\spc K}$ via the relation  
\begin{align}\label{underlineS}\underline S  (\underline \psi)  :  =  {\rm ray}     \left(    \map S  (  P_\psi)  \right) \,. 
\end{align} 
\end{cor}
   The next step is to show that the ray space transformation   $\underline S$ preserves the ray product. In preparation for this, we  introduce some notation.  For two generic Hilbert-Schmidt operators $X :  \spc H \to \spc H$ and $Y  : \spc H\to \spc H$, we denote their Hilbert-Schmidt product by 
   \begin{align} \<  X , Y  \>   := \Tr[ X^\dag  Y] \,.
   \end{align}   The ray product of two unit rays $\underline \psi$ and $\underline \phi$ can be expressed in terms of the Hilbert-Schmidt product  as 
\begin{align}\label{HSproduct}
\underline {\psi}  \cdot \underline \phi    =  \sqrt{     \<P_\psi,  P_\phi\> } \, .
\end{align}
For rank-one projectors  $P_\psi$  and $P_\phi$, the Hilbert-Schmidt product  can be equivalently expressed as 
\begin{align}\label{HSfid}
\< P_\psi,  P_\phi\>  =  F  (  P_\psi, P_\phi) \,,
\end{align} where $F(  \rho,  \sigma)    :=  \left( \Tr[ \sqrt{  \sqrt  \rho  \sigma \sqrt \rho }  ]   \right)^2 $  is the (square of the) {\em Uhlmann fidelity}  \cite{Uhl,Jozsa}, defined for any two density matrices $\rho$ and $\sigma$.   
 
The last definition we need is the definition of state space homomorphism: 
\begin{defi}
A map $\map  J  :  \St  (\spc H)  \to  \St (\spc K) $  is a {\em state space homomorphism} if it preserves convex combinations.
%, namely $J  \big(  p\,\map J (\rho) +  (1-p)  \, \map J(\sigma)\big)  =  p \,  \map J(\rho)  +  (1-p)\,   \map J(\sigma)$ for every pair of density matrices $\rho$ and $\sigma$, and for every probability $p\in  [0,1]$.  
%if it is positive and trace-preserving. 
\end{defi}

  \iffalse
State space homomorphisms are the most general maps transforming density matrices into density matrices. In general, they may not be  quantum channels, because they may not be completely positive. An example of state space homomorphism that is not a quantum channel is the partial transpose map $ \rho \mapsto \rho^T$. 
\fi

Recall that every state space homomorphism 
 can be equivalently represented in the ``Heisenberg picture'', as a map on observables.    Specifically, the Heisenberg picture of the map $\map  J  :  \St  (\spc H)  \to  \St (\spc K) $ is provided by the dual map
   $\map J^\dag: L (  \spc K) \to  L  (\spc H)$, uniquely defined by  the condition
   \begin{align}
   \<  \map J^\dag    (A)  ,   P_\psi  \>    =    \<  A  ,  \map J (P_\psi) \>  
    \qquad  \forall  A  \in  B(\spc K) \,  ,\forall  \psi   \in  \spc H \, .   
   \end{align}   
   
 \begin{lemma}\label{lem:positiveunital} A map $\map  J  :  \St  (\spc H)  \to  \St (\spc K) $ is a state space homomorphism if and only if the dual map $\map J^\dag: B(\spc K) \to B(\spc H)$ is positive  and unital. 
\end{lemma}
\Proof  
 We recall that a map is  positive if it maps non-negative operators into non-negative operators.   
  Let $P : \spc K \to \spc K$ be a non-negative operator. Then,  for every unit vector $\psi \in  \spc H$,  one has  $\<  \psi  |   \map J^\dag  (P)|\psi\>  =  \<  \map J^\dag (P)  , P_{\psi} \>  =   \<  P ,    \map J(P_\psi) \>  =  \Tr [  P \, \map J(P_\psi)]  \ge  0$, the last inequality following from the fact that $\map J(P_\psi)$ is a density matrix.  Hence, $\map J^\dag (P)$ is a non-negative operator, and $\map J^\dag$ is positive.  

Similarly, it is easy to see that $\map J^\dag$ is unital, namely $\map J^\dag  (I_{\spc K})  =  I_{\spc H}$.   Indeed, for every unit vector $\psi \in  \spc H$, one has  $\<\psi| \map J^\dag  (I_{\spc K})  |\psi\>  =  \<  \map J^\dag  (I_{\spc K}) ,  P_\psi\>  =  \<   I_{\spc K}  ,    \map J  (  P_\psi)  \>    =   \Tr[  \map J  (P_\psi)]  = 1$, because $\map J(P_\psi)$ is a density matrix.  
Summarising, the map $\map J^\dag$ is positive and unital.  

Conversely, if the map $\map J^\dag$ is positive and unital, then the map $\map J$ transforms density matrices into density matrices:   indeed, it transforms positive operators into positive operators and unit-trace operators into unit-trace operators, as one can see by running the above arguments in reverse. Moreover, since $\map J^\dag$ is linear, the map $\map J  =   \left(\map J^{\dag}\right)^\dag$ is linear, too. In particular, it is convex-linear.  Hence, $\map J$ induces a map from $\St (\spc H)$ to $\St(\spc K)$ that is compatible with convex combinations. \qed 

An important property of state space homomorphisms  is that they cannot increase the distinguishability of quantum states, or equivalently, they cannot decrease the similarity of quantum states. Taking the fidelity as the measure of similarity, we have the following 
\begin{lemma}
\label{lem:fid}
Every state space homomorphism  $\map  J : \St (\spc H) \to \St (\spc K)   $ is fidelity non-decreasing, namely  
\begin{align}
F (\map J  (\rho),  \map J  (\sigma))  \ge  F(\rho, \sigma)  \qquad \forall \rho,\sigma \in \St (\spc K)  \, .
\end{align} 
If $\map J$ is a state space symmetry,  then the equality holds. 
\end{lemma}  

\Proof  The fidelity between two generic density matrices $\rho$ and $\sigma$ can be equivalently expressed as 
 \begin{align}
F(\rho, \sigma)     &  =  \left( \max_{  \{  P_i\}}   \,  \sum_i    \sqrt{  \Tr  [  P_i  \, \rho ] \,    \Tr  [  P_i  \,\sigma]} \right)^2  \,,
\end{align} 
where the maximum is over all positive operator-valued measures (POVMs) $(  P_i)$, consisting of positive semidefinite operators $P_i\ge 0$ satisfying the normalisation condition $\sum_i  P_i  =  I$.  
Hence, one has 
\begin{align}
\nonumber &
F( \map J(\rho)  ,  \map J(\sigma)) \\  
\nonumber  & \qquad =    \left( \max_{  \{  P_i\}}   \,  \sum_i    \sqrt{  \Tr  [  \map J^\dag (P_i) \,   \rho] \,    \Tr  [ \map J^\dag ( P_i) \,    \sigma]} \right)^2  \\
 \nonumber & \qquad \ge    \left( \max_{  \{  Q_i\}}   \,  \sum_i    \sqrt{  \Tr  [  Q_i \,   \rho ] \,    \Tr  [  Q_i  \, \sigma]} \right)^2 \\
 & \qquad = F(\rho,\sigma) 
  \end{align}
  where the inequality follows from the fact that the operators $Q_i:  =  \map J^\dag  (P_i)$ form a POVM, because $\map J^\dag$ is positive and unital (Lemma \ref{lem:positiveunital}). 
   
  If $\map J$ is a state space symmetry, then one has $F (  \rho,\sigma) =  F (\map J^{-1}\circ \map J (\rho),  \map J^{-1}  \circ \map J (\sigma))   \ge  F (\map J  (\rho),  \map J (\sigma))  \ge  F (\rho, \sigma)$, whence the equality $F (\map J(\rho), \map J(\sigma))  =  F(\rho, \sigma)$.  \qed

\iffalse 

Note that a state space homomorphism with a positive inverse is  a state automorphism:  
\begin{lemma}. 
\label{lem:invertible}
Let $\map J\in\Map (\spc K)$ be an injective  state space homomorphism, and let  $\map J^{-1}  \in \Map  (\spc K)$ be its inverse.   If the inverse $\map J^{-1}$ is a positive map, then $\map J$ is a state space automorphism. 
\end{lemma}
\Proof  It is enough to show that the inverse  is trace-preserving.   Since $\map J$ is trace-preserving,  one has $\Tr [  \map J^{-1}  (X)]   =\Tr  [ ( \map J  \circ \map J^{-1}  )  (X)]   =  \Tr[X]$ for every $X\in L(\spc H)$.  Hence, $\map J^{-1}$ is trace-preserving. \qed 

\fi

We are now ready to show that every state space symmetry induces a ray space symmetry.  

\begin{lemma} 
\label{lem:symmetry}
For every state space symmetry $\map S:  \St (\spc H) \to \St (\spc K)$, the associated ray space transformation $\underline S: \underline {\spc H} \to \underline {\spc K}$, defined in Eq. (\ref{underlineS}), preserves the ray product. 
\end{lemma}  

\Proof  For two generic unit vectors $\psi$ and $\phi$,  Eqs.~(\ref{HSproduct}) and (\ref{HSfid}) imply $\underline \psi \cdot \underline \phi  =  \sqrt{\<P_\psi, P_\phi\>}  =\sqrt{  F  (P_\psi, P_\phi)}$.    On the other hand, Lemma \ref{lem:fid} guarantees the equality $F  (P_\psi, P_\phi)  =  F  (\map S  (P_\psi), \map S  (P_\phi))$.   In turn, Eqs.~(\ref{HSfid})  and (\ref{HSproduct}) imply    $ F  (\map S  (P_\psi), \map S  (P_\phi))=  \<  \map S  (  P_\psi),  \map S  (P_\phi)\>   =\left[  {\rm ray} \Big( \map S  (  P_\psi) \Big)  \cdot   {\rm ray} \Big( \map S  (  P_\phi) \Big) \right]^2$, with the notation ${\rm ray} (\cdot)$  defined as in Eq.~(\ref{raydef}).  Finally, the definition of the map $\underline S$ in Eq.~(\ref{underlineS}) yields the equality ${\rm ray} \Big( \map S  (  P_\psi) \Big)  \cdot   {\rm ray} \Big( \map S  (  P_\phi) \Big)   =  \underline S  (\underline \psi) \cdot  \underline S (\underline \phi)$. 
  Hence, we obtained the equality $\underline \psi \cdot \underline \phi    = \underline S  (\underline \psi) \cdot  \underline S (\underline \phi) $.  Since the unit vectors $\psi$ and $\phi$ are arbitrary, we conclude that $\underline S$ preserves the ray product.    \qed

We then conclude with the proof of Theorem \ref{theo:wignerdensity}.  

\medskip

{\bf Proof of Theorem \ref{theo:wignerdensity}.}  For a given space symmetry  $\map S:  \St (\spc H) \to \St (\spc K)$,  the corresponding ray space transformation  $\underline S: \underline {\spc H} \to \underline {\spc K}$ is bijective (Corollary~\ref{cor:underlineS}) and preserves the ray product (Lemma~\ref{lem:symmetry}).    Hence, it is a ray space symmetry.  

 By Wigner's theorem,   $\underline S$  has either the form $\underline S (\underline \psi)  =  \underline {  U \psi}$, or the form $\underline S (\underline \psi)  =  \underline {  U \psi^*}$, for some unitary operator $U:\spc H\to \spc K$.  In terms of rank-one projectors, this means that one has either   $\map S(P_\psi )  = % {\rm ray}^{-1}     \big(\underline S (\underline \psi)\big) = 
  U  P_\psi  U^\dag$ or $\map S(P_\psi) =%{\rm ray}^{-1}     \big(\underline S (\underline \psi)\big)  = 
  U P_{\psi^*}  U^\dag   =  U   P_\psi^T  U^\dag$.    Since  the map $\map S$ is uniquely determined by its action on the pure states, the action of $\map S$ on a generic density matrix $\rho$ is either of the form $\map S (\rho)  =  U \rho U^\dag$ or of the form $\map S(\rho)  =  U  \rho^T  U^\dag$.  \qed 

\iffalse
 \begin{lemma}
 \label{lem:twoforms}
Every state space automorphism   $\map J \in \Map (\spc K)$ is 
either of the form $\map J (X)  =  U XU^\dag$ or of the form $\map J(X)  =  U  X^T  U^\dag$, where $U\in L(\spc K)$ is a unitary operator.    
%In particular, $\map J$ is unital, namely $\map J (I)  =  I$. 
\end{lemma} 

\Proof    
Lemma \ref{lem:symmetry} shows that the restriction of  $\map J$  to $\Pur\St(\spc K)$ is a symmetry transformation, call it $S$. Corollary \ref{cor:wigner} shows that any such transformation can be extended to a linear map  $\widetilde S  \in  \Map (\spc K)$   that is either of the form $\widetilde S (X)  =  U XU^\dag$ or of the form $\widetilde S(X)  =  U  X^T  U^\dag$. % Note that $  \widetilde S$ is unital.  
Since $J$ is itself linear, we have $\map J  = \widetilde  S$. \qed 
 \fi 
 
  \section{Equivalence between the original Wigner theorem and its density matrix formulation}\label{app:equivalence}  
 
 In the previous Appendix, we have seen that the original formulation of Wigner's theorem (Theorem~\ref{theo:wigner}) implies the characterisation of state space symmetries of Theorem~\ref{theo:wignerdensity}.   We now show that the converse also holds:  Theorem~\ref{theo:wignerdensity} implies Wigner's theorem. 
   
 Let $\underline S  : \underline {\spc H} \to \underline {\spc K}$ be a ray space symmetry.    Our goal is to show that $\underline S$ is induced by a unitary or antiunitary transformation in Hilbert space. 
   Note that there is a one-to-one correspondence between the unit vectors $\psi \otimes \psi^* \in \spc H \otimes \spc H$ and the unit rays $\underline \psi \in  \underline{\spc H}$.  Hence, the transformation $\underline S$ induces a transformation  of the vectors $\psi \otimes \psi^*$, hereafter denoted by $S_{\rm prod}$.    If $\underline S   (\underline \psi)  =  \underline \psi'$, then $S_{\rm prod}  (  \psi\otimes \psi^*)   =     \psi'\otimes \psi^{\prime \,*}$. 
 
 Now, let $\{\psi_i\} \subset \spc H$  be a set of unit vectors with the property that $\{  \psi_i\otimes \psi_i^*\}$ is a spanning set for the space $\spc H\otimes \spc H$.   Then, we have 
 \begin{align}
\nonumber  \< \, S_{\rm prod}  (  \psi_i\otimes \psi_i^*)  \,   |\, S_{\rm prod}  (  \psi_j\otimes \psi_j^*)  \,  \>   & = \<  \,    \psi_i'\otimes \psi_i^{\prime \,*}   |    \psi_j'\otimes \psi_j^{\prime \,*}   \, \>\\
 \nonumber & =  |\<  \psi'_i |\psi'_j\>  |^2 \\
 \nonumber &  =  \underline {\psi_i'} \cdot \underline {\psi_j'}  \\
 \nonumber &  =  \underline {\psi_i} \cdot \underline {\psi_j}  \\
 \nonumber & =  |\<  \psi_i |\psi_j\>  |^2 \\
 \nonumber   & = \<  \,    \psi_i\otimes \psi_i^{*}   |    \psi_j\otimes \psi_j^{*}   \, \> \, .
 \end{align}
 In other words, the two sets $\{  \psi_i\otimes \psi_i^*\}$ and $\{  \psi'_i\otimes \psi_i^{\prime \,*}\}$ have the same Gram matrix. This condition guarantees that there exists a unitary transformation $W   : \spc H  \otimes \spc H  \to \spc K\otimes \spc K$ such that $\psi'_i\otimes \psi_i^{\prime \,*}    =   W     (\psi_i\otimes \psi_i^*)$  (for a proof, see e.g.~Property 3 of Ref.~\cite{chefles2004existence}).    The transformation  $W$ is uniquely defined, because the states $\{  \psi_i\otimes \psi_i^*\}$ are a basis.   Using this fact, we can guarantee the equality $\psi'\otimes \psi^{\prime \,*}    =   W     (\psi\otimes \psi^*)$  for every unit vector $\psi$.  
Hence, the transformation $\map S_{\rm prod}$ is linear.     

Now, since the product vectors $\psi\otimes \psi^*$ are in one-to-one linear correspondence with the rank-one projectors $P_\psi$, the linear transformation $S_{\rm prod}$ induces a linear transformation $\map S$ on the rank-one projectors $P_\psi$.    By linearity, the transformation  $\map S$ acts   on the set of density matrices, and is consistent with randomisations.   Moreover, since $S_{\rm prod}$ is a bijection, also $\map S$ is.  
Hence, it is a symmetry of the set of quantum states. By applying Theorem~\ref{theo:wignerdensity}, we then obtain that $\map S$ is either a unitary transformation $\map S:   \rho \mapsto U \rho U^\dag$ or an antiunitary transformation $\map S:  \rho  \mapsto U\rho^T U^\dag $.     Restricting its action on pure states,  we have that the pure state $|\psi\>\<\psi|$ is transformed either into $U|\psi\>\<\psi|U^\dag$ or into $U \, \left( |\psi\>\<\psi | \right)^T  \, U^\dag  =   U \,  |\psi^*\>\<\psi^* |   \, U^\dag$. Translating back in terms of the map $\underline S$, we obtain that $\underline S (\underline \psi) $ is either equal to $\underline{U\psi}$ or equal to $\underline{  U\psi^*}$.  Hence, $\underline S$ is either induced by a unitary transformation or by an antiunitary transformation.  This concludes the derivation of Wigner's theorem from Theorem~\ref{theo:wignerdensity}. \qed

  \section{Proof of Proposition~\ref{prop:channel2channel}}\label{app:channel2channel} 

Here we prove that every symmetry of quantum operations maps channels into channels.

  Let $\map S  :   \Op  (\spc H_{\rm in}, \spc H_{\rm out})    \to  \Op  (\spc K_{\rm in}, \spc K_{\rm out})$ be a symmetry of  quantum operations, and let $\map C  \in \Chan  (\spc H_{\rm in}, \spc H_{\rm out})$ be a quantum channel. Since the set of quantum channels is a subset of the set of quantum operations, $\map S  (\map C)$ is a quantum operation.   

Now, for every quantum operation $\map Q$ there exists another quantum operation $\map Q'$ such that $\map Q  + \map Q'$ is a quantum channel. Let $\map Q'$ be the quantum operation such that $\map C':  = \map S (\map C) + \map Q'$ is a quantum channel.   
Applying the inverse map $\map S^{-1}$ to both sides of the equation, we obtain $\map S^{-1} (\map C')  =  \map C  +  \map S^{-1}  (\map Q')$.  Since the map $\map S^{-1}$ transforms quantum operations into quantum operations, the map $\map C  +  \map S^{-1}  (\map Q')$ must be trace non-increasing.   Since both terms in the sum are completely positive, and since $\map C$ is trace preserving, we obtain the condition $\map S^{-1}  (\map Q') =  0$.  Hence, we obtained $\map S^{-1}  (\map C')  =  \map C$.  Applying the map $\map S$ on both sides, we finally conclude $\map C'   =  \map S  (\map C)$, that is, $\map S (\map C)$ is a quantum channel. Since $\map C$ was a generic channel, we conclude that $\map S$ maps channels into channels. \qed

 \section{Proof of Theorem~\ref{theo:wignerqo}}\label{app:wignerqo}
 
 In this section we first review the technical tools used in the proof, and then provide the derivation of Wigner's theorem for quantum operations.   
  
\subsubsection{Supermaps}
First of all, we observe that every symmetry of quantum operations  induces  a  linear map acting on the vector space spanned by quantum operations. 
Such higher-order linear maps are called supermaps~\cite{chiribella2008transforming,Zy08,chiribella2009theoretical,chiribella2013quantum,chiribella2013normal}, and can be used for a variety of purposes.  The special type of supermaps that transform quantum operations into quantum operations  were characterized in Ref.~\cite{chiribella2008transforming} in the finite dimensional case, and in Ref.~\cite{chiribella2013normal} in the infinite dimensional case. Another type of supermaps was defined in Ref.~\cite{Zy08}, which considered transformations of a certain convex set of completely positive maps into itself, and used them to define the dynamics in an extended version of quantum theory where the state space of a $d$-dimensional system has dimension $d^4$. 

To avoid technical complications, in the following we will restrict ourselves to the finite-dimensional case.    In the following, we will use the notation $L(\spc H,\spc K)$ for the space of linear operators on a generic Hilbert space $\spc H$ to a generic Hilbert space $\spc K$, the notation $L(\spc H):  =  L(\spc H, \spc H)$ for the linear operators on $\spc H$, and the notation $\Map  (\spc H, \spc K)$ for the set of linear maps from $L(\spc H)$ to $L(\spc K)$.   

\begin{prop}\label{prop:linearsup}
Every map $\map S:  \Op  (\spc H_{\rm in},  \spc H_{\rm out})\to  \Op  (\spc K_{\rm in},  \spc K_{\rm out})$ that is consistent with randomisations and preserves the null operation can be uniquely extended to a linear map  $\map S_{\rm lin}:  \Map  (  \spc H_{\rm in},  \spc H_{\rm out})  \to  \Map  (  \spc K_{\rm in},  \spc K_{\rm out})$. 
\end{prop} 

\Proof  Since the map $\map S$ is consistent with randomisations and preserves the null operation, we have
\begin{align}
\nonumber \map S ( p\,\map Q  ) &  =  \map S  ( p \,  \map Q  +  (1-p) \,  0  ) \\
 \nonumber &  =  p  \, \map S (\map Q)  +  (1-p) \,  \map S  (0)  \\
  &  =  p\,  \map S  (\map Q) \, ,   \label{consistentwithscaling}  
\end{align} 
for every probability $p \in  [0,1]$ and for every quantum operation $\map Q$. 

Due to Eq.~(\ref{consistentwithscaling}), any two quantum operations $\map Q $ and $\map Q'$ satisfying the condition $\lambda' \, \map Q'  =  \lambda \,  \map Q$ for positive numbers $\lambda$ and $\lambda'$, must also satisfy the condition $\lambda' \, \map S  (\map Q') =  \lambda \, \map S (\map Q)$. 

Now, we show that the map $\map S$ is linear on  conic combinations:  if $\sum_i \, \lambda_i  \,  \map Q_i  =  \sum_j  \, \lambda_j'  \, \map Q_j'$,  then 
\begin{align}\label{consistentwithcone}
 \sum_i \, \lambda_i  \,  \map S \left(\map Q_i  \right)   =   \sum_j  \, \lambda_j'  \, \map S\left( \map Q_j' \right) \,,
\end{align} 
for every set of quantum operations $\{  \map Q_i\}$  ($\{\map Q_j'\}$) and positive coefficients $\{\lambda_i\}$  ($\{\lambda_j'\}$).   

To prove Eq.~(\ref{consistentwithcone}),  we define the probability distributions $(p_i)$ and $(p_j)$ as 
\begin{align}
\nonumber p_i:  =  \frac {\lambda_i}\lambda \, , \quad   \lambda  :=  \sum_i  \lambda_i \, , \quad p_j':  =  \frac {\lambda_j'}\lambda'  \, ,\quad  \lambda'  :=  \sum_j  \lambda_j'  \, ,
\end{align}   
and the quantum operations
\begin{align}
\map Q  :  = \sum_i \,  p_i \, \map Q_i  \, ,\qquad \map Q'  :  = \sum_j \,  p_j' \, \map Q_j' \, . 
\end{align}  
With these definitions we have the equality $\lambda' \, \map Q'  =  \lambda \, \map Q$, which implies $\lambda'  \map S  (\map Q')  =  \lambda \,  \map S  (\map Q)$.  Hence, we have 
\begin{align}
\nonumber
 \sum_i  \,   \lambda_i \,  \map S  (\map Q_i)  
\nonumber &  =    \lambda \,  \sum_i  \,  p_i  \,  \map S  (\map Q_i)  \\
\nonumber &  =    \lambda \,   \map S \left(  \sum_i  \,  p_i  \,  \map Q_i \right)  \\
\nonumber &  =    \lambda \,   \map S \left(    \map Q \right) \\  
\nonumber &  =    \lambda' \,   \map S \left(    \map Q' \right)  \\ 
\nonumber &  =    \lambda' \,   \map S \left(  \sum_j  \,  p_j'  \,  \map Q_j' \right)  \\
\nonumber &  =    \lambda' \,   \sum_j  \,  p_j'  \,   \map S \left(   \map Q_j' \right)  \\ 
 & =\sum_j  \,   \lambda_j' \,  \map S  (\map Q_j') \, .  
\end{align}

Hence,  the map $\map S$ can be uniquely extended to a map $\map S_{\rm lin}$ transforming  the cone 
\begin{align}\Op_+  (\spc H_{\rm in},  \spc H_{\rm out}) :  = \{   \lambda\,  \map Q ~|~  \map Q  \in  \Op  (\spc H_{\rm in},  \spc H_{\rm out}) \, ,\lambda \ge  0  \}
\end{align} 
into the cone
\begin{align}\Op_+  (\spc K_{\rm in},  \spc K_{\rm out}) :  = \{   \lambda\,  \map Q ~|~  \map Q  \in  \Op  (\spc K_{\rm in},  \spc K_{\rm out}) \, ,\lambda \ge  0  \} \, .
\end{align} 
and the map $\map S_{\rm lin}$ is linear on conic combinations.  

To conclude, we invoke the fact that every  map from a cone $\set C$ to a cone $\set C'$ that is linear on conic combinations can be uniquely extended to a linear map from the linear span of $\set C$ to the linear span of $\set C'$~\cite{holevo2011probabilistic}. Since the linear span of the cone of quantum operations $ \Op_+  (\spc H_{\rm in},  \spc H_{\rm out})$ is the space of all maps $\Map  (  \spc H_{\rm in},  \spc H_{\rm out})$, this concludes the proof. \qed 

Since the linear extension is unique, in the following we will drop the subscript from $\map S_{\rm lin}$, and simply write $\map S$.

\subsubsection{Choi representation of linear maps and supermaps}
 
The other key ingredient in the proof of Theorem~\ref{theo:wignerqo} is  the Choi representation of linear maps~\cite{Cho75a}, widely used in quantum information theory~\cite{leung2000towards,d2001quantum,AP04,ZB04} and quantum foundations~\cite{abramsky2004categorical,chiribella2010probabilistic}. The Choi representation of a map  $  \map M :   L(\spc H_{\rm in}) \to L( \spc H_{\rm out})  $ is the linear operator ${\rm Choi} (\map M)  \in  L(\spc H_{\rm out}\otimes \spc H_{\rm in})$ defined by  
%$\Choi :    \Map  (\spc H_{\rm in}, \spc H_{\rm out})  \to  L(\spc H_{\rm out}\otimes \spc H_{\rm in})$, is defined by
\begin{equation} % {align}
\label{choi_matrix}
{\sf Choi} (\map M)  :  =  \sum_{i,j}  \map M  (  |i\>\<j|)  \otimes |i\>\<j|  \,, 
  \end{equation}
  where $\{|i\>\}$ is a fixed (but otherwise arbitrary) basis for the input Hilbert space $\spc H_{\rm in}$.   % {align}
  %%%%%%%%%
 %\iffalse 
The Choi representation is related to the Jamio\l kowski representation~\cite{Ja72}  
\begin{equation}
{\sf Jam}  (\map M)  :  =  \sum_{i,j}  \map M (|i\>\<j|) \otimes |j\>\<i| \,, 
\end{equation} 
the difference between the two representations being a partial transpose operation on the input Hilbert space.
% namely ${\sf Jam}  (\map M)= {\sf Choi} (\map M)^{T_{\rm in}}$.   
 
For a quantum operation $\map Q$,  the Choi operator is proportional to an unnormalised state, namely $  {\rm Choi} (\map Q)/d_{\rm in} $.  The relation between the set of quantum maps  
  and the set of bipartite states was investigated
  in \cite{AP04,ZB04}, and in \cite{chiribella2010probabilistic}, which extended the correspondence to a class of physical theories including  quantum theory as a special case.

The Choi representation  offers a convenient way to represent supermaps. In the Choi representation, a supermap $\map S:    \Map  (\spc H_{\rm in}, \spc H_{\rm out})   \to   \Map  (\spc K_{\rm in}, \spc K_{\rm out}) $  is associated to a linear map  $\widehat {\map S} :  L (\spc H_{\rm out}\otimes \spc H_{\rm in})  \to L (\spc K_{\rm out}\otimes \spc K_{\rm in})$ uniquely defined by 
\begin{align}
\widehat {\map S}   (   {\rm Choi}  (\map M)) :  =    {\rm Choi}   (\map S (\map M))  \qquad \forall \map M \in \Map (\spc H)\, .
\end{align} 
 
 \subsubsection{Derivation of the theorem}
The first step in the proof of Theorem \ref{theo:wignerqo} is to show that the map $\widehat {\map S}$ is a symmetry of quantum states:  
\begin{lemma}\label{lem:symm2symm}
For every symmetry of quantum operations  $\map S  :  \Op (\spc H_{\rm in} , \spc H_{\rm out}) \to \Op (\spc K_{\rm in}, \spc K_{\rm out})$, the associated map  $\hS  :   \St(\spc H_{\rm out} \otimes \spc H_{\rm in}) \to  \St( \spc K_{\rm out}\otimes \spc K_{\rm in} )$ is a state space symmetry.
\end{lemma}
In order not to break the flow of the argument, we postpone the proof of Lemma \ref{lem:symm2symm} to Appendix \ref{app:symm2symm}.     

An immediate consequence of Lemma \ref{lem:symm2symm} is that the total dimension of the spaces $\spc H_{\rm out} \otimes \spc H_{\rm in}$ and $\spc K_{\rm out}\otimes \spc K_{\rm in}$ must be the same, namely 
\begin{align}\label{prodim}
d_{\spc H_{\rm in}}  d_{\spc H_{\rm out}}   = d_{\spc K_{\rm in}}  d_{\spc K_{\rm out}} \, .
\end{align} 

The second step in the proof of Theorem \ref{theo:wignerqo} is to break down the  operation space symmetry $\map S$ into two state space symmetries.   This result is accomplished through the following lemma:  

\begin{lemma}\label{lem:break}
For every quantum operation symmetry   $\map S  :  \Map (\spc H_{\rm in} , \spc H_{\rm out}) \to \Map (\spc K_{\rm in}, \spc K_{\rm out})$, there exists a state space symmetry $\map J  :  L  (\spc H_{\rm in})\to L(\spc K_{\rm in})$ such that
\begin{align}\label{related}
   \widehat{ \map S}   (I_{\spc H_{\rm out}}\otimes \rho )  =   I_{\spc K_{\rm out}}\otimes  \map J ( \rho)     \qquad \forall \rho \in \St (\spc H_{\rm in})\, .   
\end{align} 
If $\widehat{\map S}$ is unitary, then $\map J$ is unitary.   If $\widehat{\map S}$ is antiunitary, then $\map J$ is antiunitary. 
\end{lemma}
The proof of Lemma \ref{lem:break} can be found  in Appendix \ref{app:break}.    A consequence of  Lemma \ref{lem:break} is that the input spaces $\spc H_{\rm in}$ and $\spc K_{\rm in}$ have the same dimension.  Combining this fact with Eq.~(\ref{prodim}), we obtain that also the output spaces $\spc H_{\rm out}$ and $\spc K_{\rm out}$   must have the sam dimension. In summary, 
\begin{align}
\nonumber d_{\spc H_{\rm in}}    & = d_{\spc K_{\rm in}}  :  = d_{\rm in} \\
  d_{\spc H_{\rm out}}  & =   d_{\spc K_{\rm out}}  :=  d_{\rm out} \, .
\end{align}

The Wigner theorem for quantum operations is then obtained by combining  Lemmas \ref{lem:symm2symm} and \ref{lem:break}: 
\medskip 

{\bf Proof of Theorem \ref{theo:wignerqo}.}   Let $\map J  :  \St  (\spc H_{\rm in})\to \St(\spc K_{\rm in})$  be the state space symmetry defined in Lemma~\ref{lem:break}. The goal of the proof is to show that $\widehat{\map S}$ can be broken down into the product of two state space symmetries.   

Consider first the case where  both $\widehat{\map S}$ and $\map J$ are unitary transformations. Then, write the  the maximally mixed state $I_{\spc H_{\rm out}}/d_{\rm out}$ as the marginal of the maximally entangled state $|\phi_+\>  =  \sum_i  |i\>\otimes |i\>/\sqrt {d_{{\rm out}}}  \in  \spc H_R \otimes \spc H_{\rm out}$,  where $\spc H_R$ is an auxiliary Hilbert space, isomorphic to $\spc H_{\rm out}$.     Using the notation $  P_\psi   :=  |\psi\>\<\psi|$ for a generic vector $\psi$,  we can rewrite Eq.~(\ref{related})  as 
\begin{align}
(\Tr_R  \otimes \widehat{ \map S})   (   P_{\phi_+} \otimes \rho)   =  ( \Tr_R \otimes \map I_{\spc K_{\rm out}}\otimes \map J)    (   P_{\phi_+} \otimes \rho)  \, .
\end{align}
Since Eq.~(\ref{related})  holds for every density matrix  $\rho$, it holds in particular for every pure state. Hence, we have 
\begin{align}\label{marginals}
(\Tr_R  \otimes \widehat {\map S})   (   P_{\phi_+} \otimes P_\psi)   =  ( \Tr_R \otimes \map I_{\spc K_{\rm out}}\otimes \map J)    (   P_{\phi_+}  \otimes P_\psi)  \, ,
\end{align}
for every unit vector $\psi  \in \spc H_{\rm in}$. 

Since $\widehat{\map S}$ and $\map J$ are unitary transformations, the two operators on the two sides of the equality are rank-one projectors, representing pure states. Equation (\ref{marginals}) shows that these two pure states have 
%the two pure states $   (\map I_R\otimes  \map U^\dag)  (  P_{\phi_+}  \otimes P_\psi)$ and $(\map I_R\otimes \map I_{\spc H_{\rm out}}  \otimes \map V^\dag)  (P_{\phi_+}  \otimes P_\psi)$ have 
the same marginal states once the auxiliary  system is traced out. In other words, they are two purifications of the same mixed state.    Hence, the two pure states must be equivalent up to a unitary transformation on the auxiliary system. Explicitly, one must have 
\begin{align}
\label{bbb}
(\map I_{R}  \otimes \widehat{\map S})   (  P_{ \phi_+} \otimes P_\psi)   =  ( \map W_\psi \otimes \map I_{\spc K_{\rm out}}\otimes \map J)    ( P_{\phi_+} \otimes P_\psi)  \, ,
\end{align}
for some unitary transformation  $\map W_\psi : L (\spc H_{R}) \to L (\spc H_{R})$, possibly depending on $\psi$.  

Now, we use the relation  
%\begin{align}
$(  X  \otimes I)  |\phi_+\>    =   (I\otimes X^T) \, |\phi_+\>$,
%\end{align}
valid for every operator $X  \in L(  \spc H_R)  \simeq  L(\spc H_{\rm out})$.    Using this relation, Equation (\ref{bbb}) can be rewritten as
\begin{align}
(\map I_R  \otimes \widehat{\map S})   (   P_{\phi_+} \otimes P_\psi)   =  ( \map I_R \otimes \map W^T_\psi\otimes \map J)    (  P_{ \phi_+} \otimes P_\psi)  \, ,
\end{align}
where $\map W_\psi^T$ is the transpose of $\map W_\psi$. 

From the above equation we obtain  
\begin{align}\label{desired}
\widehat{\map S}   (   P_{\varphi} \otimes P_{\psi})   =  \map W^T_\psi  (P_\varphi)  \otimes \map J (P_\psi)
\end{align}
for every pair of unit vectors $\varphi$ and $\psi$.   

Since the transformations on both sides of the equality are unitary, it is easy to show that the transformation $\map W_\psi$ must be independent of $\psi$ (see Appendix \ref{app:Wpsi} for an explicit proof).  Hence, Eq.~(\ref{desired}) becomes $\widehat{\map S}  (P_\varphi\otimes P_\psi)     =  \map W^T   (P_\varphi)  \otimes  \map J (P_\psi)$,  $\forall  \varphi, \forall \psi$ or equivalently, $\widehat{\map S}  =  \map W^T \otimes \map J$.  

Translating back from the Choi representation, we then obtain that the map $\map S$ has the form $\map S  (\map Q)  =   \map W^T  \circ     \map Q \circ \map J^T$. Defining $\map S_1:  =  \map J^T$ and $\map S_2  : =  \map W^T $ we then obtain the desired result.  

To conclude the proof, consider the case where the state space symmetries $\widehat{\map S}$ and $\map J$ are both antiunitary.   In this case, we can rewrite  them as  $\widehat{\map S}  =  \widehat {\map S'}  \circ   (\map T_{\spc H_{\rm out}} \otimes \map T_{\spc H_{\rm in}})$ and $\map J  = \map J' \circ \map T_{\spc H_{\rm in}}$, where  $\widehat{\map S}'$ and $\map J'$ are unitary transformations, and $\map T_{\spc H_{\rm in}}$ ($\map T_{\spc H_{\rm out}}$) is the transpose map on $ \spc H_{\rm in} $    ($ \spc H_{\rm out}$).     Inserting the expression of $\widehat{\map S}$ and $\map J$ into Eq.~(\ref{related}), we obtain that Eq.~(\ref{desired}) holds also for $\widehat{\map S}'$ and $\map J'$. 

Hence, the same derivation used in the unitary case holds for the maps $\widehat{\map S}'$ and $\map J'$, and we obtain $\widehat{\map S}'   =  \map W^{\prime T} \otimes \map J'$, for a suitable unitary transformation $\map W'$.    Getting back to the map $\widehat {\map S}$, we obtain $\widehat{\map S}  =  \widehat{\map S}'  \circ (  \map T_{\spc H_{\rm out}} \otimes \map T_{\spc H_{\rm in}})   =   ({\map W}^{\prime \,T} \circ \map T_{\spc H_{\rm out}}) \otimes (\map J' \circ \map T_{\spc H_{\rm in}})   =  \map S_2  \otimes \map J$, where $\map S_2  :  =  \map W^{\prime \, T}  \circ   \map T_{\spc H_{\rm out}}$ is an antiunitary symmetry. 

Translating back from the Choi representation, we then obtain that the map $\map S$ has the form $\map S  (\map Q)  =   \map S_2  \circ     \map Q \circ \map S_1$, where $\map S_1 :  =  \map J^T$ is an antiunitary transformation. \qed

  \section{Proof of Lemma \ref{lem:symm2symm}}\label{app:symm2symm}

The proof uses the notion of {\em channel-preserving homomorphism}, that is,  a supermap that maps quantum channels into quantum channels
\cite{chiribella2008transforming,Zy08,chiribella2009theoretical,chiribella2013quantum}: 

\begin{defi}
A supermap  $\map S:  \Op  (\spc H_{\rm in}, \spc H_{\rm out})    \to  \Op  (\spc K_{\rm in}, \spc K_{\rm out})$   is a {\em homomorphism of quantum operations} if it preserves convex combinations and the null operation.  An homomorphism  $\map S$ is {\em channel-preserving} if,  for every channel $\map C \in \Chan  (\spc H_{\rm in}, \spc H_{\rm out})$, one has $\map S (\map C)\in   \Chan  (\spc K_{\rm in}, \spc K_{\rm out})$.  
\end{defi} 

We now show that every channel-preserving homomorphism induces a homomorphism of quantum states: 
\begin{lemma}
\label{lem:dualmap}  
Let  $\map S  :  \Op (\spc H_{\rm in} , \spc H_{\rm out}) \to \Op (\spc K_{\rm in}, \spc K_{\rm out})$   be a channel-preserving homomorphism, let  $\hS  :  L(\spc H_{\rm out} \otimes \spc H_{\rm in} )\to L( \spc K_{\rm out}\otimes \spc K_{\rm in} )$   be the associated map in the Choi representation.  Then, there exists a state space homomorphism $\map A:  \St (\spc K_{\rm  in}) \to \St ( \spc H_{\rm in})$ such that \begin{align}\hS^\dag (  I_{\spc K_{\rm out}}  \otimes \rho )   =   I_{\spc H_{\rm in}} \otimes \map A  (\rho)  \qquad \forall \rho \in \St (\spc K_{\rm in}) \, ,
\end{align}
where $\hS^\dag$ is the adjoint of the map $\hS$.  If $\map S$ is a symmetry of quantum operations,  then $\map A $  is a  symmetry of quantum states. 
\end{lemma}

\Proof  Let us define the set $\Choi\Chan  (\spc H_{\rm in},  \spc H_{\rm out})$ to be the set of Choi operators of quantum channels in $\Chan  (\spc H_{\rm in},  \spc H_{\rm out})$.    For a non-negative  operator $A \in  L(  \spc H_{\rm out} \otimes \spc H_{\rm in})$, the relation  $\< A  ,  C\>  = 1$ holds  for every $C \in \Choi\Chan (\spc H_{\rm in}, \spc H_{\rm out})$ if and only if $A$ has the form $A  =  I_{\rm out}\otimes \rho$ for some density matrix $\rho \in \St (\spc H_{\rm in})$  \cite{chiribella2008memory}.  Now, for every density matrix $\rho\in \St (\spc K_{\rm in})$, the operator $\hS^\dag (I_{\spc K_{\rm out}}\otimes \rho)$ is non-negative   and satisfies 
\begin{align}
\nonumber \<  \widehat S^\dag (I_{\spc K_{\rm out}}\otimes \rho) ,   C \>   &  =\<   I_{\spc H_{\rm out}}\otimes \rho ,  \widehat S  ( C)\>\\&
  =1 \, ,
  \end{align}
  the last equality following from the fact that $\hS  (  C)$ is in $\Choi \Chan (\spc K_{\rm in},  \spc K_{\rm out})$, because $\map S$ is channel-preserving.   
Hence, $\hS^\dag (I_{\spc K_{\rm out}}\otimes \rho)$ must be of the form $I_{\spc H_{\rm out}}\otimes \rho'$ for some density matrix $\rho'\in \St (\spc H_{\rm in})$.   Explicitly, the density matrix $\rho'$  can be computed as 
\begin{align}
\nonumber \rho'    &:  =   \frac{\Tr_1  [  \hS^\dag  (  I_{\spc K_{\rm out}}\otimes \rho) ]}{d}\\
& =:   \map A  (\rho) \,,
\end{align} where $\Tr_1$ denotes the partial trace over the first Hilbert space.     The map $\map A$  transforms density matrices into density matrices, and it preserves convex combinations. Therefore, it is a state space homomorphism. 

Now, suppose that $\map S$ is  a symmetry of quantum operations.   Let $\widehat {\map T}  :  =\hS^{-1}$ be the inverse of $\hS$, and let $\map B  :  \St  (\spc H_{\rm in} )\to \St (\spc K_{\rm in}) $ be the state space homomorphism associated to $\widehat {\map T}$ as  in Lemma \ref{lem:dualmap}.     Hence, 
 for every density matrix $\rho\in\St (\spc K_{\rm in})$,  one has  $ I_{\spc K_{\rm out}}\otimes \rho  =   \left(   \widehat{\map T}^\dag \circ \hS^\dag\right)   (I_{\spc K_{\rm out}}\otimes \rho)    = I_{\spc K_{\rm out}}\otimes (\map B \circ \map A)  (\rho)$.   Hence, $\map B\circ \map A =\map I_{\spc K_{\rm in}}$. Similarly, one can prove that $\map A\circ \map B  = \map I_{\spc H_{\rm in}}$. Hence,  $\map A$ is a state space symmetry. \qed

Equipped with the above result, we are ready to prove Lemma~\ref{lem:symm2symm}.  

 {\bf Proof of Lemma~\ref{lem:symm2symm}.}    We need to show that the map $\hS$ is a state space symmetry whenever $\map S$ is a symmetry of quantum operations.  We first show that $\hS^\dag$ is a unital map, namely $\hS^\dag (  I_{\spc K_{\rm out}}\otimes I_{\spc K_{\rm in}})  =   I_{\spc H_{\rm out}}\otimes I_{\spc H_{\rm in}}$.  By Lemma~\ref{lem:dualmap}, we have  $\hS^\dag (I_{\spc K_{\rm out}}\otimes I_{\spc K_{\rm in}})   =  I_{\spc H_{\rm out}}  \otimes  \map A (I_{\spc K_{\rm in}})$,  where $\map A$ is a state space symmetry. By Wigner's theorem, every state space symmetry is either unitary or antiunitary, and therefore it is unital. Hence, $\map A (I_{\spc K_{\rm in}})  =  I_{\spc H_{\rm in}}$, and   $\hS^\dag (  I_{\spc K_{\rm out}}\otimes I_{\spc K_{\rm in}})  =   I_{\spc H_{\rm out}}\otimes I_{\spc H_{\rm in}}$.  This proves that $\hS^\dag$ is unital.  
 
Moreover, the map $\hS^\dag$ is positive: for every positive operator $P$ and for every unit vector $\Psi \in \spc H_{\rm out}\otimes \spc H_{\rm in}$ one has  $\<\Psi|  \hS^\dag (P)  |\Psi\>    =  \<  \hS^\dag (P)  ,  P_{\Psi}\>  =  \<   P    ,  \hS (P_\Psi)  =  \Tr [ P    \,  \hS(P_\Psi)]  \ge 0$, because  $\hS(P_\psi)$ is a positive operator.  This proves that $\hS^\dag$ is positive. 

Since $\hS^\dag$ is positive and unital, Lemma~\ref{lem:positiveunital}  implies that $\hS$ is a state space homomorphism. Applying the same argument to the inverse map $\hS^{-1}$ we obtain that $\hS$ is a state space symmetry. \qed

\iffalse
{\bf Proof of Theorem \ref{theo:symmetry}} The map $\hS$ is positive (by assumption)  and trace-preserving (by Lemma \ref{lem:tp}).  Hence, it is a state space homomorphism.  Moreover, $\hS$ has a positive inverse $\hS^{-1}$. Hence, $\hS$ is invertible, and has a positive inverse.  Then, Lemma \ref{lem:invertible} guarantees that $\hS$ is a state space symmetry.  \qed  
\fi 

 \section{Proof of Lemma \ref{lem:break}}\label{app:break}

We need to prove that   for every operation space symmetry $\map S$ there exists a state space symmetry $\map J$ such that  the condition 
\begin{align}\label{zzz} \hS   (I_{\spc H_{\rm out}}\otimes \map \rho  )  =    I_{\spc K_{\rm out}}\otimes \map J(\rho)  \qquad \forall \rho  \in  \St (\spc H_{\rm in}) \,,
\end{align} 
and  the symmetry $\map J$ is unitary (antiunitary) whenever  the symmetry $\hS$ is unitary (antiunitary).

Note that 
\begin{enumerate}
\item the map $\hS$ is a state space symmetry (Lemma \ref{lem:symm2symm}),  and 
\item Lemma \ref{lem:dualmap} already guarantees the existence  of a state space symmetry $\map A$ such that 
\begin{align}\label{aaa}\hS^\dag (I_{\spc K_{\rm out}}\otimes \rho)   =  I_{\spc H_{\rm in}}\otimes \map A (\rho) \,  \qquad \forall \rho \in \St (\spc K_{\rm in}) \, .
\end{align}
\end{enumerate} 
 Since $\hS$ and $\map A$ are state space symmetries, Wigner's theorem implies that they are either unitary or antiunitary transformations. In either case, one has $ \hS^\dag \hS  =  \map I_{\spc H_{\rm out}}  \otimes  \map I_{\spc H_{\rm in}} $ and $\map A^\dag \map A  =  \map I_{\spc K_{\rm in}}$.    Hence, Eq.~(\ref{aaa}) implies Eq.~(\ref{zzz}) with $\map J  = \map A^\dag$. 

 It only remains to show that $\map J$ is unitary (antiunitary) whenever $\hS$ is unitary (antiunitary). To this purpose, note that Equation~(\ref{zzz})  implies 
\begin{align}
\map J(\rho)   =    \frac{\Tr_{\spc K_{\rm out}}[  \hS  (  I_{\spc H_{\rm out}} \otimes \rho) ]} {d_{{\rm out}}}\qquad \forall \rho  \in  \St (\spc H_{\rm in}) \, .
\end{align}
If $\hS$ is unitary, then the map acting on $\rho$ on the right-hand-side is completely positive.  Hence, the map $\map J$ is  completely positive, which is possible only if $\map J$ is unitary. 

Similarly, if $\hS$ is antiunitary, the composition of the map  on the right-hand-side with the transpose map $\map T_{\rm \spc H_{\rm in}} $ is completely positive.   Hence, the map $\map J \circ \map T_{\rm \spc H_{\rm in}} $ is completely positive, which is possible only if $\map J$ is antiunitary. \qed

\section{Proof that the unitary transformation $\map W_\psi$ in Eq.~(\ref{desired}) is independent of $\psi$}\label{app:Wpsi} 

We provide two alternative proofs, the first based on a quantum information-theoretic argument, and the second on explicit matrix calculations. 

{\em Quantum information-theoretic proof.} Equation~(\ref{desired}) states that the unitary transformations $\hS$, $\map W_\psi$, and $\map J$ satisfy the condition
\begin{align}
\widehat{\map S}   (   P_{\varphi} \otimes P_{\psi})   =  \map W^T_\psi  (P_\varphi)  \otimes \map J (P_\psi)
\end{align}
for every unit vector $\varphi$.    A physical way to read the equation is to imagine that the reversible process $\hS$ is applied to a system, initially in the state $P_\psi$,  coupled to an ancilla in the state $P_\varphi$.  
The result of the process is to implement a fixed unitary transformation $\map J$ on the system, and another unitary transformation $\map W_\psi$ on the ancilla. Since the transformation $\map J$ is unitary, it can be reversed by applying the inverse process $\map J^{-1}$. If this is done, the system is brought back to the state $P_\psi$, while the ancilla is in a new state $\map W_\psi  (P_\varphi)$, possibly depending on $\psi$. However, the so-called No Information Without Disturbance Theorem implies that the state of the ancilla should be independent of $\psi$, for otherwise one could extract information about $\psi$ by measuring the ancilla, without affecting the state of the system (see \cite{chiribella2014distinguishability} for a general proof of the No Information Without Disturbance Theorem which does not require the details of the quantum framework). Hence, we conclude that,  for every state $P_\varphi$, the state $\map W^T_\psi  (\varphi)$ should be independent of $\psi$. In other words, the transformation $\map W^T_\psi$ should be independent of $\psi$.

{\em Matrix algebra proof.}  Let $\psi$ and $\psi'$ be two different unit vectors.  Since the transformation $\hS$ is unitary, we have 
\begin{align}
\nonumber  \< \, \widehat{\map S}   (   P_{\varphi} \otimes P_{\psi}) \, , \,   \widehat{\map S}   (   P_{\varphi} \otimes P_{\psi'}) \>        &=    \<   P_{\varphi} \otimes P_{\psi}   ,   P_{\varphi} \otimes P_{\psi'}  \>  \\
&  = \<P_\psi,  P_{\psi'}\>  \, .
\end{align}
 Using Equation~(\ref{desired})  for $\psi$ and $\psi'$, we obtain  
\begin{align}
\nonumber \<P_\psi,  P_\psi'\>     &  = \< \, \widehat{\map S}   (   P_{\varphi} \otimes P_{\psi}) \, , \,   \widehat{\map S}   (   P_{\varphi} \otimes P_{\psi'}) \>   \\
\nonumber &   =  \<    \map W_\psi^T  (P_\varphi)  \otimes   \map J ( P_\psi)  ,   \map W_{\psi'}^T (  P_\varphi) \otimes \map J (P_{\psi'}) \>  \\
\nonumber & =  \<  \map W_\psi^T  (\varphi),   \map W_{\psi'}^T   (P_\varphi)  \>  ~  \<   \map J  (P_\psi) ,  \map J (P_{\psi'}) \>  \\
& =  \<  P_\varphi,     \overline {\map  W}_\psi \map W_{\psi'}^T   (P_\varphi)  \>  ~  \<   P_\psi , P_{\psi'} \>\, ,  \label{thiseq}
\end{align}
the last equation following from the fact that  $\map J$ is a state space symmetry, and therefore preserves the ray product. 

Equation~(\ref{thiseq})  must hold for every choice of unit vectors $\varphi, \psi $, and $\psi'$.   Choosing $\psi$ and $\psi'$ to be non-orthogonal, we have $\< P_\psi,  P_{\psi'} \>  \not  = 0$, and Eq.~(\ref{thiseq})  becomes 
\begin{align}
\<  P_\varphi,     \overline {\map  W}_\psi \map W_{\psi'}^T   (P_\varphi)  \>   =  1   \qquad \forall  |\varphi\>  \in \spc H_{\rm out} \, ,
\end{align}
which is equivalent to $  \overline {\map  W}_\psi \map W_{\psi'}^T   = \map I_{\spc H_{\rm out}} $. Hence, $ \map  W_\psi   = \map W_{\psi'}$ whenever $\psi$ and $\psi'$ are non-orthogonal.  

Now, for every two unit vectors $\psi_1$ and $\psi_2$ there is a third unit vector $\psi$ that is neither orthogonal to $\psi_1$ nor to $\psi_2$.  Hence, one must have $ \map  W_{\psi_1}   = \map W_{\psi}$ and $ \map  W_{\psi_2}   = \map W_{\psi}$, which implies $ \map  W_{\psi_1}   = \map W_{\psi_1}$.  In summary, the unitary $\map W_\psi$ is independent of $\psi$.

\section{Double antiunitary transformations  are not completely positive}\label{app:dt}

An active interpretation of double antiunitary operations is incompatible with a general notion of physical transformation,  captured by the framework of quantum supermaps~\cite{chiribella2008transforming,chiribella2009theoretical,chiribella2013quantum,chiribella2013normal,chiribella2016optimal,ebler2018enhanced,rosset2018resource,miyazaki2019complex,quintino2019probabilistic,kristjansson2020resource}, also called quantum superchannels in later works~\cite{gour2019comparison,gour2019quantify,wang2019resource,xu2019coherence,burniston2020necessary,saxena2020dynamical,chen2020entanglement,liu2020operational} (see also~\cite{Zy08} for a related notion of supermap transforming a certain set of completely positive maps into itself). 

Quantum supermaps represent the most general transformations that can in principle be applied locally to the reduced dynamics of  subsystems.
 Under such requirement, Ref.~~\cite{chiribella2008transforming} showed  that the most general quantum supermap on the set of quantum oeprations consists in the insertion of the input quantum operation $\map Q$ between two quantum channels, $\map A  $ and $\map B$, thus obtaining the new quantum operation $\map Q'  =  \map B \circ (  \map Q\otimes \map I_A)  \circ \map A$, where $\map I_A$ denotes the identity channel on an auxiliary quantum system, output by channel $\map A$,  input into channel $\map B$, and unaffected by the quantum operation $\map Q$.    This result was later extended to infinite dimensions in Ref.~\cite{chiribella2013normal}.   
 
 Our Wigner theorem for quantum operations implies that the only reversible quantum supermaps are those for which the channels $\map A$ and $\map B$ are unitary, and the auxiliary system $A$ is trivial ({\em i.~e.~}one-dimensional).

 The same conclusion was reached by Ref.~\cite{castro2018dynamics} for quantum supermaps that are generated by a continuous-time evolution.  Our Wigner theorem shows that the assumption of continuity is not necessary in the case of quantum supermaps on quantum operations: reversibility and the applicability to subsystems are already  enough to single out the double unitary form for any transformation of quantum operations that can be implemented actively in standard quantum mechanics. 
 
 %% change here (KZ)
In contrast, double antiunitary symmetries cannot be applied locally to the evolution of subsystems.   Mathematically, they are associated to linear maps that are not completely positive.  In the following, we provide a proof of this fact.

First, note that every double antiunitary transformations  is unitarily equivalent to the double transpose map  $\tau_{{\rm in}/{\rm out}}:   \map Q  \mapsto   \tau_{\rm out} \circ\map Q  \circ \tau_{\rm in}$. Since unitary equivalence just amounts to a change of basis, in the following we will restrict our attention to the map  $\tau_{{\rm in}/{\rm out}}$. 

Under the action of the double transpose,   a generic quantum operation  $\map Q$ with Kraus operators   $\{\map Q_i\}$ is transformed  into a quantum operation  $\map Q'$ with  Kraus operators   $\{\map Q^*_i\}$, obtained by complex conjugation in a fixed basis.

Now, let us examine the transformation $\map Q \mapsto \map Q'$ in the Choi representation~\cite{Cho75a}.  The Choi operator of $\map Q$ is 
\begin{align}
\nonumber \Choi  (\map Q  )    &:   =  \sum_{m,n}   \,  \map Q  (|m\>\<n|) \otimes |m\>\<n|  \\
  &  =  \sum_i  \,(Q_i\otimes I_{\spc H_{\rm in}})   \,  |I_{\rm \spc H_{\rm in}} \kk\bb I_{\rm \spc H_{\rm in}}  |    (Q_i^\dag \otimes I) \, ,
\end{align}
where we used the notation  $|I_{\rm \spc H_{\rm in}} \kk  : = \sum_m  \, |m\> \otimes |m\>$.   
The Choi operator of $\map Q'$ is 
\begin{align}
\nonumber \Choi  (\map Q'  )  &=  \sum_i  \,(Q_i^*\otimes I_{\spc H_{\rm in}})   \,  |I_{\rm \spc H_{\rm in}} \kk\bb I_{\rm \spc H_{\rm in}}  |    (Q_i^T \otimes I) \\
\nonumber &  =   [\Choi  (\map Q )]^*\\
&  =  [\Choi  (\map Q )]^T \, ,
\end{align}
the last equality following from the fact that $\Choi  (\map Q )$ is self-adjoint. 
Hence, at the level of Choi operators, the  double transpose map $\tau_{{\rm in}/{\rm out}}$ is  just the transpose. Since the transpose is known to be not completely positive, we conclude that the double transpose map is not a completely positive transformation.  Physically, the failure of complete positivity implies that it is not possible to apply the map  $\tau_{{\rm in}/{\rm out}}$ on a generic bipartite quantum operation~\cite{chiribella2008transforming}.

\section{Choi representation of the transformations $\Theta$ and $\Theta'$}\label{app:bisto}

Here we consider the transformations $\Theta:  \map Q \to \map Q^\dag$ and $\Theta'  :  \map Q  \to \map Q^T$, which are the natural extensions of the   time reversal transformations $  U\mapsto U^\dag$ and $U\mapsto U^T$, defined at the level of unitary dynamics.

Let us consider first the Choi representation of the transformation $\Theta':    \map Q  \to \map Q^T$. 

For  generic completely positive map $\map Q:  L(\spc H_{\rm in}) \to L(\spc H_{\rm out})$,  written in the Kraus representation $\map Q  (\rho) :=  \sum_i Q_i  \rho  Q_i^\dag$, the Choi operator is given by  
\begin{align}
\nonumber \Choi  (\map Q  )    &:   =  \sum_{m,n}   \,  \map Q  (|m\>\<n|) \otimes |m\>\<n|  \\
  &  =  \sum_i  \,(Q_i\otimes I_{\spc H_{\rm in}})   \,  |I_{\rm \spc H_{\rm in}} \kk\bb I_{\rm \spc H_{\rm in}}  |    (Q_i^\dag \otimes I) \, ,
\end{align}
where we used the notation  $|I_{\rm \spc H_{\rm in}} \kk  : = \sum_m  \, |m\> \otimes |m\>$. 
 
Using the condition 
\begin{align} (A \otimes I_{\rm \spc H_{\rm in}}  )| I_{\rm \spc H_{\rm in}}  \kk  =   (I_{\rm \spc H_{\rm out}}  \otimes A^T)  |I_{\rm \spc H_{\rm out}}\kk \, ,
\end{align}
valid for every operator $A :  \spc H_{\rm in} \to \sc H_{\rm out}$, we obtain the relation 
\begin{align}
\nonumber \Choi (\map Q)   &=   \sum_i  \,   (I_{\rm \spc H_{\rm out}}  \otimes Q_i^T)   \,  |I_{\rm \spc H_{\rm out}} \kk\bb I_{\rm \spc H_{\rm out}}  |    (Q_i^* \otimes I_{\spc H_{\rm out}}) \\
  &  =   {\tt SWAP}_{\rm in/out}  \,  \Choi  \left(\map Q^T\right) \, {\tt SWAP}_{\rm in/out} \, ,
\end{align}
where $\tt SWAP_{\rm in/out}:  \spc H_{\rm in} \otimes \spc H_{\rm out} \to \spc H_{\rm out} \otimes \spc H_{\rm in} $ is the swap operator, defined by the condition ${\tt SWAP}  (  \phi  \otimes \psi)  =  \psi \otimes \phi$, $\forall \phi, \forall \psi$.   

Hence, at the level of Choi operators, the  map $\Theta'  : \map Q \mapsto \map Q^T$ is represented by a unitary transformation. Explicitly, the transformation is 
\begin{align}  
\widehat{{\Theta'}} :      Q   \mapsto     {\tt SWAP}_{\rm out/in}   \, Q \,     {\tt SWAP}_{\rm out/in} \, ,      
\end{align}
with ${\tt SWAP}_{\rm out/in}  :  =  {\tt SWAP}_{\rm in/out}^{-1}$.  This unitary transformation corresponds to the exchange between the input and output spaces of the Choi operator \cite{BZ}.

Let us now consider the Choi representation of the time reversal  $\Theta:  \map Q  \to \map Q^\dag$. 
By definition, we have 
\begin{align}
\nonumber \Choi (\map Q^\dag)  &  =   \sum_i  \,(Q_i^\dag\otimes I_{\spc H_{\rm in}})   \,  |I_{\rm \spc H_{\rm in}} \kk\bb I_{\rm \spc H_{\rm in}}  |    (Q_i \otimes I)\\
 \nonumber   &  = \left[ \sum_i  \,(Q_i^T\otimes I_{\spc H_{\rm in}})   \,  |I_{\rm \spc H_{\rm in}} \kk\bb I_{\rm \spc H_{\rm in}}  |    (Q_i^* \otimes I)\right]^*  \\
\nonumber   &  =   \left[\Choi  (\map Q^T)\right]^*  \\
\nonumber    & =    \left[\Choi  (\map Q^T)\right]^T \\
\nonumber    & =    \left[  {\tt SWAP}_{\rm out/in}   \, Q \,     {\tt SWAP}_{\rm out/in}   \right]^T \, .
\end{align}
Hence, at the level of Choi representation, the  map $\Theta'  : \map Q \mapsto \map Q^T$ is represented by  the map 
\begin{align}
\widehat{\Theta}:      Q  \mapsto   \left[  {\tt SWAP}_{\rm out/in}   \, Q \,     {\tt SWAP}_{\rm out/in}   \right]^T\, , 
\end{align}
namely the antiunitary transformation  resulting from the composition of the swap operation with the transpose.

\section{Proof of Theorem~\ref{theo:nogo}}\label{app:nogo}

The proof is by contradiction: we assume that the symmetry   $\map S$  exists, and show that its existence leads to a logical contradiction. 

Let us start from the case of symmetries satisfying the condition  $\map S  (\map U)   = \map U^\dag$ for every unitary channel $\map U$.   Since $\map S$ is symmetry of quantum operations, its action on a generic unitary channel $\map U$  is either $\map S (\map U)    = \map W  \circ \map U \circ \map V $ or $\map S(\map U)    =   \map W  \circ \tau_{\rm in/out}  (\map U)\circ \map V$, where $\tau_{\rm in/out}$ is the double transpose.

Let us inspect these two cases separately. In the first case,  the transformation $\map S (\map U)    = \map W  \circ \map U \circ \map V $ is a sequence of quantum channels.  Refs.~\cite{chiribella2016optimal,quintino2019probabilistic} showed that the transformation $\map U \mapsto \map U^\dag$ cannot be implemented by inserting the channel $\map U$ in a sequence of quantum channels.  For completeness, we provide an alternative proof here.  The condition  $ \map W  \circ \map U \circ \map V =  \map U^\dag  $    implies   $WU V =      e^{i\gamma_U}  \,  U^\dag$ for every unitary operator $U$.  Choosing $U=  I$ reveals that one should have  $V  = e^{i\gamma_I} W^\dag$.   Choosing $U=  W$, one obtains  $ W e^{i\gamma_I}  =   e^{i\gamma _W}   W^\dag$.
  Hence, one must have   $W^\dag  U e^{i\gamma _W}   W^\dag  =      e^{i\gamma_U}  \,  U^\dag$, which implies that $ W^\dag U$ is proportional to $  (W^{\dag}  U)^\dag$.    In other words, $W^\dag U$ must be proportional to a Hermitian operator. But since $W^\dag U$ is a generic unitary operator, this condition cannot be satisfied.   
    
  In the second case, the channel $\tau_{\rm in/out}  (\map U)$  has the form $\rho  \mapsto  U^* \rho U^T$. Hence, the condition  $ \map W  \circ \tau_{\rm in/out}  (\map U)\circ \map V  =  \map U^\dag$ implies $W\overline U V =      e^{i\gamma_U}  \,  U^\dag$ for every unitary operator $U$. Choosing $U=  I$ reveals that one should have  $V  = e^{i\gamma_I} W^\dag$.   Choosing $ U=  \overline W$, one obtains  $ W e^{i\gamma_I}  =   e^{i\gamma _{\overline W}}   W^T$.
  Hence, one must have   $W^T  \overline U e^{i\gamma _W}   W^\dag  =      e^{i\gamma_U}  \,  U^\dag$, which implies that $ W^T \overline U$ is proportional to $  ( W^{T} \overline U)^T$.    In other words, $W^T \overline U$ must be proportional to a symmetric operator. But since $W^T \overline U$ is a generic unitary operator, this condition cannot be satisfied.   
  
Summarising, no symmetry of the set of quantum operations can satisfy the condition    $\map S  (\map U)   = \map U^\dag$ for every unitary channel $\map U$.   Let us now consider  symmetries satisfying the condition  $\map S  (\map U)   = \map U^T$ for every unitary channel $\map U$.   Note that the relation $\map U^\dag  =  \tau_{\rm in/out}  (\map U^T)$ holds for every unitary channel $\map U$.     If there existed a symmetry $\map S$ such that $\map S (\map U)  =  \map U^T$ for every unitary channel, then one could combine the symmetry $\map S$ with the double transpose, thus obtaining a symmetry $\map S'  =   \tau_{\rm in/out}   \circ \map S$ such that $\map S' (\map U)  =  \map U^\dag$ for every unitary channel, in contradiction with the previous part of the proof. 
    \qed

\bibliography{references}

\begin{thebibliography}{101}%
\makeatletter
\providecommand \@ifxundefined [1]{%
 \@ifx{#1\undefined}
}%
\providecommand \@ifnum [1]{%
 \ifnum #1\expandafter \@firstoftwo
 \else \expandafter \@secondoftwo
 \fi
}%
\providecommand \@ifx [1]{%
 \ifx #1\expandafter \@firstoftwo
 \else \expandafter \@secondoftwo
 \fi
}%
\providecommand \natexlab [1]{#1}%
\providecommand \enquote  [1]{``#1''}%
\providecommand \bibnamefont  [1]{#1}%
\providecommand \bibfnamefont [1]{#1}%
\providecommand \citenamefont [1]{#1}%
\providecommand \href@noop [0]{\@secondoftwo}%
\providecommand \href [0]{\begingroup \@sanitize@url \@href}%
\providecommand \@href[1]{\@@startlink{#1}\@@href}%
\providecommand \@@href[1]{\endgroup#1\@@endlink}%
\providecommand \@sanitize@url [0]{\catcode `\\12\catcode `\$12\catcode
  `\&12\catcode `\#12\catcode `\^12\catcode `\_12\catcode `\%12\relax}%
\providecommand \@@startlink[1]{}%
\providecommand \@@endlink[0]{}%
\providecommand \url  [0]{\begingroup\@sanitize@url \@url }%
\providecommand \@url [1]{\endgroup\@href {#1}{\urlprefix }}%
\providecommand \urlprefix  [0]{URL }%
\providecommand \Eprint [0]{\href }%
\providecommand \doibase [0]{http://dx.doi.org/}%
\providecommand \selectlanguage [0]{\@gobble}%
\providecommand \bibinfo  [0]{\@secondoftwo}%
\providecommand \bibfield  [0]{\@secondoftwo}%
\providecommand \translation [1]{[#1]}%
\providecommand \BibitemOpen [0]{}%
\providecommand \bibitemStop [0]{}%
\providecommand \bibitemNoStop [0]{.\EOS\space}%
\providecommand \EOS [0]{\spacefactor3000\relax}%
\providecommand \BibitemShut  [1]{\csname bibitem#1\endcsname}%
\let\auto@bib@innerbib\@empty
%</preamble>
\bibitem [{\citenamefont {Weyl}(1950)}]{weyl1950theory}%
  \BibitemOpen
  \bibfield  {author} {\bibinfo {author} {\bibfnamefont {Hermann}\ \bibnamefont
  {Weyl}},\ }\href@noop {} {\emph {\bibinfo {title} {The theory of groups and
  quantum mechanics}}}\ (\bibinfo  {publisher} {Dover, New York},\ \bibinfo
  {year} {1950})\BibitemShut {NoStop}%
\bibitem [{\citenamefont {Wigner}(1959)}]{wigner1959group}%
  \BibitemOpen
  \bibfield  {author} {\bibinfo {author} {\bibfnamefont {Eugene~Paul}\
  \bibnamefont {Wigner}},\ }\href@noop {} {\emph {\bibinfo {title} {Group
  theory and its application to the quantum mechanics of atomic spectra}}}\
  (\bibinfo  {publisher} {Academic Press},\ \bibinfo {year} {1959})\BibitemShut
  {NoStop}%
\bibitem [{\citenamefont {Cornwell}(1997)}]{cornwell1997group}%
  \BibitemOpen
  \bibfield  {author} {\bibinfo {author} {\bibfnamefont {John~F}\ \bibnamefont
  {Cornwell}},\ }\href@noop {} {\emph {\bibinfo {title} {Group theory in
  physics: An introduction}}}\ (\bibinfo  {publisher} {Academic Press},\
  \bibinfo {year} {1997})\BibitemShut {NoStop}%
\bibitem [{\citenamefont {Tasaki}(2020)}]{tasaki2020physics}%
  \BibitemOpen
  \bibfield  {author} {\bibinfo {author} {\bibfnamefont {Hal}\ \bibnamefont
  {Tasaki}},\ }\href@noop {} {\emph {\bibinfo {title} {Physics and mathematics
  of quantum many-body systems}}}\ (\bibinfo  {publisher} {Springer},\ \bibinfo
  {year} {2020})\BibitemShut {NoStop}%
\bibitem [{\citenamefont {Hayashi}(2017)}]{hayashi2017group}%
  \BibitemOpen
  \bibfield  {author} {\bibinfo {author} {\bibfnamefont {Masahito}\
  \bibnamefont {Hayashi}},\ }\href@noop {} {\emph {\bibinfo {title} {A Group
  Theoretic Approach to Quantum Information}}}\ (\bibinfo  {publisher}
  {Springer},\ \bibinfo {year} {2017})\BibitemShut {NoStop}%
\bibitem [{\citenamefont {Weinberg}(1995)}]{weinberg1995quantum}%
  \BibitemOpen
  \bibfield  {author} {\bibinfo {author} {\bibfnamefont {Steven}\ \bibnamefont
  {Weinberg}},\ }\href@noop {} {\emph {\bibinfo {title} {The quantum theory of
  fields}}},\ Vol.~\bibinfo {volume} {2}\ (\bibinfo  {publisher} {Cambridge
  University Press},\ \bibinfo {year} {1995})\BibitemShut {NoStop}%
\bibitem [{\citenamefont {Wigner}(1960)}]{Wi60}%
  \BibitemOpen
  \bibfield  {author} {\bibinfo {author} {\bibfnamefont {Eugene}\ \bibnamefont
  {Wigner}},\ }\bibfield  {title} {\enquote {\bibinfo {title} {Normal form of
  antiunitary operators},}\ }\href@noop {} {\bibfield  {journal} {\bibinfo
  {journal} {J. Math. Phys.}\ }\textbf {\bibinfo {volume} {1}},\ \bibinfo
  {pages} {409} (\bibinfo {year} {1960})}\BibitemShut {NoStop}%
\bibitem [{\citenamefont {Chiribella}\ \emph
  {et~al.}(2008{\natexlab{a}})\citenamefont {Chiribella}, \citenamefont
  {D'Ariano},\ and\ \citenamefont {Perinotti}}]{chiribella2008transforming}%
  \BibitemOpen
  \bibfield  {author} {\bibinfo {author} {\bibfnamefont {Giulio}\ \bibnamefont
  {Chiribella}}, \bibinfo {author} {\bibfnamefont {Giacomo~Mauro}\ \bibnamefont
  {D'Ariano}}, \ and\ \bibinfo {author} {\bibfnamefont {Paolo}\ \bibnamefont
  {Perinotti}},\ }\bibfield  {title} {\enquote {\bibinfo {title} {Transforming
  quantum operations: quantum supermaps},}\ }\href@noop {} {\bibfield
  {journal} {\bibinfo  {journal} {EPL (Europhysics Letters)}\ }\textbf
  {\bibinfo {volume} {83}},\ \bibinfo {pages} {30004} (\bibinfo {year}
  {2008}{\natexlab{a}})}\BibitemShut {NoStop}%
\bibitem [{\citenamefont {Chiribella}\ \emph
  {et~al.}(2009{\natexlab{a}})\citenamefont {Chiribella}, \citenamefont
  {D'Ariano},\ and\ \citenamefont {Perinotti}}]{chiribella2009theoretical}%
  \BibitemOpen
  \bibfield  {author} {\bibinfo {author} {\bibfnamefont {Giulio}\ \bibnamefont
  {Chiribella}}, \bibinfo {author} {\bibfnamefont {Giacomo~Mauro}\ \bibnamefont
  {D'Ariano}}, \ and\ \bibinfo {author} {\bibfnamefont {Paolo}\ \bibnamefont
  {Perinotti}},\ }\bibfield  {title} {\enquote {\bibinfo {title} {Theoretical
  framework for quantum networks},}\ }\href@noop {} {\bibfield  {journal}
  {\bibinfo  {journal} {Physical Review A}\ }\textbf {\bibinfo {volume} {80}},\
  \bibinfo {pages} {022339} (\bibinfo {year} {2009}{\natexlab{a}})}\BibitemShut
  {NoStop}%
\bibitem [{\citenamefont {Chiribella}\ \emph
  {et~al.}(2013{\natexlab{a}})\citenamefont {Chiribella}, \citenamefont
  {D'Ariano}, \citenamefont {Perinotti},\ and\ \citenamefont
  {Valiron}}]{chiribella2013quantum}%
  \BibitemOpen
  \bibfield  {author} {\bibinfo {author} {\bibfnamefont {Giulio}\ \bibnamefont
  {Chiribella}}, \bibinfo {author} {\bibfnamefont {Giacomo~Mauro}\ \bibnamefont
  {D'Ariano}}, \bibinfo {author} {\bibfnamefont {Paolo}\ \bibnamefont
  {Perinotti}}, \ and\ \bibinfo {author} {\bibfnamefont {Benoit}\ \bibnamefont
  {Valiron}},\ }\bibfield  {title} {\enquote {\bibinfo {title} {Quantum
  computations without definite causal structure},}\ }\href@noop {} {\bibfield
  {journal} {\bibinfo  {journal} {Physical Review A}\ }\textbf {\bibinfo
  {volume} {88}},\ \bibinfo {pages} {022318} (\bibinfo {year}
  {2013}{\natexlab{a}})}\BibitemShut {NoStop}%
\bibitem [{\citenamefont {Chiribella}\ \emph
  {et~al.}(2013{\natexlab{b}})\citenamefont {Chiribella}, \citenamefont
  {Toigo},\ and\ \citenamefont {Umanit{\`a}}}]{chiribella2013normal}%
  \BibitemOpen
  \bibfield  {author} {\bibinfo {author} {\bibfnamefont {Giulio}\ \bibnamefont
  {Chiribella}}, \bibinfo {author} {\bibfnamefont {Alessandro}\ \bibnamefont
  {Toigo}}, \ and\ \bibinfo {author} {\bibfnamefont {Veronica}\ \bibnamefont
  {Umanit{\`a}}},\ }\bibfield  {title} {\enquote {\bibinfo {title} {Normal
  completely positive maps on the space of quantum operations},}\ }\href@noop
  {} {\bibfield  {journal} {\bibinfo  {journal} {Open Systems and Information
  Dynamics}\ }\textbf {\bibinfo {volume} {20}},\ \bibinfo {pages} {1350003}
  (\bibinfo {year} {2013}{\natexlab{b}})}\BibitemShut {NoStop}%
\bibitem [{\citenamefont {Bisio}\ and\ \citenamefont
  {Perinotti}(2019)}]{bisio2019theoretical}%
  \BibitemOpen
  \bibfield  {author} {\bibinfo {author} {\bibfnamefont {Alessandro}\
  \bibnamefont {Bisio}}\ and\ \bibinfo {author} {\bibfnamefont {Paolo}\
  \bibnamefont {Perinotti}},\ }\bibfield  {title} {\enquote {\bibinfo {title}
  {Theoretical framework for higher-order quantum theory},}\ }\href@noop {}
  {\bibfield  {journal} {\bibinfo  {journal} {Proceedings of the Royal Society
  A}\ }\textbf {\bibinfo {volume} {475}},\ \bibinfo {pages} {20180706}
  (\bibinfo {year} {2019})}\BibitemShut {NoStop}%
\bibitem [{\citenamefont {{\.Z}yczkowski}(2008)}]{Zy08}%
  \BibitemOpen
  \bibfield  {author} {\bibinfo {author} {\bibfnamefont {Karol}\ \bibnamefont
  {{\.Z}yczkowski}},\ }\bibfield  {title} {\enquote {\bibinfo {title} {Quartic
  quantum theory: an extension of the standard quantum mechanics},}\
  }\href@noop {} {\bibfield  {journal} {\bibinfo  {journal} {Journal of Physics
  A: Mathematical and Theoretical}\ }\textbf {\bibinfo {volume} {41}},\
  \bibinfo {pages} {355302} (\bibinfo {year} {2008})}\BibitemShut {NoStop}%
\bibitem [{\citenamefont {Chiribella}\ \emph
  {et~al.}(2009{\natexlab{b}})\citenamefont {Chiribella}, \citenamefont
  {D'Ariano}, \citenamefont {Perinotti},\ and\ \citenamefont
  {Valiron}}]{chiribella2009beyond}%
  \BibitemOpen
  \bibfield  {author} {\bibinfo {author} {\bibfnamefont {G}~\bibnamefont
  {Chiribella}}, \bibinfo {author} {\bibfnamefont {GM}~\bibnamefont
  {D'Ariano}}, \bibinfo {author} {\bibfnamefont {P}~\bibnamefont {Perinotti}},
  \ and\ \bibinfo {author} {\bibfnamefont {B}~\bibnamefont {Valiron}},\
  }\bibfield  {title} {\enquote {\bibinfo {title} {Beyond quantum computers},}\
  }\href@noop {} {\bibfield  {journal} {\bibinfo  {journal} {arXiv preprint
  arXiv:0912.0195}\ } (\bibinfo {year} {2009}{\natexlab{b}})}\BibitemShut
  {NoStop}%
\bibitem [{\citenamefont {Oreshkov}\ \emph {et~al.}(2012)\citenamefont
  {Oreshkov}, \citenamefont {Costa},\ and\ \citenamefont
  {Brukner}}]{oreshkov2012quantum}%
  \BibitemOpen
  \bibfield  {author} {\bibinfo {author} {\bibfnamefont {Ognyan}\ \bibnamefont
  {Oreshkov}}, \bibinfo {author} {\bibfnamefont {Fabio}\ \bibnamefont {Costa}},
  \ and\ \bibinfo {author} {\bibfnamefont {{\v{C}}aslav}\ \bibnamefont
  {Brukner}},\ }\bibfield  {title} {\enquote {\bibinfo {title} {Quantum
  correlations with no causal order},}\ }\href@noop {} {\bibfield  {journal}
  {\bibinfo  {journal} {Nature Communications}\ }\textbf {\bibinfo {volume}
  {3}},\ \bibinfo {pages} {1--8} (\bibinfo {year} {2012})}\BibitemShut
  {NoStop}%
\bibitem [{\citenamefont {Perinotti}(2017)}]{perinotti2017causal}%
  \BibitemOpen
  \bibfield  {author} {\bibinfo {author} {\bibfnamefont {Paolo}\ \bibnamefont
  {Perinotti}},\ }\bibfield  {title} {\enquote {\bibinfo {title} {Causal
  structures and the classification of higher order quantum computations},}\
  }in\ \href@noop {} {\emph {\bibinfo {booktitle} {Time in physics}}}\
  (\bibinfo  {publisher} {Springer},\ \bibinfo {year} {2017})\ pp.\ \bibinfo
  {pages} {103--127}\BibitemShut {NoStop}%
\bibitem [{\citenamefont {Castro-Ruiz}\ \emph {et~al.}(2018)\citenamefont
  {Castro-Ruiz}, \citenamefont {Giacomini},\ and\ \citenamefont
  {Brukner}}]{castro2018dynamics}%
  \BibitemOpen
  \bibfield  {author} {\bibinfo {author} {\bibfnamefont {Esteban}\ \bibnamefont
  {Castro-Ruiz}}, \bibinfo {author} {\bibfnamefont {Flaminia}\ \bibnamefont
  {Giacomini}}, \ and\ \bibinfo {author} {\bibfnamefont {{\v{C}}aslav}\
  \bibnamefont {Brukner}},\ }\bibfield  {title} {\enquote {\bibinfo {title}
  {Dynamics of quantum causal structures},}\ }\href@noop {} {\bibfield
  {journal} {\bibinfo  {journal} {Physical Review X}\ }\textbf {\bibinfo
  {volume} {8}},\ \bibinfo {pages} {011047} (\bibinfo {year}
  {2018})}\BibitemShut {NoStop}%
\bibitem [{\citenamefont {Rosset}\ \emph {et~al.}(2018)\citenamefont {Rosset},
  \citenamefont {Buscemi},\ and\ \citenamefont {Liang}}]{rosset2018resource}%
  \BibitemOpen
  \bibfield  {author} {\bibinfo {author} {\bibfnamefont {Denis}\ \bibnamefont
  {Rosset}}, \bibinfo {author} {\bibfnamefont {Francesco}\ \bibnamefont
  {Buscemi}}, \ and\ \bibinfo {author} {\bibfnamefont {Yeong-Cherng}\
  \bibnamefont {Liang}},\ }\bibfield  {title} {\enquote {\bibinfo {title}
  {Resource theory of quantum memories and their faithful verification with
  minimal assumptions},}\ }\href@noop {} {\bibfield  {journal} {\bibinfo
  {journal} {Physical Review X}\ }\textbf {\bibinfo {volume} {8}},\ \bibinfo
  {pages} {021033} (\bibinfo {year} {2018})}\BibitemShut {NoStop}%
\bibitem [{\citenamefont {Ebler}\ \emph {et~al.}(2018)\citenamefont {Ebler},
  \citenamefont {Salek},\ and\ \citenamefont {Chiribella}}]{ebler2018enhanced}%
  \BibitemOpen
  \bibfield  {author} {\bibinfo {author} {\bibfnamefont {Daniel}\ \bibnamefont
  {Ebler}}, \bibinfo {author} {\bibfnamefont {Sina}\ \bibnamefont {Salek}}, \
  and\ \bibinfo {author} {\bibfnamefont {Giulio}\ \bibnamefont {Chiribella}},\
  }\bibfield  {title} {\enquote {\bibinfo {title} {Enhanced communication with
  the assistance of indefinite causal order},}\ }\href@noop {} {\bibfield
  {journal} {\bibinfo  {journal} {Physical Review Letters}\ }\textbf {\bibinfo
  {volume} {120}},\ \bibinfo {pages} {120502} (\bibinfo {year}
  {2018})}\BibitemShut {NoStop}%
\bibitem [{\citenamefont {Gour}\ and\ \citenamefont
  {Winter}(2019)}]{gour2019quantify}%
  \BibitemOpen
  \bibfield  {author} {\bibinfo {author} {\bibfnamefont {Gilad}\ \bibnamefont
  {Gour}}\ and\ \bibinfo {author} {\bibfnamefont {Andreas}\ \bibnamefont
  {Winter}},\ }\bibfield  {title} {\enquote {\bibinfo {title} {How to quantify
  a dynamical quantum resource},}\ }\href@noop {} {\bibfield  {journal}
  {\bibinfo  {journal} {Physical Review Letters}\ }\textbf {\bibinfo {volume}
  {123}},\ \bibinfo {pages} {150401} (\bibinfo {year} {2019})}\BibitemShut
  {NoStop}%
\bibitem [{\citenamefont {Wang}\ and\ \citenamefont
  {Wilde}(2019)}]{wang2019resource}%
  \BibitemOpen
  \bibfield  {author} {\bibinfo {author} {\bibfnamefont {Xin}\ \bibnamefont
  {Wang}}\ and\ \bibinfo {author} {\bibfnamefont {Mark~M}\ \bibnamefont
  {Wilde}},\ }\bibfield  {title} {\enquote {\bibinfo {title} {Resource theory
  of asymmetric distinguishability for quantum channels},}\ }\href@noop {}
  {\bibfield  {journal} {\bibinfo  {journal} {Physical Review Research}\
  }\textbf {\bibinfo {volume} {1}},\ \bibinfo {pages} {033169} (\bibinfo {year}
  {2019})}\BibitemShut {NoStop}%
\bibitem [{\citenamefont {Xu}(2019)}]{xu2019coherence}%
  \BibitemOpen
  \bibfield  {author} {\bibinfo {author} {\bibfnamefont {Jianwei}\ \bibnamefont
  {Xu}},\ }\bibfield  {title} {\enquote {\bibinfo {title} {Coherence of quantum
  channels},}\ }\href@noop {} {\bibfield  {journal} {\bibinfo  {journal}
  {Physical Review A}\ }\textbf {\bibinfo {volume} {100}},\ \bibinfo {pages}
  {052311} (\bibinfo {year} {2019})}\BibitemShut {NoStop}%
\bibitem [{\citenamefont {Theurer}\ \emph {et~al.}(2019)\citenamefont
  {Theurer}, \citenamefont {Egloff}, \citenamefont {Zhang},\ and\ \citenamefont
  {Plenio}}]{theurer2019quantifying}%
  \BibitemOpen
  \bibfield  {author} {\bibinfo {author} {\bibfnamefont {Thomas}\ \bibnamefont
  {Theurer}}, \bibinfo {author} {\bibfnamefont {Dario}\ \bibnamefont {Egloff}},
  \bibinfo {author} {\bibfnamefont {Lijian}\ \bibnamefont {Zhang}}, \ and\
  \bibinfo {author} {\bibfnamefont {Martin~B}\ \bibnamefont {Plenio}},\
  }\bibfield  {title} {\enquote {\bibinfo {title} {Quantifying operations with
  an application to coherence},}\ }\href@noop {} {\bibfield  {journal}
  {\bibinfo  {journal} {Physical Review Letters}\ }\textbf {\bibinfo {volume}
  {122}},\ \bibinfo {pages} {190405} (\bibinfo {year} {2019})}\BibitemShut
  {NoStop}%
\bibitem [{\citenamefont {Liu}\ and\ \citenamefont
  {Yuan}(2020)}]{liu2020operational}%
  \BibitemOpen
  \bibfield  {author} {\bibinfo {author} {\bibfnamefont {Yunchao}\ \bibnamefont
  {Liu}}\ and\ \bibinfo {author} {\bibfnamefont {Xiao}\ \bibnamefont {Yuan}},\
  }\bibfield  {title} {\enquote {\bibinfo {title} {Operational resource theory
  of quantum channels},}\ }\href@noop {} {\bibfield  {journal} {\bibinfo
  {journal} {Physical Review Research}\ }\textbf {\bibinfo {volume} {2}},\
  \bibinfo {pages} {012035(R)} (\bibinfo {year} {2020})}\BibitemShut {NoStop}%
\bibitem [{\citenamefont {Takagi}\ \emph {et~al.}(2020)\citenamefont {Takagi},
  \citenamefont {Wang},\ and\ \citenamefont {Hayashi}}]{takagi2020application}%
  \BibitemOpen
  \bibfield  {author} {\bibinfo {author} {\bibfnamefont {Ryuji}\ \bibnamefont
  {Takagi}}, \bibinfo {author} {\bibfnamefont {Kun}\ \bibnamefont {Wang}}, \
  and\ \bibinfo {author} {\bibfnamefont {Masahito}\ \bibnamefont {Hayashi}},\
  }\bibfield  {title} {\enquote {\bibinfo {title} {Application of the resource
  theory of channels to communication scenarios},}\ }\href@noop {} {\bibfield
  {journal} {\bibinfo  {journal} {Physical Review Letters}\ }\textbf {\bibinfo
  {volume} {124}},\ \bibinfo {pages} {120502} (\bibinfo {year}
  {2020})}\BibitemShut {NoStop}%
\bibitem [{\citenamefont {Kristj{\'{a}}nsson}\ \emph
  {et~al.}(2020)\citenamefont {Kristj{\'{a}}nsson}, \citenamefont {Chiribella},
  \citenamefont {Salek}, \citenamefont {Ebler},\ and\ \citenamefont
  {Wilson}}]{kristjansson2020resource}%
  \BibitemOpen
  \bibfield  {author} {\bibinfo {author} {\bibfnamefont {Hl{\'{e}}r}\
  \bibnamefont {Kristj{\'{a}}nsson}}, \bibinfo {author} {\bibfnamefont
  {Giulio}\ \bibnamefont {Chiribella}}, \bibinfo {author} {\bibfnamefont
  {Sina}\ \bibnamefont {Salek}}, \bibinfo {author} {\bibfnamefont {Daniel}\
  \bibnamefont {Ebler}}, \ and\ \bibinfo {author} {\bibfnamefont {Matthew}\
  \bibnamefont {Wilson}},\ }\bibfield  {title} {\enquote {\bibinfo {title}
  {Resource theories of communication},}\ }\href {\doibase
  10.1088/1367-2630/ab8ef7} {\bibfield  {journal} {\bibinfo  {journal} {New
  Journal of Physics}\ }\textbf {\bibinfo {volume} {22}},\ \bibinfo {pages}
  {073014} (\bibinfo {year} {2020})}\BibitemShut {NoStop}%
\bibitem [{\citenamefont {Saxena}\ \emph {et~al.}(2020)\citenamefont {Saxena},
  \citenamefont {Chitambar},\ and\ \citenamefont {Gour}}]{saxena2020dynamical}%
  \BibitemOpen
  \bibfield  {author} {\bibinfo {author} {\bibfnamefont {Gaurav}\ \bibnamefont
  {Saxena}}, \bibinfo {author} {\bibfnamefont {Eric}\ \bibnamefont
  {Chitambar}}, \ and\ \bibinfo {author} {\bibfnamefont {Gilad}\ \bibnamefont
  {Gour}},\ }\bibfield  {title} {\enquote {\bibinfo {title} {Dynamical resource
  theory of quantum coherence},}\ }\href@noop {} {\bibfield  {journal}
  {\bibinfo  {journal} {Physical Review Research}\ }\textbf {\bibinfo {volume}
  {2}},\ \bibinfo {pages} {023298} (\bibinfo {year} {2020})}\BibitemShut
  {NoStop}%
\bibitem [{\citenamefont {Bisio}\ \emph {et~al.}(2009)\citenamefont {Bisio},
  \citenamefont {Chiribella}, \citenamefont {D'Ariano}, \citenamefont
  {Facchini},\ and\ \citenamefont {Perinotti}}]{bisio2009optimal}%
  \BibitemOpen
  \bibfield  {author} {\bibinfo {author} {\bibfnamefont {Alessandro}\
  \bibnamefont {Bisio}}, \bibinfo {author} {\bibfnamefont {Giulio}\
  \bibnamefont {Chiribella}}, \bibinfo {author} {\bibfnamefont {Giacomo~Mauro}\
  \bibnamefont {D'Ariano}}, \bibinfo {author} {\bibfnamefont {Stefano}\
  \bibnamefont {Facchini}}, \ and\ \bibinfo {author} {\bibfnamefont {Paolo}\
  \bibnamefont {Perinotti}},\ }\bibfield  {title} {\enquote {\bibinfo {title}
  {Optimal quantum tomography of states, measurements, and transformations},}\
  }\href@noop {} {\bibfield  {journal} {\bibinfo  {journal} {Physical Review
  Letters}\ }\textbf {\bibinfo {volume} {102}},\ \bibinfo {pages} {010404}
  (\bibinfo {year} {2009})}\BibitemShut {NoStop}%
\bibitem [{\citenamefont {Chiribella}\ \emph
  {et~al.}(2008{\natexlab{b}})\citenamefont {Chiribella}, \citenamefont
  {D'Ariano},\ and\ \citenamefont {Perinotti}}]{chiribella2008optimal}%
  \BibitemOpen
  \bibfield  {author} {\bibinfo {author} {\bibfnamefont {Giulio}\ \bibnamefont
  {Chiribella}}, \bibinfo {author} {\bibfnamefont {Giacomo~Mauro}\ \bibnamefont
  {D'Ariano}}, \ and\ \bibinfo {author} {\bibfnamefont {Paolo}\ \bibnamefont
  {Perinotti}},\ }\bibfield  {title} {\enquote {\bibinfo {title} {Optimal
  cloning of unitary transformation},}\ }\href@noop {} {\bibfield  {journal}
  {\bibinfo  {journal} {Physical Review Letters}\ }\textbf {\bibinfo {volume}
  {101}},\ \bibinfo {pages} {180504} (\bibinfo {year}
  {2008}{\natexlab{b}})}\BibitemShut {NoStop}%
\bibitem [{\citenamefont {Bisio}\ \emph {et~al.}(2011)\citenamefont {Bisio},
  \citenamefont {D'Ariano}, \citenamefont {Perinotti},\ and\ \citenamefont
  {Sedl{\'a}k}}]{bisio2011cloning}%
  \BibitemOpen
  \bibfield  {author} {\bibinfo {author} {\bibfnamefont {Alessandro}\
  \bibnamefont {Bisio}}, \bibinfo {author} {\bibfnamefont {Giacomo~Mauro}\
  \bibnamefont {D'Ariano}}, \bibinfo {author} {\bibfnamefont {Paolo}\
  \bibnamefont {Perinotti}}, \ and\ \bibinfo {author} {\bibfnamefont {Michal}\
  \bibnamefont {Sedl{\'a}k}},\ }\bibfield  {title} {\enquote {\bibinfo {title}
  {Cloning of a quantum measurement},}\ }\href@noop {} {\bibfield  {journal}
  {\bibinfo  {journal} {Physical Review A}\ }\textbf {\bibinfo {volume} {84}},\
  \bibinfo {pages} {042330} (\bibinfo {year} {2011})}\BibitemShut {NoStop}%
\bibitem [{\citenamefont {Bisio}\ \emph {et~al.}(2010)\citenamefont {Bisio},
  \citenamefont {Chiribella}, \citenamefont {D'Ariano}, \citenamefont
  {Facchini},\ and\ \citenamefont {Perinotti}}]{bisio2010optimal}%
  \BibitemOpen
  \bibfield  {author} {\bibinfo {author} {\bibfnamefont {Alessandro}\
  \bibnamefont {Bisio}}, \bibinfo {author} {\bibfnamefont {Giulio}\
  \bibnamefont {Chiribella}}, \bibinfo {author} {\bibfnamefont {Giacomo~Mauro}\
  \bibnamefont {D'Ariano}}, \bibinfo {author} {\bibfnamefont {Stefano}\
  \bibnamefont {Facchini}}, \ and\ \bibinfo {author} {\bibfnamefont {Paolo}\
  \bibnamefont {Perinotti}},\ }\bibfield  {title} {\enquote {\bibinfo {title}
  {Optimal quantum learning of a unitary transformation},}\ }\href@noop {}
  {\bibfield  {journal} {\bibinfo  {journal} {Physical Review A}\ }\textbf
  {\bibinfo {volume} {81}},\ \bibinfo {pages} {032324} (\bibinfo {year}
  {2010})}\BibitemShut {NoStop}%
\bibitem [{\citenamefont {Sedl{\'a}k}\ and\ \citenamefont
  {Ziman}(2020)}]{sedlak2020probabilistic}%
  \BibitemOpen
  \bibfield  {author} {\bibinfo {author} {\bibfnamefont {Michal}\ \bibnamefont
  {Sedl{\'a}k}}\ and\ \bibinfo {author} {\bibfnamefont {M{\'a}rio}\
  \bibnamefont {Ziman}},\ }\bibfield  {title} {\enquote {\bibinfo {title}
  {Probabilistic storage and retrieval of qubit phase gates},}\ }\href@noop {}
  {\bibfield  {journal} {\bibinfo  {journal} {Physical Review A}\ }\textbf
  {\bibinfo {volume} {102}},\ \bibinfo {pages} {032618} (\bibinfo {year}
  {2020})}\BibitemShut {NoStop}%
\bibitem [{\citenamefont {Chiribella}\ and\ \citenamefont
  {Ebler}(2016)}]{chiribella2016optimal}%
  \BibitemOpen
  \bibfield  {author} {\bibinfo {author} {\bibfnamefont {Giulio}\ \bibnamefont
  {Chiribella}}\ and\ \bibinfo {author} {\bibfnamefont {Daniel}\ \bibnamefont
  {Ebler}},\ }\bibfield  {title} {\enquote {\bibinfo {title} {Optimal quantum
  networks and one-shot entropies},}\ }\href@noop {} {\bibfield  {journal}
  {\bibinfo  {journal} {New Journal of Physics}\ }\textbf {\bibinfo {volume}
  {18}},\ \bibinfo {pages} {093053} (\bibinfo {year} {2016})}\BibitemShut
  {NoStop}%
\bibitem [{\citenamefont {Miyazaki}\ \emph {et~al.}(2019)\citenamefont
  {Miyazaki}, \citenamefont {Soeda},\ and\ \citenamefont
  {Murao}}]{miyazaki2019complex}%
  \BibitemOpen
  \bibfield  {author} {\bibinfo {author} {\bibfnamefont {Jisho}\ \bibnamefont
  {Miyazaki}}, \bibinfo {author} {\bibfnamefont {Akihito}\ \bibnamefont
  {Soeda}}, \ and\ \bibinfo {author} {\bibfnamefont {Mio}\ \bibnamefont
  {Murao}},\ }\bibfield  {title} {\enquote {\bibinfo {title} {Complex
  conjugation supermap of unitary quantum maps and its universal implementation
  protocol},}\ }\href@noop {} {\bibfield  {journal} {\bibinfo  {journal}
  {Physical Review Research}\ }\textbf {\bibinfo {volume} {1}},\ \bibinfo
  {pages} {013007} (\bibinfo {year} {2019})}\BibitemShut {NoStop}%
\bibitem [{\citenamefont {Quintino}\ \emph {et~al.}(2019)\citenamefont
  {Quintino}, \citenamefont {Dong}, \citenamefont {Shimbo}, \citenamefont
  {Soeda},\ and\ \citenamefont {Murao}}]{quintino2019probabilistic}%
  \BibitemOpen
  \bibfield  {author} {\bibinfo {author} {\bibfnamefont {Marco~T{\'u}lio}\
  \bibnamefont {Quintino}}, \bibinfo {author} {\bibfnamefont {Qingxiuxiong}\
  \bibnamefont {Dong}}, \bibinfo {author} {\bibfnamefont {Atsushi}\
  \bibnamefont {Shimbo}}, \bibinfo {author} {\bibfnamefont {Akihito}\
  \bibnamefont {Soeda}}, \ and\ \bibinfo {author} {\bibfnamefont {Mio}\
  \bibnamefont {Murao}},\ }\bibfield  {title} {\enquote {\bibinfo {title}
  {Probabilistic exact universal quantum circuits for transforming unitary
  operations},}\ }\href@noop {} {\bibfield  {journal} {\bibinfo  {journal}
  {Physical Review A}\ }\textbf {\bibinfo {volume} {100}},\ \bibinfo {pages}
  {062339} (\bibinfo {year} {2019})}\BibitemShut {NoStop}%
\bibitem [{\citenamefont {Gour}\ and\ \citenamefont
  {Scandolo}(2020)}]{gour2020dynamical}%
  \BibitemOpen
  \bibfield  {author} {\bibinfo {author} {\bibfnamefont {Gilad}\ \bibnamefont
  {Gour}}\ and\ \bibinfo {author} {\bibfnamefont {Carlo~Maria}\ \bibnamefont
  {Scandolo}},\ }\bibfield  {title} {\enquote {\bibinfo {title} {Dynamical
  entanglement},}\ }\href@noop {} {\bibfield  {journal} {\bibinfo  {journal}
  {Physical Review Letters}\ }\textbf {\bibinfo {volume} {125}},\ \bibinfo
  {pages} {180505} (\bibinfo {year} {2020})}\BibitemShut {NoStop}%
\bibitem [{\citenamefont {Chen}\ and\ \citenamefont
  {Chitambar}(2020)}]{chen2020entanglement}%
  \BibitemOpen
  \bibfield  {author} {\bibinfo {author} {\bibfnamefont {Senrui}\ \bibnamefont
  {Chen}}\ and\ \bibinfo {author} {\bibfnamefont {Eric}\ \bibnamefont
  {Chitambar}},\ }\bibfield  {title} {\enquote {\bibinfo {title}
  {Entanglement-breaking superchannels},}\ }\href@noop {} {\bibfield  {journal}
  {\bibinfo  {journal} {Quantum}\ }\textbf {\bibinfo {volume} {4}},\ \bibinfo
  {pages} {299} (\bibinfo {year} {2020})}\BibitemShut {NoStop}%
\bibitem [{\citenamefont {Ben~Dana}\ \emph {et~al.}(2017)\citenamefont
  {Ben~Dana}, \citenamefont {Garc\'{\i}a~D\'{\i}az}, \citenamefont {Mejatty},\
  and\ \citenamefont {Winter}}]{dana2017resource}%
  \BibitemOpen
  \bibfield  {author} {\bibinfo {author} {\bibfnamefont {Khaled}\ \bibnamefont
  {Ben~Dana}}, \bibinfo {author} {\bibfnamefont {Mar\'{\i}a}\ \bibnamefont
  {Garc\'{\i}a~D\'{\i}az}}, \bibinfo {author} {\bibfnamefont {Mohamed}\
  \bibnamefont {Mejatty}}, \ and\ \bibinfo {author} {\bibfnamefont {Andreas}\
  \bibnamefont {Winter}},\ }\bibfield  {title} {\enquote {\bibinfo {title}
  {Resource theory of coherence: Beyond states},}\ }\href@noop {} {\bibfield
  {journal} {\bibinfo  {journal} {Physical Review A}\ }\textbf {\bibinfo
  {volume} {95}},\ \bibinfo {pages} {062327} (\bibinfo {year}
  {2017})}\BibitemShut {NoStop}%
\bibitem [{\citenamefont {Kraus}\ \emph {et~al.}(1983)\citenamefont {Kraus},
  \citenamefont {B{\"o}hm}, \citenamefont {Dollard},\ and\ \citenamefont
  {WH}}]{kraus1983states}%
  \BibitemOpen
  \bibfield  {author} {\bibinfo {author} {\bibfnamefont {Karl}\ \bibnamefont
  {Kraus}}, \bibinfo {author} {\bibfnamefont {Arno}\ \bibnamefont {B{\"o}hm}},
  \bibinfo {author} {\bibfnamefont {John~D}\ \bibnamefont {Dollard}}, \ and\
  \bibinfo {author} {\bibfnamefont {William}\ \bibnamefont {WH}},\ }\bibfield
  {title} {\enquote {\bibinfo {title} {States, effects, and operations:
  fundamental notions of quantum theory. lectures in mathematical physics at
  the university of texas at austin},}\ }\href@noop {} {\bibfield  {journal}
  {\bibinfo  {journal} {Lecture Notes in Physics}\ }\textbf {\bibinfo {volume}
  {190}} (\bibinfo {year} {1983})}\BibitemShut {NoStop}%
\bibitem [{\citenamefont {Heinosaari}\ and\ \citenamefont
  {Ziman}(2011)}]{heinosaari2011mathematical}%
  \BibitemOpen
  \bibfield  {author} {\bibinfo {author} {\bibfnamefont {Teiko}\ \bibnamefont
  {Heinosaari}}\ and\ \bibinfo {author} {\bibfnamefont {M{\'a}rio}\
  \bibnamefont {Ziman}},\ }\href@noop {} {\emph {\bibinfo {title} {The
  mathematical language of quantum theory: from uncertainty to entanglement}}}\
  (\bibinfo  {publisher} {Cambridge University Press},\ \bibinfo {year}
  {2011})\BibitemShut {NoStop}%
\bibitem [{\citenamefont {Schwinger}(1951)}]{schwinger1951theory}%
  \BibitemOpen
  \bibfield  {author} {\bibinfo {author} {\bibfnamefont {Julian}\ \bibnamefont
  {Schwinger}},\ }\bibfield  {title} {\enquote {\bibinfo {title} {The theory of
  quantized fields. i},}\ }\href@noop {} {\bibfield  {journal} {\bibinfo
  {journal} {Physical Review}\ }\textbf {\bibinfo {volume} {82}},\ \bibinfo
  {pages} {914} (\bibinfo {year} {1951})}\BibitemShut {NoStop}%
\bibitem [{\citenamefont {L{\"u}ders}(1954)}]{luders1954equivalence}%
  \BibitemOpen
  \bibfield  {author} {\bibinfo {author} {\bibfnamefont {Gerhart}\ \bibnamefont
  {L{\"u}ders}},\ }\bibfield  {title} {\enquote {\bibinfo {title} {On the
  equivalence of invariance under time reversal and under particle-antiparticle
  conjugation for relativistic field theories},}\ }\href@noop {} {\bibfield
  {journal} {\bibinfo  {journal} {Dan. Mat. Fys. Medd.}\ }\textbf {\bibinfo
  {volume} {28}},\ \bibinfo {pages} {1--17} (\bibinfo {year}
  {1954})}\BibitemShut {NoStop}%
\bibitem [{\citenamefont {Pauli}(1955)}]{pauli1955niels}%
  \BibitemOpen
  \bibfield  {author} {\bibinfo {author} {\bibfnamefont {Wolfgang}\
  \bibnamefont {Pauli}},\ }in\ \href@noop {} {\emph {\bibinfo {booktitle}
  {Niels Bohr and the Development of Physics}}}\ (\bibinfo  {publisher}
  {McGraw-Hill, New York},\ \bibinfo {year} {1955})\ pp.\ \bibinfo {pages}
  {30--51}\BibitemShut {NoStop}%
\bibitem [{\citenamefont {Bartlett}\ \emph {et~al.}(2007)\citenamefont
  {Bartlett}, \citenamefont {Rudolph},\ and\ \citenamefont
  {Spekkens}}]{bartlett2007reference}%
  \BibitemOpen
  \bibfield  {author} {\bibinfo {author} {\bibfnamefont {Stephen~D}\
  \bibnamefont {Bartlett}}, \bibinfo {author} {\bibfnamefont {Terry}\
  \bibnamefont {Rudolph}}, \ and\ \bibinfo {author} {\bibfnamefont {Robert~W}\
  \bibnamefont {Spekkens}},\ }\bibfield  {title} {\enquote {\bibinfo {title}
  {Reference frames, superselection rules, and quantum information},}\
  }\href@noop {} {\bibfield  {journal} {\bibinfo  {journal} {Reviews of Modern
  Physics}\ }\textbf {\bibinfo {volume} {79}},\ \bibinfo {pages} {555}
  (\bibinfo {year} {2007})}\BibitemShut {NoStop}%
\bibitem [{\citenamefont {Aurell}\ \emph {et~al.}(2015)\citenamefont {Aurell},
  \citenamefont {Zakrzewski},\ and\ \citenamefont {\.{Z}yczkowski}}]{AZZ}%
  \BibitemOpen
  \bibfield  {author} {\bibinfo {author} {\bibfnamefont {Erik}\ \bibnamefont
  {Aurell}}, \bibinfo {author} {\bibfnamefont {Jakub}\ \bibnamefont
  {Zakrzewski}}, \ and\ \bibinfo {author} {\bibfnamefont {Karol}\ \bibnamefont
  {\.{Z}yczkowski}},\ }\bibfield  {title} {\enquote {\bibinfo {title} {Time
  reversals of irreversible quantum maps},}\ }\href@noop {} {\bibfield
  {journal} {\bibinfo  {journal} {Journal of Physics A: Mathematical and
  Theoretical}\ }\textbf {\bibinfo {volume} {48}},\ \bibinfo {pages} {38FT01}
  (\bibinfo {year} {2015})}\BibitemShut {NoStop}%
\bibitem [{\citenamefont {Karimipour}\ \emph {et~al.}(2020)\citenamefont
  {Karimipour}, \citenamefont {Benatti},\ and\ \citenamefont
  {Floreanini}}]{KBF20}%
  \BibitemOpen
  \bibfield  {author} {\bibinfo {author} {\bibfnamefont {Vahid}\ \bibnamefont
  {Karimipour}}, \bibinfo {author} {\bibfnamefont {Fabio}\ \bibnamefont
  {Benatti}}, \ and\ \bibinfo {author} {\bibfnamefont {Roberto}\ \bibnamefont
  {Floreanini}},\ }\bibfield  {title} {\enquote {\bibinfo {title} {Quasi
  inversion of qubit channels},}\ }\href@noop {} {\bibfield  {journal}
  {\bibinfo  {journal} {Phys. Rev.}\ }\textbf {\bibinfo {volume} {A~101}},\
  \bibinfo {pages} {032109} (\bibinfo {year} {2020})}\BibitemShut {NoStop}%
\bibitem [{\citenamefont {Shahbeigi}\ \emph {et~al.}(2021)\citenamefont
  {Shahbeigi}, \citenamefont {Sadri}, \citenamefont {Moradi}, \citenamefont
  {{\.Z}yczkowski},\ and\ \citenamefont {Karimipour}}]{SSMZK21}%
  \BibitemOpen
  \bibfield  {author} {\bibinfo {author} {\bibfnamefont {Fereshte}\
  \bibnamefont {Shahbeigi}}, \bibinfo {author} {\bibfnamefont {Koorosh}\
  \bibnamefont {Sadri}}, \bibinfo {author} {\bibfnamefont {Morteza}\
  \bibnamefont {Moradi}}, \bibinfo {author} {\bibfnamefont {Karol}\
  \bibnamefont {{\.Z}yczkowski}}, \ and\ \bibinfo {author} {\bibfnamefont
  {Vahid}\ \bibnamefont {Karimipour}},\ }\bibfield  {title} {\enquote {\bibinfo
  {title} {Quasi-inversion of quantum and classical channels in finite
  dimensions},}\ }\href@noop {} {\bibfield  {journal} {\bibinfo  {journal}
  {arXiv preprint arXiv:2104.06062}\ } (\bibinfo {year} {2021})}\BibitemShut
  {NoStop}%
\bibitem [{\citenamefont {Aharonov}\ \emph {et~al.}(1964)\citenamefont
  {Aharonov}, \citenamefont {Bergmann},\ and\ \citenamefont
  {Lebowitz}}]{aharonov1964time}%
  \BibitemOpen
  \bibfield  {author} {\bibinfo {author} {\bibfnamefont {Yakir}\ \bibnamefont
  {Aharonov}}, \bibinfo {author} {\bibfnamefont {Peter~G}\ \bibnamefont
  {Bergmann}}, \ and\ \bibinfo {author} {\bibfnamefont {Joel~L}\ \bibnamefont
  {Lebowitz}},\ }\bibfield  {title} {\enquote {\bibinfo {title} {Time symmetry
  in the quantum process of measurement},}\ }\href@noop {} {\bibfield
  {journal} {\bibinfo  {journal} {Physical Review}\ }\textbf {\bibinfo {volume}
  {134}},\ \bibinfo {pages} {B1410} (\bibinfo {year} {1964})}\BibitemShut
  {NoStop}%
\bibitem [{\citenamefont {Reznik}\ and\ \citenamefont
  {Aharonov}(1995)}]{reznik1995time}%
  \BibitemOpen
  \bibfield  {author} {\bibinfo {author} {\bibfnamefont {Benni}\ \bibnamefont
  {Reznik}}\ and\ \bibinfo {author} {\bibfnamefont {Yakir}\ \bibnamefont
  {Aharonov}},\ }\bibfield  {title} {\enquote {\bibinfo {title} {Time-symmetric
  formulation of quantum mechanics},}\ }\href@noop {} {\bibfield  {journal}
  {\bibinfo  {journal} {Physical Review A}\ }\textbf {\bibinfo {volume} {52}},\
  \bibinfo {pages} {2538} (\bibinfo {year} {1995})}\BibitemShut {NoStop}%
\bibitem [{\citenamefont {Aharonov}\ \emph {et~al.}(2010)\citenamefont
  {Aharonov}, \citenamefont {Popescu},\ and\ \citenamefont
  {Tollaksen}}]{aharonov2010time}%
  \BibitemOpen
  \bibfield  {author} {\bibinfo {author} {\bibfnamefont {Yakir}\ \bibnamefont
  {Aharonov}}, \bibinfo {author} {\bibfnamefont {Sandu}\ \bibnamefont
  {Popescu}}, \ and\ \bibinfo {author} {\bibfnamefont {Jeff}\ \bibnamefont
  {Tollaksen}},\ }\bibfield  {title} {\enquote {\bibinfo {title} {A
  time-symmetric formulation of quantum mechanics},}\ }\href@noop {} {\bibfield
   {journal} {\bibinfo  {journal} {Physics Today}\ }\textbf {\bibinfo {volume}
  {63}},\ \bibinfo {pages} {27} (\bibinfo {year} {2010})}\BibitemShut {NoStop}%
\bibitem [{\citenamefont {Coecke}\ and\ \citenamefont
  {Lal}(2012)}]{coecke2012time}%
  \BibitemOpen
  \bibfield  {author} {\bibinfo {author} {\bibfnamefont {Bob}\ \bibnamefont
  {Coecke}}\ and\ \bibinfo {author} {\bibfnamefont {Raymond}\ \bibnamefont
  {Lal}},\ }\bibfield  {title} {\enquote {\bibinfo {title} {Time asymmetry of
  probabilities versus relativistic causal structure: An arrow of time},}\
  }\href@noop {} {\bibfield  {journal} {\bibinfo  {journal} {Physical Review
  Letters}\ }\textbf {\bibinfo {volume} {108}},\ \bibinfo {pages} {200403}
  (\bibinfo {year} {2012})}\BibitemShut {NoStop}%
\bibitem [{\citenamefont {Oreshkov}\ and\ \citenamefont
  {Cerf}(2015)}]{oreshkov2015operational}%
  \BibitemOpen
  \bibfield  {author} {\bibinfo {author} {\bibfnamefont {Ognyan}\ \bibnamefont
  {Oreshkov}}\ and\ \bibinfo {author} {\bibfnamefont {Nicolas~J}\ \bibnamefont
  {Cerf}},\ }\bibfield  {title} {\enquote {\bibinfo {title} {Operational
  formulation of time reversal in quantum theory},}\ }\href@noop {} {\bibfield
  {journal} {\bibinfo  {journal} {Nature Physics}\ }\textbf {\bibinfo {volume}
  {11}},\ \bibinfo {pages} {853--858} (\bibinfo {year} {2015})}\BibitemShut
  {NoStop}%
\bibitem [{\citenamefont {Leifer}\ and\ \citenamefont
  {Pusey}(2017)}]{leifer2017time}%
  \BibitemOpen
  \bibfield  {author} {\bibinfo {author} {\bibfnamefont {Matthew~S}\
  \bibnamefont {Leifer}}\ and\ \bibinfo {author} {\bibfnamefont {Matthew~F}\
  \bibnamefont {Pusey}},\ }\bibfield  {title} {\enquote {\bibinfo {title} {Is a
  time symmetric interpretation of quantum theory possible without
  retrocausality?}}\ }\href@noop {} {\bibfield  {journal} {\bibinfo  {journal}
  {Proceedings of the Royal Society A: Mathematical, Physical and Engineering
  Sciences}\ }\textbf {\bibinfo {volume} {473}},\ \bibinfo {pages} {20160607}
  (\bibinfo {year} {2017})}\BibitemShut {NoStop}%
\bibitem [{\citenamefont {Coecke}\ \emph {et~al.}(2017)\citenamefont {Coecke},
  \citenamefont {Gogioso},\ and\ \citenamefont {Selby}}]{coecke2017time}%
  \BibitemOpen
  \bibfield  {author} {\bibinfo {author} {\bibfnamefont {Bob}\ \bibnamefont
  {Coecke}}, \bibinfo {author} {\bibfnamefont {Stefano}\ \bibnamefont
  {Gogioso}}, \ and\ \bibinfo {author} {\bibfnamefont {John~H}\ \bibnamefont
  {Selby}},\ }\bibfield  {title} {\enquote {\bibinfo {title} {The time-reverse
  of any causal theory is eternal noise},}\ }\href@noop {} {\bibfield
  {journal} {\bibinfo  {journal} {arXiv preprint arXiv:1711.05511}\ } (\bibinfo
  {year} {2017})}\BibitemShut {NoStop}%
\bibitem [{\citenamefont {Di~Biagio}\ \emph {et~al.}(2020)\citenamefont
  {Di~Biagio}, \citenamefont {Don{\`a}},\ and\ \citenamefont
  {Rovelli}}]{di2020quantum}%
  \BibitemOpen
  \bibfield  {author} {\bibinfo {author} {\bibfnamefont {Andrea}\ \bibnamefont
  {Di~Biagio}}, \bibinfo {author} {\bibfnamefont {Pietro}\ \bibnamefont
  {Don{\`a}}}, \ and\ \bibinfo {author} {\bibfnamefont {Carlo}\ \bibnamefont
  {Rovelli}},\ }\bibfield  {title} {\enquote {\bibinfo {title} {Quantum
  information and the arrow of time},}\ }\href@noop {} {\bibfield  {journal}
  {\bibinfo  {journal} {arXiv preprint arXiv:2010.05734}\ } (\bibinfo {year}
  {2020})}\BibitemShut {NoStop}%
\bibitem [{\citenamefont {Chiribella}\ and\ \citenamefont
  {Liu}(2020)}]{chiribella2020quantum}%
  \BibitemOpen
  \bibfield  {author} {\bibinfo {author} {\bibfnamefont {Giulio}\ \bibnamefont
  {Chiribella}}\ and\ \bibinfo {author} {\bibfnamefont {Zixuan}\ \bibnamefont
  {Liu}},\ }\bibfield  {title} {\enquote {\bibinfo {title} {The quantum time
  flip},}\ }\href@noop {} {\bibfield  {journal} {\bibinfo  {journal} {arXiv
  preprint arXiv:2012.03859}\ } (\bibinfo {year} {2020})}\BibitemShut {NoStop}%
\bibitem [{\citenamefont {Hardy}(2021)}]{hardy2021time}%
  \BibitemOpen
  \bibfield  {author} {\bibinfo {author} {\bibfnamefont {Lucien}\ \bibnamefont
  {Hardy}},\ }\bibfield  {title} {\enquote {\bibinfo {title} {Time symmetry in
  operational theories},}\ }\href@noop {} {\bibfield  {journal} {\bibinfo
  {journal} {arXiv preprint arXiv:2104.00071}\ } (\bibinfo {year}
  {2021})}\BibitemShut {NoStop}%
\bibitem [{\citenamefont {Griffiths}(2003)}]{griffiths2003consistent}%
  \BibitemOpen
  \bibfield  {author} {\bibinfo {author} {\bibfnamefont {Robert~B}\
  \bibnamefont {Griffiths}},\ }\href@noop {} {\emph {\bibinfo {title}
  {Consistent quantum theory}}}\ (\bibinfo  {publisher} {Cambridge University
  Press},\ \bibinfo {year} {2003})\BibitemShut {NoStop}%
\bibitem [{\citenamefont {Chiribella}\ \emph {et~al.}(2010)\citenamefont
  {Chiribella}, \citenamefont {D'Ariano},\ and\ \citenamefont
  {Perinotti}}]{chiribella2010probabilistic}%
  \BibitemOpen
  \bibfield  {author} {\bibinfo {author} {\bibfnamefont {Giulio}\ \bibnamefont
  {Chiribella}}, \bibinfo {author} {\bibfnamefont {Giacomo~Mauro}\ \bibnamefont
  {D'Ariano}}, \ and\ \bibinfo {author} {\bibfnamefont {Paolo}\ \bibnamefont
  {Perinotti}},\ }\bibfield  {title} {\enquote {\bibinfo {title} {Probabilistic
  theories with purification},}\ }\href@noop {} {\bibfield  {journal} {\bibinfo
   {journal} {Physical Review A}\ }\textbf {\bibinfo {volume} {81}},\ \bibinfo
  {pages} {062348} (\bibinfo {year} {2010})}\BibitemShut {NoStop}%
\bibitem [{\citenamefont {Chiribella}\ \emph {et~al.}(2011)\citenamefont
  {Chiribella}, \citenamefont {D’Ariano},\ and\ \citenamefont
  {Perinotti}}]{chiribella2011informational}%
  \BibitemOpen
  \bibfield  {author} {\bibinfo {author} {\bibfnamefont {Giulio}\ \bibnamefont
  {Chiribella}}, \bibinfo {author} {\bibfnamefont {Giacomo~Mauro}\ \bibnamefont
  {D’Ariano}}, \ and\ \bibinfo {author} {\bibfnamefont {Paolo}\ \bibnamefont
  {Perinotti}},\ }\bibfield  {title} {\enquote {\bibinfo {title} {Informational
  derivation of quantum theory},}\ }\href@noop {} {\bibfield  {journal}
  {\bibinfo  {journal} {Physical Review A}\ }\textbf {\bibinfo {volume} {84}},\
  \bibinfo {pages} {012311} (\bibinfo {year} {2011})}\BibitemShut {NoStop}%
\bibitem [{\citenamefont {Chiribella}\ \emph {et~al.}(2016)\citenamefont
  {Chiribella}, \citenamefont {D’Ariano},\ and\ \citenamefont
  {Perinotti}}]{chiribella2016quantum}%
  \BibitemOpen
  \bibfield  {author} {\bibinfo {author} {\bibfnamefont {Giulio}\ \bibnamefont
  {Chiribella}}, \bibinfo {author} {\bibfnamefont {Giacomo~Mauro}\ \bibnamefont
  {D’Ariano}}, \ and\ \bibinfo {author} {\bibfnamefont {Paolo}\ \bibnamefont
  {Perinotti}},\ }\bibfield  {title} {\enquote {\bibinfo {title} {Quantum from
  principles},}\ }in\ \href@noop {} {\emph {\bibinfo {booktitle} {Quantum
  theory: informational foundations and foils}}}\ (\bibinfo  {publisher}
  {Springer},\ \bibinfo {year} {2016})\ pp.\ \bibinfo {pages}
  {171--221}\BibitemShut {NoStop}%
\bibitem [{\citenamefont {D'Ariano}\ \emph {et~al.}(2017)\citenamefont
  {D'Ariano}, \citenamefont {Chiribella},\ and\ \citenamefont
  {Perinotti}}]{d2017quantum}%
  \BibitemOpen
  \bibfield  {author} {\bibinfo {author} {\bibfnamefont {Giacomo~Mauro}\
  \bibnamefont {D'Ariano}}, \bibinfo {author} {\bibfnamefont {Giulio}\
  \bibnamefont {Chiribella}}, \ and\ \bibinfo {author} {\bibfnamefont {Paolo}\
  \bibnamefont {Perinotti}},\ }\href@noop {} {\emph {\bibinfo {title} {Quantum
  theory from first principles: an informational approach}}}\ (\bibinfo
  {publisher} {Cambridge University Press},\ \bibinfo {year}
  {2017})\BibitemShut {NoStop}%
\bibitem [{\citenamefont {Horodecki}\ and\ \citenamefont
  {Oppenheim}(2013)}]{horodecki2013fundamental}%
  \BibitemOpen
  \bibfield  {author} {\bibinfo {author} {\bibfnamefont {Micha{\l}}\
  \bibnamefont {Horodecki}}\ and\ \bibinfo {author} {\bibfnamefont {Jonathan}\
  \bibnamefont {Oppenheim}},\ }\bibfield  {title} {\enquote {\bibinfo {title}
  {Fundamental limitations for quantum and nanoscale thermodynamics},}\
  }\href@noop {} {\bibfield  {journal} {\bibinfo  {journal} {Nature
  communications}\ }\textbf {\bibinfo {volume} {4}},\ \bibinfo {pages} {1--6}
  (\bibinfo {year} {2013})}\BibitemShut {NoStop}%
\bibitem [{\citenamefont {Gour}\ \emph {et~al.}(2015)\citenamefont {Gour},
  \citenamefont {M{\"u}ller}, \citenamefont {Narasimhachar}, \citenamefont
  {Spekkens},\ and\ \citenamefont {Halpern}}]{gour2015resource}%
  \BibitemOpen
  \bibfield  {author} {\bibinfo {author} {\bibfnamefont {Gilad}\ \bibnamefont
  {Gour}}, \bibinfo {author} {\bibfnamefont {Markus~P}\ \bibnamefont
  {M{\"u}ller}}, \bibinfo {author} {\bibfnamefont {Varun}\ \bibnamefont
  {Narasimhachar}}, \bibinfo {author} {\bibfnamefont {Robert~W}\ \bibnamefont
  {Spekkens}}, \ and\ \bibinfo {author} {\bibfnamefont {Nicole~Yunger}\
  \bibnamefont {Halpern}},\ }\bibfield  {title} {\enquote {\bibinfo {title}
  {The resource theory of informational nonequilibrium in thermodynamics},}\
  }\href@noop {} {\bibfield  {journal} {\bibinfo  {journal} {Physics Reports}\
  }\textbf {\bibinfo {volume} {583}},\ \bibinfo {pages} {1--58} (\bibinfo
  {year} {2015})}\BibitemShut {NoStop}%
\bibitem [{\citenamefont {Chiribella}\ and\ \citenamefont
  {Scandolo}(2017)}]{chiribella2017microcanonical}%
  \BibitemOpen
  \bibfield  {author} {\bibinfo {author} {\bibfnamefont {Giulio}\ \bibnamefont
  {Chiribella}}\ and\ \bibinfo {author} {\bibfnamefont {Carlo~Maria}\
  \bibnamefont {Scandolo}},\ }\bibfield  {title} {\enquote {\bibinfo {title}
  {Microcanonical thermodynamics in general physical theories},}\ }\href@noop
  {} {\bibfield  {journal} {\bibinfo  {journal} {New Journal of Physics}\
  }\textbf {\bibinfo {volume} {19}},\ \bibinfo {pages} {123043} (\bibinfo
  {year} {2017})}\BibitemShut {NoStop}%
\bibitem [{\citenamefont {Uhlmann}(2016)}]{Uhl16}%
  \BibitemOpen
  \bibfield  {author} {\bibinfo {author} {\bibfnamefont {Armin}\ \bibnamefont
  {Uhlmann}},\ }\bibfield  {title} {\enquote {\bibinfo {title} {Anti-
  (conjugate) linearity},}\ }\href@noop {} {\bibfield  {journal} {\bibinfo
  {journal} {Science China Physics, Mechanics and Astronomy}\ }\textbf
  {\bibinfo {volume} {59}},\ \bibinfo {pages} {630301} (\bibinfo {year}
  {2016})}\BibitemShut {NoStop}%
\bibitem [{\citenamefont {Holevo}(2012)}]{holevo2012quantum}%
  \BibitemOpen
  \bibfield  {author} {\bibinfo {author} {\bibfnamefont {Alexander~S}\
  \bibnamefont {Holevo}},\ }\href@noop {} {\emph {\bibinfo {title} {Quantum
  systems, channels, information: a mathematical introduction}}},\
  Vol.~\bibinfo {volume} {16}\ (\bibinfo  {publisher} {Walter de Gruyter},\
  \bibinfo {year} {2012})\BibitemShut {NoStop}%
\bibitem [{\citenamefont {Cappellini}\ \emph {et~al.}(2007)\citenamefont
  {Cappellini}, \citenamefont {Sommers},\ and\ \citenamefont
  {\.{Z}yczkowski}}]{capellini2008}%
  \BibitemOpen
  \bibfield  {author} {\bibinfo {author} {\bibfnamefont {Valerio}\ \bibnamefont
  {Cappellini}}, \bibinfo {author} {\bibfnamefont {Hans-J\"urgen}\ \bibnamefont
  {Sommers}}, \ and\ \bibinfo {author} {\bibfnamefont {Karol}\ \bibnamefont
  {\.{Z}yczkowski}},\ }\bibfield  {title} {\enquote {\bibinfo {title}
  {Subnormalized states and trace-nonincreasing maps},}\ }\href {\doibase
  10.1063/1.2738359} {\bibfield  {journal} {\bibinfo  {journal} {Journal of
  Mathematical Physics}\ }\textbf {\bibinfo {volume} {48}},\ \bibinfo {pages}
  {052110} (\bibinfo {year} {2007})},\ \Eprint
  {http://arxiv.org/abs/https://doi.org/10.1063/1.2738359}
  {https://doi.org/10.1063/1.2738359} \BibitemShut {NoStop}%
\bibitem [{\citenamefont {Holevo}\ and\ \citenamefont
  {Werner}(2001)}]{holevo2001evaluating}%
  \BibitemOpen
  \bibfield  {author} {\bibinfo {author} {\bibfnamefont {Alexander~S}\
  \bibnamefont {Holevo}}\ and\ \bibinfo {author} {\bibfnamefont {Reinhard~F}\
  \bibnamefont {Werner}},\ }\bibfield  {title} {\enquote {\bibinfo {title}
  {Evaluating capacities of bosonic gaussian channels},}\ }\href@noop {}
  {\bibfield  {journal} {\bibinfo  {journal} {Physical Review A}\ }\textbf
  {\bibinfo {volume} {63}},\ \bibinfo {pages} {032312} (\bibinfo {year}
  {2001})}\BibitemShut {NoStop}%
\bibitem [{\citenamefont {Yang}\ \emph {et~al.}(2017)\citenamefont {Yang},
  \citenamefont {Chiribella},\ and\ \citenamefont {Hu}}]{yang2017units}%
  \BibitemOpen
  \bibfield  {author} {\bibinfo {author} {\bibfnamefont {Yuxiang}\ \bibnamefont
  {Yang}}, \bibinfo {author} {\bibfnamefont {Giulio}\ \bibnamefont
  {Chiribella}}, \ and\ \bibinfo {author} {\bibfnamefont {Qinheping}\
  \bibnamefont {Hu}},\ }\bibfield  {title} {\enquote {\bibinfo {title} {Units
  of rotational information},}\ }\href@noop {} {\bibfield  {journal} {\bibinfo
  {journal} {New Journal of Physics}\ }\textbf {\bibinfo {volume} {19}},\
  \bibinfo {pages} {123003} (\bibinfo {year} {2017})}\BibitemShut {NoStop}%
\bibitem [{\citenamefont {Landau}\ and\ \citenamefont
  {Streater}(1993)}]{landau1993birkhoff}%
  \BibitemOpen
  \bibfield  {author} {\bibinfo {author} {\bibfnamefont {L~J}\ \bibnamefont
  {Landau}}\ and\ \bibinfo {author} {\bibfnamefont {R~F}\ \bibnamefont
  {Streater}},\ }\bibfield  {title} {\enquote {\bibinfo {title} {On
  {B}irkhoff's theorem for doubly stochastic completely positive maps of matrix
  algebras},}\ }\href@noop {} {\bibfield  {journal} {\bibinfo  {journal}
  {Linear Algebra and its Applications}\ }\textbf {\bibinfo {volume} {193}},\
  \bibinfo {pages} {107--127} (\bibinfo {year} {1993})}\BibitemShut {NoStop}%
\bibitem [{\citenamefont {Mendl}\ and\ \citenamefont
  {Wolf}(2009)}]{mendl2009unital}%
  \BibitemOpen
  \bibfield  {author} {\bibinfo {author} {\bibfnamefont {Christian~B}\
  \bibnamefont {Mendl}}\ and\ \bibinfo {author} {\bibfnamefont {Michael~M}\
  \bibnamefont {Wolf}},\ }\bibfield  {title} {\enquote {\bibinfo {title}
  {Unital quantum channels--convex structure and revivals of birkhoff's
  theorem},}\ }\href@noop {} {\bibfield  {journal} {\bibinfo  {journal}
  {Communications in Mathematical Physics}\ }\textbf {\bibinfo {volume}
  {289}},\ \bibinfo {pages} {1057--1086} (\bibinfo {year} {2009})}\BibitemShut
  {NoStop}%
\bibitem [{\citenamefont {Petz}(1988)}]{petz1988sufficiency}%
  \BibitemOpen
  \bibfield  {author} {\bibinfo {author} {\bibfnamefont {D{\'e}nes}\
  \bibnamefont {Petz}},\ }\bibfield  {title} {\enquote {\bibinfo {title}
  {Sufficiency of channels over von neumann algebras},}\ }\href@noop {}
  {\bibfield  {journal} {\bibinfo  {journal} {The Quarterly Journal of
  Mathematics}\ }\textbf {\bibinfo {volume} {39}},\ \bibinfo {pages} {97--108}
  (\bibinfo {year} {1988})}\BibitemShut {NoStop}%
\bibitem [{\citenamefont {Barnum}\ and\ \citenamefont
  {Knill}(2002)}]{barnum2002reversing}%
  \BibitemOpen
  \bibfield  {author} {\bibinfo {author} {\bibfnamefont {Howard}\ \bibnamefont
  {Barnum}}\ and\ \bibinfo {author} {\bibfnamefont {Emanuel}\ \bibnamefont
  {Knill}},\ }\bibfield  {title} {\enquote {\bibinfo {title} {Reversing quantum
  dynamics with near-optimal quantum and classical fidelity},}\ }\href@noop {}
  {\bibfield  {journal} {\bibinfo  {journal} {Journal of Mathematical Physics}\
  }\textbf {\bibinfo {volume} {43}},\ \bibinfo {pages} {2097--2106} (\bibinfo
  {year} {2002})}\BibitemShut {NoStop}%
\bibitem [{\citenamefont {Hayden}\ \emph {et~al.}(2004)\citenamefont {Hayden},
  \citenamefont {Jozsa}, \citenamefont {Petz},\ and\ \citenamefont
  {Winter}}]{hayden2004structure}%
  \BibitemOpen
  \bibfield  {author} {\bibinfo {author} {\bibfnamefont {Patrick}\ \bibnamefont
  {Hayden}}, \bibinfo {author} {\bibfnamefont {Richard}\ \bibnamefont {Jozsa}},
  \bibinfo {author} {\bibfnamefont {Denes}\ \bibnamefont {Petz}}, \ and\
  \bibinfo {author} {\bibfnamefont {Andreas}\ \bibnamefont {Winter}},\
  }\bibfield  {title} {\enquote {\bibinfo {title} {Structure of states which
  satisfy strong subadditivity of quantum entropy with equality},}\ }\href@noop
  {} {\bibfield  {journal} {\bibinfo  {journal} {Communications in Mathematical
  Physics}\ }\textbf {\bibinfo {volume} {246}},\ \bibinfo {pages} {359--374}
  (\bibinfo {year} {2004})}\BibitemShut {NoStop}%
\bibitem [{\citenamefont {Ng}\ and\ \citenamefont
  {Mandayam}(2010)}]{ng2010simple}%
  \BibitemOpen
  \bibfield  {author} {\bibinfo {author} {\bibfnamefont {Hui~Khoon}\
  \bibnamefont {Ng}}\ and\ \bibinfo {author} {\bibfnamefont {Prabha}\
  \bibnamefont {Mandayam}},\ }\bibfield  {title} {\enquote {\bibinfo {title}
  {Simple approach to approximate quantum error correction based on the
  transpose channel},}\ }\href@noop {} {\bibfield  {journal} {\bibinfo
  {journal} {Physical Review A}\ }\textbf {\bibinfo {volume} {81}},\ \bibinfo
  {pages} {062342} (\bibinfo {year} {2010})}\BibitemShut {NoStop}%
\bibitem [{\citenamefont {Fawzi}\ and\ \citenamefont
  {Renner}(2015)}]{fawzi2015quantum}%
  \BibitemOpen
  \bibfield  {author} {\bibinfo {author} {\bibfnamefont {Omar}\ \bibnamefont
  {Fawzi}}\ and\ \bibinfo {author} {\bibfnamefont {Renato}\ \bibnamefont
  {Renner}},\ }\bibfield  {title} {\enquote {\bibinfo {title} {Quantum
  conditional mutual information and approximate {M}arkov chains},}\
  }\href@noop {} {\bibfield  {journal} {\bibinfo  {journal} {Communications in
  Mathematical Physics}\ }\textbf {\bibinfo {volume} {340}},\ \bibinfo {pages}
  {575--611} (\bibinfo {year} {2015})}\BibitemShut {NoStop}%
\bibitem [{\citenamefont {Beigi}\ \emph {et~al.}(2016)\citenamefont {Beigi},
  \citenamefont {Datta},\ and\ \citenamefont {Leditzky}}]{beigi2016decoding}%
  \BibitemOpen
  \bibfield  {author} {\bibinfo {author} {\bibfnamefont {Salman}\ \bibnamefont
  {Beigi}}, \bibinfo {author} {\bibfnamefont {Nilanjana}\ \bibnamefont
  {Datta}}, \ and\ \bibinfo {author} {\bibfnamefont {Felix}\ \bibnamefont
  {Leditzky}},\ }\bibfield  {title} {\enquote {\bibinfo {title} {Decoding
  quantum information via the petz recovery map},}\ }\href@noop {} {\bibfield
  {journal} {\bibinfo  {journal} {Journal of Mathematical Physics}\ }\textbf
  {\bibinfo {volume} {57}},\ \bibinfo {pages} {082203} (\bibinfo {year}
  {2016})}\BibitemShut {NoStop}%
\bibitem [{\citenamefont {Junge}\ \emph {et~al.}(2018)\citenamefont {Junge},
  \citenamefont {Renner}, \citenamefont {Sutter}, \citenamefont {Wilde},\ and\
  \citenamefont {Winter}}]{junge2018universal}%
  \BibitemOpen
  \bibfield  {author} {\bibinfo {author} {\bibfnamefont {Marius}\ \bibnamefont
  {Junge}}, \bibinfo {author} {\bibfnamefont {Renato}\ \bibnamefont {Renner}},
  \bibinfo {author} {\bibfnamefont {David}\ \bibnamefont {Sutter}}, \bibinfo
  {author} {\bibfnamefont {Mark~M}\ \bibnamefont {Wilde}}, \ and\ \bibinfo
  {author} {\bibfnamefont {Andreas}\ \bibnamefont {Winter}},\ }\bibfield
  {title} {\enquote {\bibinfo {title} {Universal recovery maps and approximate
  sufficiency of quantum relative entropy},}\ }in\ \href@noop {} {\emph
  {\bibinfo {booktitle} {Annales Henri Poincar{\'e}}}},\ Vol.~\bibinfo {volume}
  {19}\ (\bibinfo {organization} {Springer},\ \bibinfo {year} {2018})\ pp.\
  \bibinfo {pages} {2955--2978}\BibitemShut {NoStop}%
\bibitem [{\citenamefont {Choi}(1975)}]{Cho75a}%
  \BibitemOpen
  \bibfield  {author} {\bibinfo {author} {\bibfnamefont {Man-Duen}\
  \bibnamefont {Choi}},\ }\bibfield  {title} {\enquote {\bibinfo {title}
  {Completely positive linear maps on complex matrices},}\ }\href@noop {}
  {\bibfield  {journal} {\bibinfo  {journal} {Lin. Alg. Appl.}\ }\textbf
  {\bibinfo {volume} {10}},\ \bibinfo {pages} {285} (\bibinfo {year}
  {1975})}\BibitemShut {NoStop}%
\bibitem [{\citenamefont {Bengtsson}\ and\ \citenamefont
  {\.{Z}yczkowski}(2017)}]{BZ}%
  \BibitemOpen
  \bibfield  {author} {\bibinfo {author} {\bibfnamefont {Ingemar}\ \bibnamefont
  {Bengtsson}}\ and\ \bibinfo {author} {\bibfnamefont {Karol}\ \bibnamefont
  {\.{Z}yczkowski}},\ }\href@noop {} {\emph {\bibinfo {title} {Geometry of
  Quantum States, II ed.}}}\ (\bibinfo  {publisher} {Cambridge University
  Press},\ \bibinfo {year} {2017})\BibitemShut {NoStop}%
\bibitem [{\citenamefont {Doebner}\ and\ \citenamefont
  {Goldin}(1992)}]{doebner1992general}%
  \BibitemOpen
  \bibfield  {author} {\bibinfo {author} {\bibfnamefont {H-D}\ \bibnamefont
  {Doebner}}\ and\ \bibinfo {author} {\bibfnamefont {Gerald~A}\ \bibnamefont
  {Goldin}},\ }\bibfield  {title} {\enquote {\bibinfo {title} {On a general
  nonlinear {S}chr{\"o}dinger equation admitting diffusion currents},}\
  }\href@noop {} {\bibfield  {journal} {\bibinfo  {journal} {Physics Letters
  A}\ }\textbf {\bibinfo {volume} {162}},\ \bibinfo {pages} {397--401}
  (\bibinfo {year} {1992})}\BibitemShut {NoStop}%
\bibitem [{\citenamefont {Doebner}\ \emph {et~al.}(1999)\citenamefont
  {Doebner}, \citenamefont {Goldin},\ and\ \citenamefont
  {Nattermann}}]{doebner1999gauge}%
  \BibitemOpen
  \bibfield  {author} {\bibinfo {author} {\bibfnamefont {H-D}\ \bibnamefont
  {Doebner}}, \bibinfo {author} {\bibfnamefont {GA}~\bibnamefont {Goldin}}, \
  and\ \bibinfo {author} {\bibfnamefont {P}~\bibnamefont {Nattermann}},\
  }\bibfield  {title} {\enquote {\bibinfo {title} {Gauge transformations in
  quantum mechanics and the unification of nonlinear {S}chr{\"o}dinger
  equations},}\ }\href@noop {} {\bibfield  {journal} {\bibinfo  {journal}
  {Journal of Mathematical Physics}\ }\textbf {\bibinfo {volume} {40}},\
  \bibinfo {pages} {49--63} (\bibinfo {year} {1999})}\BibitemShut {NoStop}%
\bibitem [{\citenamefont {Goldin}(2008)}]{goldin2008nonlinear}%
  \BibitemOpen
  \bibfield  {author} {\bibinfo {author} {\bibfnamefont {GA}~\bibnamefont
  {Goldin}},\ }\bibfield  {title} {\enquote {\bibinfo {title} {Nonlinear
  quantum mechanics: Results and open questions},}\ }\href@noop {} {\bibfield
  {journal} {\bibinfo  {journal} {Physics of Atomic Nuclei}\ }\textbf {\bibinfo
  {volume} {71}},\ \bibinfo {pages} {884} (\bibinfo {year} {2008})}\BibitemShut
  {NoStop}%
\bibitem [{\citenamefont {Cavalcanti}\ \emph {et~al.}(2012)\citenamefont
  {Cavalcanti}, \citenamefont {Menicucci},\ and\ \citenamefont
  {Pienaar}}]{cavalcanti2012preparation}%
  \BibitemOpen
  \bibfield  {author} {\bibinfo {author} {\bibfnamefont {Eric~G}\ \bibnamefont
  {Cavalcanti}}, \bibinfo {author} {\bibfnamefont {Nicolas~C}\ \bibnamefont
  {Menicucci}}, \ and\ \bibinfo {author} {\bibfnamefont {Jacques~L}\
  \bibnamefont {Pienaar}},\ }\bibfield  {title} {\enquote {\bibinfo {title}
  {The preparation problem in nonlinear extensions of quantum theory},}\
  }\href@noop {} {\bibfield  {journal} {\bibinfo  {journal} {arXiv preprint
  arXiv:1206.2725}\ } (\bibinfo {year} {2012})}\BibitemShut {NoStop}%
\bibitem [{\citenamefont {Crooks}(2008)}]{crooks2008quantum}%
  \BibitemOpen
  \bibfield  {author} {\bibinfo {author} {\bibfnamefont {Gavin~E}\ \bibnamefont
  {Crooks}},\ }\bibfield  {title} {\enquote {\bibinfo {title} {Quantum
  operation time reversal},}\ }\href@noop {} {\bibfield  {journal} {\bibinfo
  {journal} {Physical Review A}\ }\textbf {\bibinfo {volume} {77}},\ \bibinfo
  {pages} {034101} (\bibinfo {year} {2008})}\BibitemShut {NoStop}%
\bibitem [{\citenamefont {Saunders}\ \emph {et~al.}(2010)\citenamefont
  {Saunders}, \citenamefont {Barrett}, \citenamefont {Kent},\ and\
  \citenamefont {Wallace}}]{saunders2010many}%
  \BibitemOpen
  \bibfield  {author} {\bibinfo {author} {\bibfnamefont {Simon}\ \bibnamefont
  {Saunders}}, \bibinfo {author} {\bibfnamefont {Jonathan}\ \bibnamefont
  {Barrett}}, \bibinfo {author} {\bibfnamefont {Adrian}\ \bibnamefont {Kent}},
  \ and\ \bibinfo {author} {\bibfnamefont {David}\ \bibnamefont {Wallace}},\
  }\href@noop {} {\emph {\bibinfo {title} {Many worlds?: Everett, quantum
  theory, \& reality}}}\ (\bibinfo  {publisher} {Oxford University Press},\
  \bibinfo {year} {2010})\BibitemShut {NoStop}%
\bibitem [{\citenamefont {Uhlmann}(1976)}]{Uhl}%
  \BibitemOpen
  \bibfield  {author} {\bibinfo {author} {\bibfnamefont {Armin}\ \bibnamefont
  {Uhlmann}},\ }\bibfield  {title} {\enquote {\bibinfo {title} {The
  ``transition probability" in the space of a $*$-algebra},}\ }\href@noop {}
  {\bibfield  {journal} {\bibinfo  {journal} {Rep. Math. Phys.}\ }\textbf
  {\bibinfo {volume} {9}},\ \bibinfo {pages} {273--279} (\bibinfo {year}
  {1976})}\BibitemShut {NoStop}%
\bibitem [{\citenamefont {Jozsa}(1994)}]{Jozsa}%
  \BibitemOpen
  \bibfield  {author} {\bibinfo {author} {\bibfnamefont {Richard}\ \bibnamefont
  {Jozsa}},\ }\bibfield  {title} {\enquote {\bibinfo {title} {Fidelity for
  mixed quantum states},}\ }\href@noop {} {\bibfield  {journal} {\bibinfo
  {journal} {J. Modern Optics}\ }\textbf {\bibinfo {volume} {41}},\ \bibinfo
  {pages} {2315--2323} (\bibinfo {year} {1994})}\BibitemShut {NoStop}%
\bibitem [{\citenamefont {Chefles}\ \emph {et~al.}(2004)\citenamefont
  {Chefles}, \citenamefont {Jozsa},\ and\ \citenamefont
  {Winter}}]{chefles2004existence}%
  \BibitemOpen
  \bibfield  {author} {\bibinfo {author} {\bibfnamefont {Anthony}\ \bibnamefont
  {Chefles}}, \bibinfo {author} {\bibfnamefont {Richard}\ \bibnamefont
  {Jozsa}}, \ and\ \bibinfo {author} {\bibfnamefont {Andreas}\ \bibnamefont
  {Winter}},\ }\bibfield  {title} {\enquote {\bibinfo {title} {On the existence
  of physical transformations between sets of quantum states},}\ }\href@noop {}
  {\bibfield  {journal} {\bibinfo  {journal} {International Journal of Quantum
  Information}\ }\textbf {\bibinfo {volume} {2}},\ \bibinfo {pages} {11--21}
  (\bibinfo {year} {2004})}\BibitemShut {NoStop}%
\bibitem [{\citenamefont {Holevo}(2011)}]{holevo2011probabilistic}%
  \BibitemOpen
  \bibfield  {author} {\bibinfo {author} {\bibfnamefont {Alexander~S}\
  \bibnamefont {Holevo}},\ }\href@noop {} {\emph {\bibinfo {title}
  {Probabilistic and statistical aspects of quantum theory}}},\ Vol.~\bibinfo
  {volume} {1}\ (\bibinfo  {publisher} {Springer Science \& Business Media},\
  \bibinfo {year} {2011})\BibitemShut {NoStop}%
\bibitem [{\citenamefont {Leung}(2000)}]{leung2000towards}%
  \BibitemOpen
  \bibfield  {author} {\bibinfo {author} {\bibfnamefont {Debbie~W}\
  \bibnamefont {Leung}},\ }\bibfield  {title} {\enquote {\bibinfo {title}
  {Towards robust quantum computation},}\ }\href@noop {} {\bibfield  {journal}
  {\bibinfo  {journal} {arXiv preprint cs/0012017}\ } (\bibinfo {year}
  {2000})}\BibitemShut {NoStop}%
\bibitem [{\citenamefont {D'Ariano}\ and\ \citenamefont
  {Lo~Presti}(2001)}]{d2001quantum}%
  \BibitemOpen
  \bibfield  {author} {\bibinfo {author} {\bibfnamefont {Giacomo~Mauro}\
  \bibnamefont {D'Ariano}}\ and\ \bibinfo {author} {\bibfnamefont
  {Paoloplacido}\ \bibnamefont {Lo~Presti}},\ }\bibfield  {title} {\enquote
  {\bibinfo {title} {Quantum tomography for measuring experimentally the matrix
  elements of an arbitrary quantum operation},}\ }\href@noop {} {\bibfield
  {journal} {\bibinfo  {journal} {Physical Review Letters}\ }\textbf {\bibinfo
  {volume} {86}},\ \bibinfo {pages} {4195} (\bibinfo {year}
  {2001})}\BibitemShut {NoStop}%
\bibitem [{\citenamefont {Arrighi}\ and\ \citenamefont
  {Patricot}(2004)}]{AP04}%
  \BibitemOpen
  \bibfield  {author} {\bibinfo {author} {\bibfnamefont {Pablo}\ \bibnamefont
  {Arrighi}}\ and\ \bibinfo {author} {\bibfnamefont {Christophe}\ \bibnamefont
  {Patricot}},\ }\bibfield  {title} {\enquote {\bibinfo {title} {On quantum
  operations as quantum states},}\ }\href@noop {} {\bibfield  {journal}
  {\bibinfo  {journal} {Annals Phys.}\ }\textbf {\bibinfo {volume} {311}},\
  \bibinfo {pages} {26} (\bibinfo {year} {2004})}\BibitemShut {NoStop}%
\bibitem [{\citenamefont {{\.Z}yczkowski}\ and\ \citenamefont
  {Bengtsson}(2004)}]{ZB04}%
  \BibitemOpen
  \bibfield  {author} {\bibinfo {author} {\bibfnamefont {Karol}\ \bibnamefont
  {{\.Z}yczkowski}}\ and\ \bibinfo {author} {\bibfnamefont {Ingemar}\
  \bibnamefont {Bengtsson}},\ }\bibfield  {title} {\enquote {\bibinfo {title}
  {On duality between quantum states and quantum maps},}\ }\href@noop {}
  {\bibfield  {journal} {\bibinfo  {journal} {Open Syst. Inf. Dyn.}\ }\textbf
  {\bibinfo {volume} {11}},\ \bibinfo {pages} {3} (\bibinfo {year}
  {2004})}\BibitemShut {NoStop}%
\bibitem [{\citenamefont {Abramsky}\ and\ \citenamefont
  {Coecke}(2004)}]{abramsky2004categorical}%
  \BibitemOpen
  \bibfield  {author} {\bibinfo {author} {\bibfnamefont {Samson}\ \bibnamefont
  {Abramsky}}\ and\ \bibinfo {author} {\bibfnamefont {Bob}\ \bibnamefont
  {Coecke}},\ }\bibfield  {title} {\enquote {\bibinfo {title} {A categorical
  semantics of quantum protocols},}\ }in\ \href@noop {} {\emph {\bibinfo
  {booktitle} {Proceedings of the 19th Annual IEEE Symposium on Logic in
  Computer Science, 2004.}}}\ (\bibinfo {organization} {IEEE},\ \bibinfo {year}
  {2004})\ pp.\ \bibinfo {pages} {415--425}\BibitemShut {NoStop}%
\bibitem [{\citenamefont {Jamio{\l}kowski}(1972)}]{Ja72}%
  \BibitemOpen
  \bibfield  {author} {\bibinfo {author} {\bibfnamefont {Andrzej}\ \bibnamefont
  {Jamio{\l}kowski}},\ }\bibfield  {title} {\enquote {\bibinfo {title} {Linear
  transformations which preserve trace and positive semi-definiteness of
  operators},}\ }\href@noop {} {\bibfield  {journal} {\bibinfo  {journal} {Rep.
  Math. Phys.}\ }\textbf {\bibinfo {volume} {3}},\ \bibinfo {pages} {275}
  (\bibinfo {year} {1972})}\BibitemShut {NoStop}%
\bibitem [{\citenamefont {Chiribella}\ \emph
  {et~al.}(2008{\natexlab{c}})\citenamefont {Chiribella}, \citenamefont
  {D'Ariano},\ and\ \citenamefont {Perinotti}}]{chiribella2008memory}%
  \BibitemOpen
  \bibfield  {author} {\bibinfo {author} {\bibfnamefont {Giulio}\ \bibnamefont
  {Chiribella}}, \bibinfo {author} {\bibfnamefont {Giacomo~M}\ \bibnamefont
  {D'Ariano}}, \ and\ \bibinfo {author} {\bibfnamefont {Paolo}\ \bibnamefont
  {Perinotti}},\ }\bibfield  {title} {\enquote {\bibinfo {title} {Memory
  effects in quantum channel discrimination},}\ }\href@noop {} {\bibfield
  {journal} {\bibinfo  {journal} {Physical Review Letters}\ }\textbf {\bibinfo
  {volume} {101}},\ \bibinfo {pages} {180501} (\bibinfo {year}
  {2008}{\natexlab{c}})}\BibitemShut {NoStop}%
\bibitem [{\citenamefont
  {Chiribella}(2014)}]{chiribella2014distinguishability}%
  \BibitemOpen
  \bibfield  {author} {\bibinfo {author} {\bibfnamefont {Giulio}\ \bibnamefont
  {Chiribella}},\ }\bibfield  {title} {\enquote {\bibinfo {title}
  {Distinguishability and copiability of programs in general process
  theories},}\ }\href@noop {} {\bibfield  {journal} {\bibinfo  {journal}
  {International Journal of Software and Informatics}\ }\textbf {\bibinfo
  {volume} {8}},\ \bibinfo {pages} {209} (\bibinfo {year} {2014})}\BibitemShut
  {NoStop}%
\bibitem [{\citenamefont {Gour}(2019)}]{gour2019comparison}%
  \BibitemOpen
  \bibfield  {author} {\bibinfo {author} {\bibfnamefont {Gilad}\ \bibnamefont
  {Gour}},\ }\bibfield  {title} {\enquote {\bibinfo {title} {Comparison of
  quantum channels by superchannels},}\ }\href@noop {} {\bibfield  {journal}
  {\bibinfo  {journal} {IEEE Transactions on Information Theory}\ }\textbf
  {\bibinfo {volume} {65}},\ \bibinfo {pages} {5880--5904} (\bibinfo {year}
  {2019})}\BibitemShut {NoStop}%
\bibitem [{\citenamefont {Burniston}\ \emph {et~al.}(2020)\citenamefont
  {Burniston}, \citenamefont {Grabowecky}, \citenamefont {Scandolo},
  \citenamefont {Chiribella},\ and\ \citenamefont
  {Gour}}]{burniston2020necessary}%
  \BibitemOpen
  \bibfield  {author} {\bibinfo {author} {\bibfnamefont {John}\ \bibnamefont
  {Burniston}}, \bibinfo {author} {\bibfnamefont {Michael}\ \bibnamefont
  {Grabowecky}}, \bibinfo {author} {\bibfnamefont {Carlo~Maria}\ \bibnamefont
  {Scandolo}}, \bibinfo {author} {\bibfnamefont {Giulio}\ \bibnamefont
  {Chiribella}}, \ and\ \bibinfo {author} {\bibfnamefont {Gilad}\ \bibnamefont
  {Gour}},\ }\bibfield  {title} {\enquote {\bibinfo {title} {Necessary and
  sufficient conditions on measurements of quantum channels},}\ }\href@noop {}
  {\bibfield  {journal} {\bibinfo  {journal} {Proceedings of the Royal Society
  A}\ }\textbf {\bibinfo {volume} {476}},\ \bibinfo {pages} {20190832}
  (\bibinfo {year} {2020})}\BibitemShut {NoStop}%
\end{thebibliography}%

\end{document}